\documentclass[12pt, twocolumn]{aastex61}

\usepackage{multirow}
\usepackage{ulem}
\renewcommand{\vec}[1]{ {\mathbf #1} }

\accepted{\today}

\shorttitle{NLFFF--MHD CASTING}
\shortauthors{Fleishman et al.}

\begin{document}


\title{ Casting the Coronal Magnetic Field Reconstruction Tools  in 3D Using MHD Bifrost Model}

\author{Gregory Fleishman$^1$}
%
%
\author{Sergey Anfinogentov$^2$}
%
\author{Maria Loukitcheva$^{1,3,4}$}
%
%
\author{Ivan Mysh'yakov$^2$}
%
%
\author{Alexey Stupishin$^3$}

\altaffiliation{$^1$Physics Department, Center for Solar-Terrestrial Research, New Jersey Institute of Technology
Newark, NJ, 07102-1982}

\altaffiliation{$^2$Institute of Solar-Terrestrial Physics  (ISZF), Lermontov st., 126a, Irkutsk, 664033  Russia}

\altaffiliation{$^3$Saint Petersburg State University, 7/9 Universitetskaya nab., St.
Petersburg, 199034 Russia}

\altaffiliation{$^4$Max-Planck-Institut f\"ur Sonnensystemforschung, Justus-von-Liebig-Weg 3, 37077 G\"ottingen, Germany}


\begin{abstract}
Quantifying coronal magnetic field remains a central problem in solar physics. Nowadays the coronal magnetic field is often modelled using nonlinear force-free field (NLFFF) reconstructions, whose accuracy has not yet been comprehensively assessed. Here we perform a detailed casting of the NLFFF reconstruction tools, such as $\pi$-disambiguation, photospheric field preprocessing, and volume reconstruction methods using a 3D snapshot of the publicly available full-fledged radiative MHD model. Specifically, from the MHD model we know the magnetic field vector in the entire 3D domain, which enables us to perform "voxel-by-voxel" comparison of the restored and the true magnetic field in the 3D model volume. Our tests show that the available $\pi$-disambiguation methods often fail at the quiet sun areas dominated by small-scale magnetic elements, while they work well at the AR photosphere and (even better) chromosphere. The preprocessing of the photospheric magnetic field, although does produce a more force-free boundary condition, also results in some effective `elevation' of the magnetic field components. This `elevation' height is different for the longitudinal and transverse components, which results in a systematic error in absolute heights in the reconstructed magnetic data cube. The extrapolations performed starting from actual AR photospheric magnetogram are free from this systematic error, while have other metrics comparable with those for extrapolations from the preprocessed magnetograms. This finding favors the use of extrapolations from the original photospheric magnetogram without preprocessing. Our tests further suggest that extrapolations from a force-free chromospheric boundary produce measurably better results, than those from the photospheric boundary.

\end{abstract}


\section{Introduction}


The magnetic structure of the solar corona plays a key role in all of the solar activity, yet direct measurements of the coronal magnetic field are extremely difficult to make. Instead, the field strength and direction are measured at the photospheric (or possibly chromospheric) boundary; specifically, the vector fields are measured from full-Stokes polarized intensity of Zeeman sensitive spectral lines with circular polarization providing line-of-sight field strength and linear polarization providing the transverse field. Then, to assess the coronal magnetic field, these measured photospheric fields are extended into the corona through potential or force-free field extrapolations. Starting with \cite{1981SoPh...69..343S}, 
these extrapolations have reached nowadays a high level of sophistication \citep[see reviews by][]{1989SSRv...51...11S, 1997SoPh..174..129A, 2008JGRA..113.3S02W, 2012LRSP....9....5W, 2014A&ARv..22...78W}. 

Although most of the extrapolation methods work well on reproducing available analytical NLFFF solutions \citep[e.g.,][]{1990ApJ...352..343L}, which are static and relatively simple morphologically, the assessment of the methods on realistic dynamical cases having a lot of spatial complexity is still limited \citep{2009ApJ...696.1780D}. Since a straightforward 'voxel-by-voxel' comparison between the NLFFF reconstructed data cube and real data cannot be performed given the lack of the coronal magnetic field diagnostics, a number of indirect tests is typically employed. These can be comparison of a subset of a selected magnetic field lines with the EUV loops or computing various metrics describing the field \lq force-freeness\rq. However, comparison with the bright EUV loops allows for only a morphological comparison and only for a small fraction of the volume, where such loops are present, but does not allow quantitative assessment of the magnetic field vector components. Force-free metrics computed for various instances of the extrapolated data cubes are comparably imperfect and, thus, they do not offer an easy way of favoring one or another extrapolation method. \citet{2015ApJ...811..107D} considered the effect of the spatial resolution of the base magnetogram on the quality of the coronal magnetic field reconstruction with alternative extrapolation methods. They found that the result of an extrapolation is sensitive to the spatial resolution of the input base map and the alternative codes yield noticeably different magnetic data cubes. Thus, the casting of the competing extrapolation methods using available coronal diagnostics remains inconclusive, since no true comparison of the reconstructed and real magnetic field can be performed in 3D domain. Nevertheless, direct tests of the NLFFF coronal reconstruction tool performance with the use of modern, highly realistic spatially complex 3-dimensional (3D) MHD models of a fragment of solar atmosphere can be performed and, thus, are called for.

In this study we employ a publicly available outcome \citep[\textit{en024048{\_}hion}, \url{http://sdc.uio.no/search/simulations},][]{2016A&A...585A...4C}  of an advanced 3D radiation magnetohydrodynamic (RMHD) code Bifrost \citep{2011A&A...531A.154G} to perform comprehensive tests of various steps used for the coronal magnetic modeling: (i) $\pi$-disambiguation of the measured transverse magnetic field; (ii) preprocessing of the photospheric magnetic field; and (iii) NLFFF extrapolation from the bottom boundary up to the coronal volume. In each step we make a comparison between available alternative codes: either two different methods or two alternative implementations of the same methods for the steps mentioned above.


\section{RMHD Model Data Cube}

Bifrost is a flexible and massively parallel code for general purposes developed by \citet{2011A&A...531A.154G}. The Bifrost-based RMHD Model of \citet{2016A&A...585A...4C}, \textit{en024048{\_}hion}, represents realistic simulations of the solar outer atmosphere with a magnetic field topology similar to an enhanced network area on the Sun. The simulation data cubes used in this work were obtained from the Hinode Science Data Centre
Europe for the publicly available timespan of 1590~s.

\subsection{Overview of the Data Cube}
\label{S_dcube_overview}

The simulation covers a physical extent of $24~\times~24~\times~16.8$~Mm, with a grid of $504~\times~504~\times~496$ cells, extending from --2.4~Mm below the photosphere, which corresponds to $Z=0$~Mm, to 14.4~Mm above the photosphere. It encompasses the upper convection zone, photosphere, chromosphere, and the lower corona. The horizontal axes have an equidistant grid spacing of 48~km, the vertical grid spacing is non-uniform, with a spacing of 19~km between $Z = -1$~Mm and $Z = 5$~Mm. The spacing increases toward the top of the computational domain to a maximum of 98~km. The magnetic topology is defined by two opposite polarity
patches separated by 8~Mm with an average unsigned strength of  $\sim$50~G in the photosphere, representing two patches of quiet-Sun network.
The simulation includes optically thick radiative transfer in the photosphere and low chromosphere, parameterized radiative losses in the upper chromosphere, transition region and corona, thermal conduction along magnetic field lines, and an equation of state that includes the effects of non-equilibrium ionization of hydrogen. The Bifrost code solves the equations of resistive MHD on a staggered Cartesian grid but the published data cubes for \textit{en024048{\_}hion} have the variables specified at the cell centers.

One of the simulation snapshots, the snapshot 385, has already been used in a series of papers to investigate the formation of the IRIS diagnostics \citep{2013ApJ...772...89L, 2013ApJ...772...90L, 2013ApJ...778..143P, 2015ApJ...806...14P, 2015ApJ...811...80R, 2015ApJ...811...81R, 2015ApJ...813...34L}, the formation of other chromospheric lines \citep{2012ApJ...749..136L,2015ApJ...802..136L,2013ApJ...764L..11D,2012ApJ...758L..43S,2015ApJ...803...65S}, as well as to model continuum free-free emission from solar chromosphere \citep{2015A&A...575A..15L} with the perspective of ALMA observations \citep{2015ASPC..499..349L, 2017arXiv170206018L}.


As has been said, the available  \textit{en024048{\_}hion} data cubes are specified on a nonuniform grid in the vertical direction, although most of the NLFFF approaches employ a regular grid, which has a number of computational advances including uniform precision of the spatial derivatives. In order to make the \textit{en024048{\_}hion} model compatible with the extrapolated data cubes,
the original model grid was transformed to a uniform grid (with $48$~km step along all three axes) by applying linear interpolation in vertical direction.

\begin{figure}
    \centering
    \includegraphics[width=0.98\columnwidth]{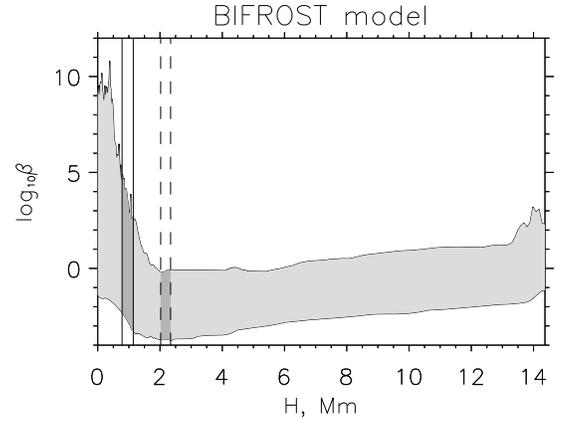}\\
    \caption{\label{f_beta_vs_height} Distribution of maximal and minimal values of plasma $\beta$ at different altitudes. Vertical solid lines mark altitude range where $\beta$-photosphere levels for different bin factors are located. Dashed vertical lines mark range of chromosphere levels. }
\end{figure}

Given that the applicability of the force-free magnetic modeling relies on the dominant role of the Lorentz force in the overall balance of forces, we start with analysis of height distribution of the plasma parameter $\beta=p_{\rm kin}/p_{\rm B}$ in the model volume. In Figure~\ref{f_beta_vs_height} we plot the range of the $\beta$ values as a function of the model height using the original nonuniform grid and demarcate a number of characteristic levels for further references. Level $H=0$ corresponds to the nominal photosphere in the model data cube. Since the model has been created to describe an enhanced network of the quiet sun, rather than a typical AR, the $\beta$ values at the level of the nominal photosphere are, not surprisingly, much larger than those for an AR \citep{2001SoPh..203...71G}. Such a boundary condition, having an overall weak magnetic field with small-scale inserts of a larger field, represents a big challenge for the currently available methods of the coronal force-free modeling designed for ARs with relatively large-scale areas of the strong magnetic field, where the plasma parameter $\beta$ although larger than one, is not exceptionally large.

Therefore, to emulate a photospheric boundary representative of a typical AR, we select a higher level, where the distribution of the plasma $\beta$ is similar to that in a typical AR \citep{2001SoPh..203...71G}; we call this level a `typical AR $\beta$ photosphere,' or just $\beta$-photosphere for short.
Specifically, we define this layer as the lowest layer above the nominal photosphere, where 95$\%$ of the nodes have $\beta<10^{2}$.
This level roughly corresponds to $H_{\beta-ph}\approx1$~Mm; the exact height varies between the data cubes obtained  with different binning factors (see section~\ref{S_dcube_setup}). Then, we define a chromospheric interface as the lowest layer where 95$\%$ of the nodes have $\beta<10^{-1}$,
which is supposed to offer an ideal boundary condition for the force-free modeling. This level roughly corresponds to the height $H_{chr}\approx2.5$~Mm. We return to more specific definition of these levels in the next section, while introducing rebinned data cubes with a lower spatial resolution.





\subsection{Setup for the Testing}
\label{S_dcube_setup}

From the original magnetic data cube we initially created a new, uniform data cube with full resolution over $x,~y$ coordinates and a regular grid in $z$ direction with the voxel height of 48~km equal to the pixel sizes in $x,~y$ direction, while entirely discarding the subphotospheric part of the data cube. This yields a new full-resolution data cube with a regular grid having cubic voxels, convenient for further manipulation and analysis, to which we refer to as  `original regular' data cube hereafter.

This new original data cube was then rebinned to produce lower-resolution data cubes\footnote{ All these data cubes with regular spacing as well as standard deviations are available at our project web-site: \url{http://www.ioffe.ru/LEA/SF_AR/files/Magnetic_data_cubes/Bifrost/index.html}} with the binning factors $n=2,~3,~4,~6,~7,~9$. Apparently, a lower-resolution voxel includes $n^3$ original voxels. Therefore,  for each new voxel we assign the mean values of the magnetic field components to represent these values in the center of the new voxel such as
\begin{equation}\label{Eq_B_mean_def}
    \overline{B}_{\alpha}=\frac{1}{n^3}\sum\limits_{i=1}^{n^3} B_{\alpha}[i], \qquad \alpha=x,~y,~{\rm or}~z,
\end{equation}
and the standard deviations $\delta B_{\alpha}$, where
\begin{equation}\label{Eq_B_var_def}
   \delta B_{\alpha}^2=\frac{1}{n^3-1}\sum\limits_{i=1}^{n^3} (B_{\alpha}[i]-\overline{B}_{\alpha})^2, \qquad \alpha=x,~y,~{\rm or}~z
   .
\end{equation}



\begin{figure}[!h]\centering
\includegraphics[width=1.\columnwidth]{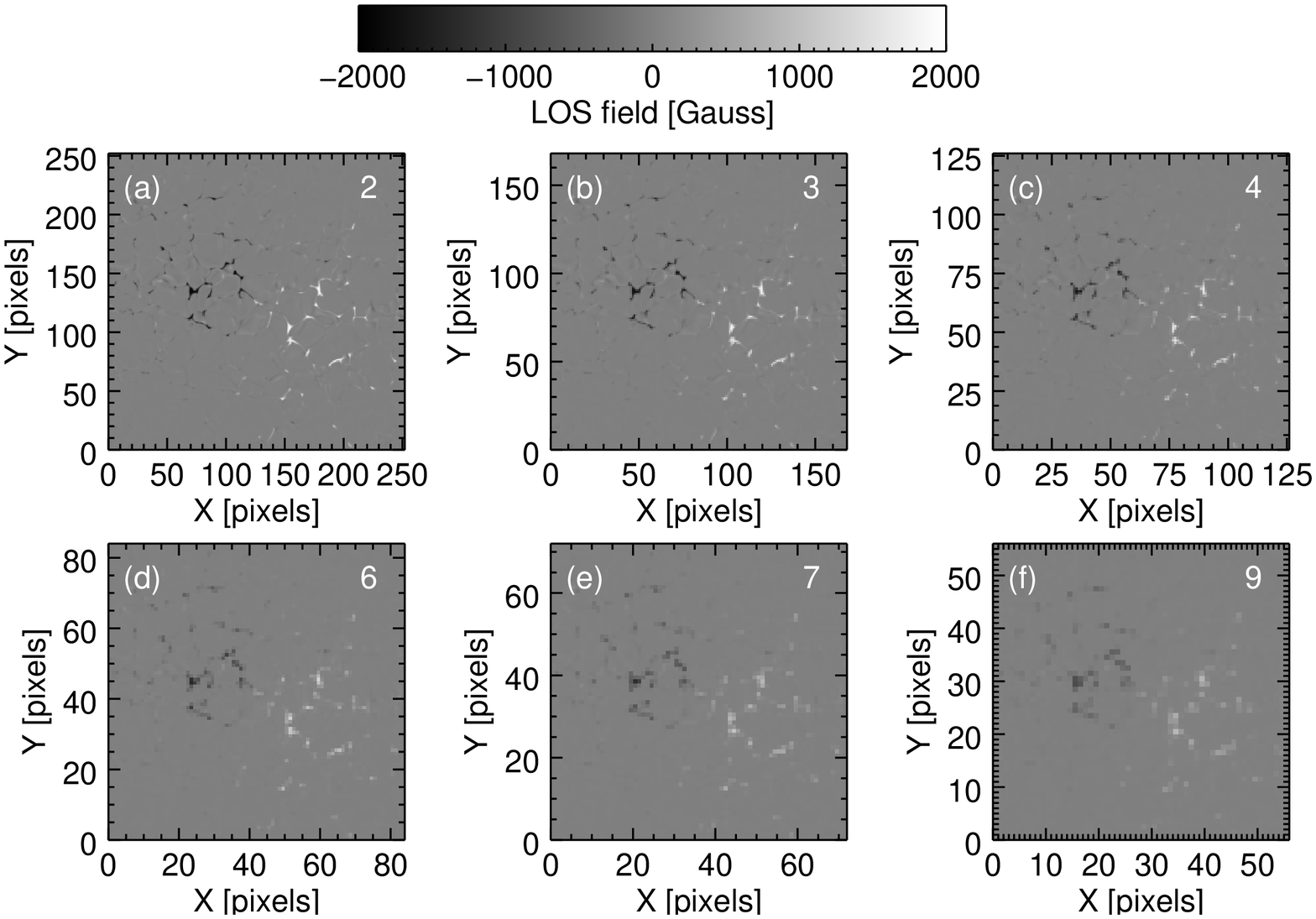}\\
\includegraphics[width=0.96\columnwidth]{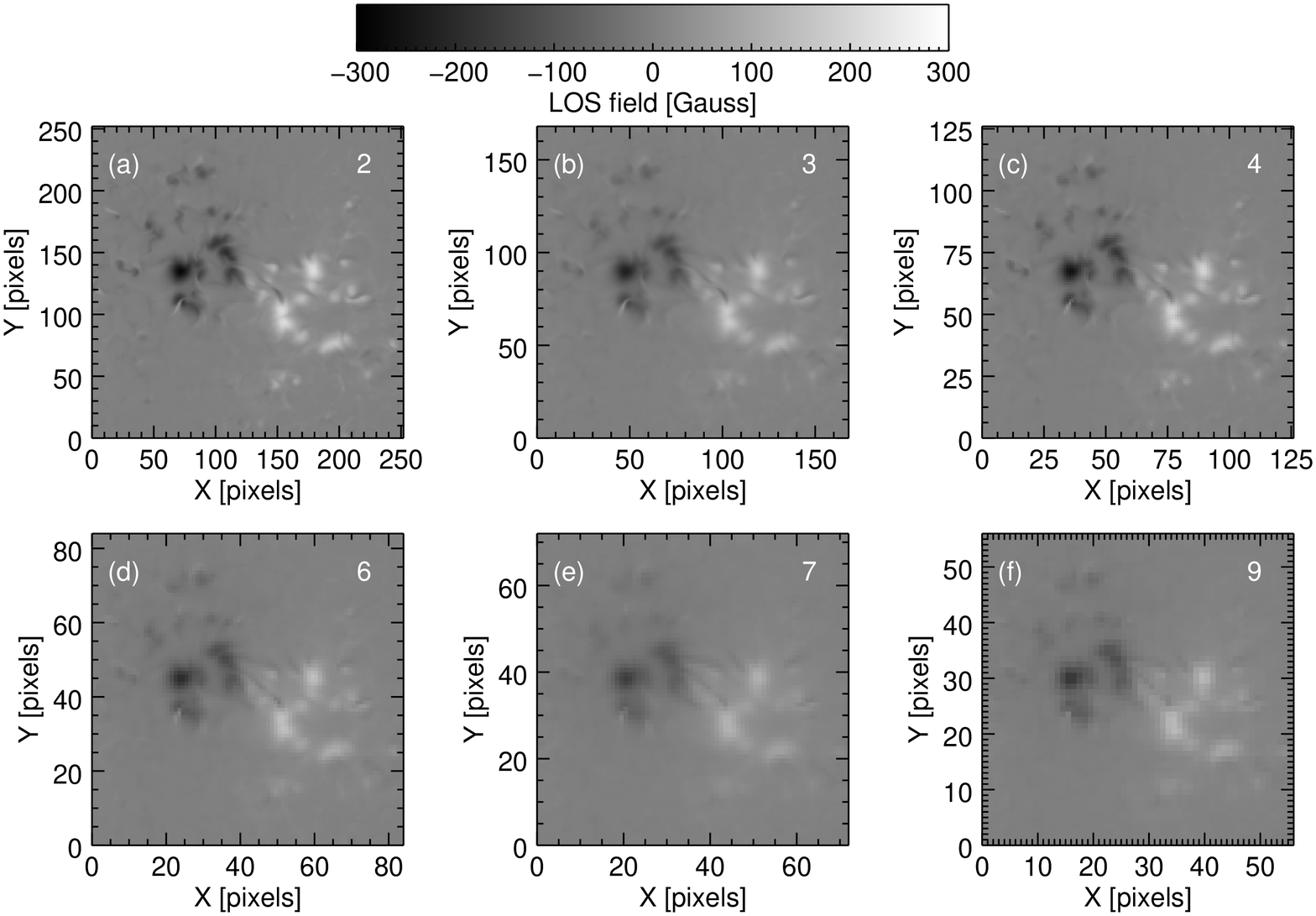}\\
\includegraphics[width=0.96\columnwidth]{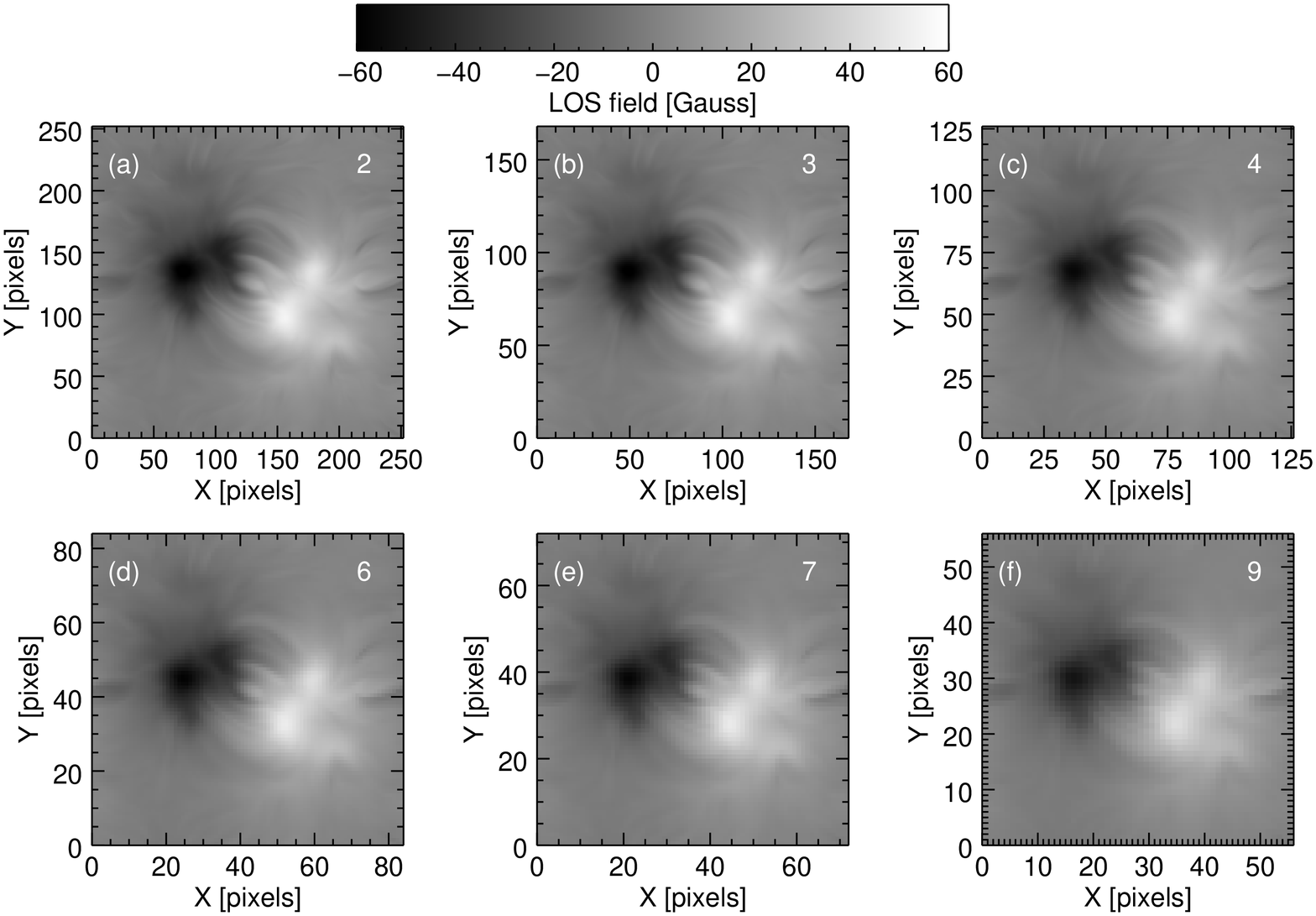}
\caption{\label{f_bifrost_385_rebin_ph0} Line-of-sight ($B_z$) magnetograms at the level of nominal photosphere (top), $\beta$-photosphere (middle), and chromosphere (bottom) in the magnetic data cubes obtained from the original Bifrost model by rebinning the volume by factors of 2, 3, 4, 6, 7, and 9.  Each set of six panels uses the same scale shown above the set. 
 }
\end{figure}

This set of the data cubes with different resolutions represents the input for our tests. The distributions of $z$-component of the magnetic field for different binning factors at the nominal and $\beta$-photospheres and at the chromosphere are shown in Figure~\ref{f_bifrost_385_rebin_ph0}.

%

\subsection{ Volume Properties of the Regular Data Cubes}
\label{S_dcube_analysis}

There are a number of numerical characteristics, similar to those introduced by \citet{2000ApJ...540.1150W,2006SoPh..235..161S}, which are used to evaluate to what extent the reconstructed magnetic field matches the force-free criterion:
\begin{equation}\label{IM_E01}
    \theta
    =
    arcsin
    \left(
        \frac{
            \sum_{i}^{N}
            \sigma_{i}
        }
        {
            N
        }
    \right)
    ,
    \quad
    \theta_{j}
    =
    arcsin
    \left(
        \frac{
            \sum_{i}^{N}
            \left|
                \textit{\textbf{j}}
            \right|_{i}
            \sigma_{i}
        }
        {
            \sum_{i}^{N}
            \left|
                \textit{\textbf{j}}
            \right|_{i}
        }
    \right)
    ,
    \quad $$$$
    \sigma_{i}
    =
    \frac{
    \left|
        \textit{\textbf{j}}
        \times
        \textit{\textbf{B}}
    \right|_{i}
    }
    {
        \left|
            \textit{\textbf{j}}
        \right|_{i}
        \left|
            \textit{\textbf{B}}
        \right|_{i}
    },
\end{equation}
and the divergence-free criterion:
\begin{equation}\label{IM_E02}
    f
    =
    \frac{
        1
    }
    {
        N
    }
    \sum_{i}^{N}
    \frac{
        \left|
            \nabla
            \textit{\textbf{B}}
        \right|_{i}
    }
    {
        6
        \left|
            \textit{\textbf{B}}
        \right|_{i}
    }
    dx,
\end{equation}
where $\textit{\textbf{B}}$ is the model field;  $\textbf{\textit{j}}$ is the corresponding electric current density,  the summation is performed over the voxels of the computational grid (excluding boundaries), \textit{dx} is the grid spacing; $\sigma_i$ is the sine of the angle between the magnetic field and the current density at the $i$-th node of the computational grid; $\theta$ is the angle averaged over all nodes, in the ideal case of the force-free field it must be zero; $\theta_j$ is a similar metrics but weighted with the electric current that means that contributions form strong currents dominate this metrics; $f$ is a parameter characterizing the accuracy with which the magnetic field is divergence-free.

Here we compute these metrics for the original regular data cube and for the data cubes obtained after the rebinning. 
Table \ref{IM_T01} presents the metrics, calculated for the data cube set. For each data cube we present three sets of metrics. The first one is calculated for the entire data cube volume starting from the nominal photosphere. The two others are for subdomains that start at either $\beta$-photosphere or 
chromosphere levels. 
Independently of the binning factor, each metrics displays  the same trend as a function of the subdomain used for the computation: from the full data cube to the chromosphere-limited subdomain,  $\theta$ shows a minor decrease, while $\theta_{j}$ decreases significantly, which is not surprising.  Indeed, $\theta_{j}$ is weighted with the current density, while the strongest currents concentrate near the bottom boundary. If this bottom boundary is located below the chromosphere, then the contribution from the non-force-free field region dominates. Above the chromosphere the field corresponds to a force-free configuration with a reasonably high accuracy, though not exactly. It is interesting, that the divergence-free parameter $f$ also improves when only the subdomain located above the chromosphere is taken into account, which, perhaps, originates from regridding  the original nonuniform vertical grid to the uniform one used here. In addition, the prominent small-scale structure of magnetic field at the lower levels results in  more significant errors in differential schemes used here to compute spatial derivatives. 

All metrics are getting worse with the increase of the binning factor. The reason for that is the unavoidable increase of  numerical errors of derivative computation.
Indeed, rebinning the model data causes distortions and, thus, an increase of the  divergence-free  metrics $f$. 

\begin{deluxetable*}{ c c c c c c }

    \tabletypesize{ \small }
    \tablewidth{ 0pt }
    \tablecaption{ Numerical characteristics of different domains depending on binning factor.\label{IM_T01} }
    \tablehead{
        \colhead{ Binning factor  } & 
        \colhead{ Volume above} & \colhead{layer \# } & \colhead{ $\theta^{\circ}$}  & \colhead{ $\theta _{j}^{\circ}$}  & \colhead{  {$f\times 10^{6}$} }
    }

    \startdata
        \multirow{3}{2em}{ 1 } & Nominal photosphere &  0 & 19.23 & 48.60 &  568 \\
                               & $\beta$-photosphere & 17 & 17.21 & 18.65 &  158 \\
                               & chromosphere        & 42 & 16.23 &  5.59 &  107 \\
        \\
        \multirow{3}{2em}{ 2 } & Nominal photosphere &  0 & 18.59 & 46.00 & 1573 \\
                               & $\beta$-photosphere &  8 & 16.65 & 18.57 &  401 \\
                               & chromosphere        & 21 & 15.55 &  5.97 &  245 \\
        \\
        \multirow{3}{2em}{ 3 } & Nominal photosphere &  0 & 18.48 & 43.93 & 2594 \\
                               & $\beta$-photosphere &  6 & 16.32 & 14.70 &  609 \\
                               & chromosphere        & 14 & 15.37 &  6.41 &  410 \\
        \\
        \multirow{3}{2em}{ 4 } & Nominal photosphere &  0 & 18.62 & 42.41 & 3315 \\
                               & $\beta$-photosphere &  4 & 16.67 & 16.53 &  934 \\
                               & chromosphere        & 11 & 15.49 &  6.84 &  583 \\
        \\
        \multirow{3}{2em}{ 6 } & Nominal photosphere &  0 & 19.12 & 39.53 & 3690 \\
                               & $\beta$-photosphere &  3 & 16.94 & 13.53 & 1315 \\
                               & chromosphere        &  7 & 15.87 &  7.60 &  951 \\
        \\
        \multirow{3}{2em}{ 7 } & Nominal photosphere &  0 & 19.40 & 38.24 & 3671 \\
                               & $\beta$-photosphere &  3 & 16.93 & 11.37 & 1415 \\
                               & chromosphere        &  6 & 16.12 &  7.94 & 1135 \\
        \\
        \multirow{3}{2em}{ 9 } & Nominal photosphere &  0 & 20.03 & 36.13 & 4057 \\
                               & $\beta$-photosphere &  2 & 17.82 & 13.63 & 1939 \\
                               & chromosphere        &  5 & 16.64 &  8.58 & 1482 \\
    \enddata
    \tablecomments{ In the "layer" column for each particular layer showed its level in the corresponding rebinned grid. BIFROST photosphere always correspond to zero level. }

\end{deluxetable*}


%
%
%

\section{$\pi$-Disambiguation Tests}
\label{S_pi_dis}

Reconstruction of the coronal magnetic field starts typically from the magnetic vector data on the photospheric or, less common, chromospheric levels of the solar atmosphere. The magnetic vector measurements are performed using spectropolarimetry of Zeeman splitting of magnetic-sensitive optical or IR lines with circular polarization providing the line-of-sight component, while linear polarization providing the transverse component  of the magnetic field vector. Apparently, only the azimuth angle, but not direction of the transverse component is being measured. This uncertainty of the transverse component direction is commonly called $\pi$-ambiguity (180$^\circ$-ambiguity), thus a method resolving this $\pi$-ambiguity is called for.

\subsection{  $\pi$-disambiguation methods}

A number of disambiguation methods have been proposed \citep{1994SoPh..155..235M, 2005ApJ...629L..69G, 2013SoPh..283..195G, 2013SoPh..282..107C, 2014SoPh..289.1499R}, which differ in their approaches, performance, and speed. The detailed comparison of different methods can be found in comprehensive reviews \citep{2006SoPh..237..267M, 2009SoPh..260...83L}. For our tests we selected the minimum energy (ME) solution \citep{1994SoPh..155..235M}, the so-called super fast and quality disambiguation (SFQ)  method  \citep{2014SoPh..289.1499R} also known as new disambiguation method (NDA), and an acute angle (AA) method, which makes a choice by comparison of the azimuth with that of a `reference' field  \citep{2006SoPh..237..267M}. The reference field can be a potential or linear force-free field, for example, but we employ a simple version of the AA method that utilizes a naive approach where the true transverse field direction is  selected to be closer to the potential field.

The ME disambiguation\footnote{\url{http://www.cora.nwra.com/AMBIG/}} is based on simulated annealing \citep{1953JChPh..21.1087M} and is regarded as providing the highest disambiguation quality among other methods. However this method is rather computationally expensive.
Our implementation of the ME method follows \citet{1994SoPh..155..235M} with an improvement suggested by \citet{2009SoPh..260...83L}, namely with the initial `temperature' varying from pixel to pixel depending on maximum temperature among all possible combinations of transversal field orientation of the neighbor pixels. In addition, this implementation adopts that for any pixel keeping its transverse field orientation reasonably long (typically, during 100 consecutive iterations), the ambiguity is treated ``resolved'' and the algorithm do not  change its state over the remaining iterations.  Thus, the number of analyzed pixels decreases with the number of iteration, which reduces the annealing time.

The SFQ method \citep{2014SoPh..289.1499R} includes two steps: a preliminary disambiguation and a ``cleaning'' procedure. The former is a  local comparison of the observed ambiguous field  with the reference (potential) one. The comparison procedure is similar to that in the AA method.  The discontinuities in this first order approximate solution are then cleaned out with an iterative procedure. On each iteration the  transverse magnetic field is compared with its  local average and inverted in those pixels where the alternative value is closer to the smoothed field.


\subsection{Performance metrics for the $\pi$-disambiguation methods }

To test the quality of  $\pi$-disambiguation methods we used the following metrics  \citep{2006SoPh..237..267M}: pixel error $E_\mathrm{pix} $ and flux error $E_\mathrm{flux}$ :
$$ E_\mathrm{pix} = \frac{N_\mathrm{err}}{N}, $$
$$ E_\mathrm{flux} = \frac{\sum_{\mathrm{err}}B_\tau(x,y)}{\sum_{\mathrm{all}} B_\tau(x,y)}, $$
where $N$ is the total number of pixels and $N_\mathrm{err}$ is the number of pixels where disambiguation method fails, $B_\tau(x,y)$ is the absolute value of the transversal magnetic field in pixel $(x,y)$, $\sum_{\mathrm{err}}$ and $\sum_{\mathrm{all}}$  are summations over those pixels where the disambiguation failed and  over all pixels in the magnetogram, respectively, as well as the typical computation time.

\newpage

\subsection{Comparison of the alternative approaches.}

The $\pi$-disambiguation methods described above have been applied to all reference layers and all rebinned data cubes. Here we summarize there performance.

		\begin{deluxetable}{c|ccc|ccc|cc}
			\tablecolumns{9}
			\tablewidth{0pc}
			\tabletypesize{\footnotesize}
			\tablecaption{$\pi$-disambiguation results for the nominal photosphere  \label{table_disambig_nom_phot}}
			\tablehead{\colhead{} &  \multicolumn{3}{c}{\textbf{Pixel error, \%}} & \multicolumn{3}{c}{\textbf{Flux error, \%}}& \multicolumn{2}{c}{\textbf{Computation time}}\\
				\colhead{Bin } & \colhead{SFQ } & \colhead{ME} & \colhead{AA} & \colhead{SFQ } & \colhead{ME} & \colhead{AA} &  \colhead{SFQ } & \colhead{ME}
			}
			\startdata
			2  &	19.5  &	 20.5  	&30.3&	18.8  &	14.8  &	28.4 &	2.9s  & 13m57s\\
			3  &	19.0  &	23.9  	&30.3&	19.2  &	16.4  &	27.7 &	1.4s  &	6m30s  \\
			4  &	19.2  &	27.6  	&29.6&	20.4  &	18.7  &	26.6 &	 0.7s   &	4m04s \\
			6  &	14.3  &	34.2  	&28.0&	14.0  &	27.0  &	25.3 &	0.38s  &	1m46s	\\
			7  &	12.3  & 33.0	&26.8&	12.8  &	26.2  &	23.9 & 0.31s	& 1m18s	\\
			9  &	8.6   &	33.0  	&23.5& 9.1  	&	27.8& 19.0 & 0.18s &48s	\\
			\enddata

		\end{deluxetable}

	\begin{deluxetable}{c|ccc|ccc|cc}
		\tablecolumns{9}
		\tablewidth{0pc}
		\tabletypesize{\footnotesize}
		\tablecaption{$\pi$-disambiguation results for the  AR $\beta$ photosphere  \label{table_disambig_beta_phot}}

\tablehead{\colhead{} &  \multicolumn{3}{c}{\textbf{Pixel error, \%}} & \multicolumn{3}{c}{\textbf{Flux error, \%}}& \multicolumn{2}{c}{\textbf{Computation time}}\\
			\colhead{Bin } & \colhead{SFQ } & \colhead{ME} & \colhead{AA} & \colhead{SFQ } & \colhead{ME} & \colhead{AA} &  \colhead{SFQ } & \colhead{ME}
		}

		\startdata
		2  & 1.7  & 5.8		&8.1&0.6	&2.3& 3.5&	1.1s 	   & 8m38s\\
		3  & 1.3 &	3.1		&6.3& 0.4	& 1.0& 2.5 &0.37s	   & 3m39s \\
		4  & 1.3 &2.3	 	&6.7&0.4	&0.71& 2.7&	0.18s	   & 2m11s\\
		6  & 0.76 &	0.4		&5.7&0.15	&0.045	&2.1&  0.09s  &	3m56s\\
		7  & 0.46 &	0.21	&5.8& 0.11	& 0.016 &2.0& 0.09s	  & 3m39s\\
		9  & 0.63 &	0.35	&5.6&0.21	&0.027&2.0 &  0.02s  &26s	\\
		\enddata
		
	\end{deluxetable}
	
		\begin{deluxetable}{c|ccc|ccc|cc}
			\tablecolumns{9}
			\tablewidth{0pc}
			\tabletypesize{\footnotesize}
			\tablecaption{$\pi$-disambiguation results for the chromosphere  \label{table_disambig_chromo}}
			\tablehead{\colhead{} &  \multicolumn{3}{c}{\textbf{Pixel error, \%}} & \multicolumn{3}{c}{\textbf{Flux error, \%}}& \multicolumn{2}{c}{\textbf{Computation time}}\\
				\colhead{Bin } & \colhead{SFQ } & \colhead{ME} & \colhead{AA} & \colhead{SFQ } & \colhead{ME} & \colhead{AA} &  \colhead{SFQ } & \colhead{ME}
			}
			\startdata
			2  & 0.005&0     &5.0&0.0001&0       & 1.5& 0.73s	& 6m03s\\
			3  &0.007&0      &4.9&0.0003&0      & 1.4&0.3s      & 2m35s\\
			4  &0.03 &0.013&4.6&0.006&0.0004&1.3&0.13s     & 1m28s\\
			6  & 0.06&0.014&4.7&0.014&0.0005&1.4&0.07s      &40s	\\
			7  & 0.06&0       &4.6& 0.02&0         &1.3&0.05s     &30s\\
			9  & 0.03&0      &4.4&	0.01&0         &1.3&0.04s     &19s\\
			\enddata
			
		\end{deluxetable}

\begin{figure}\centering
	\includegraphics[width=0.99\columnwidth]{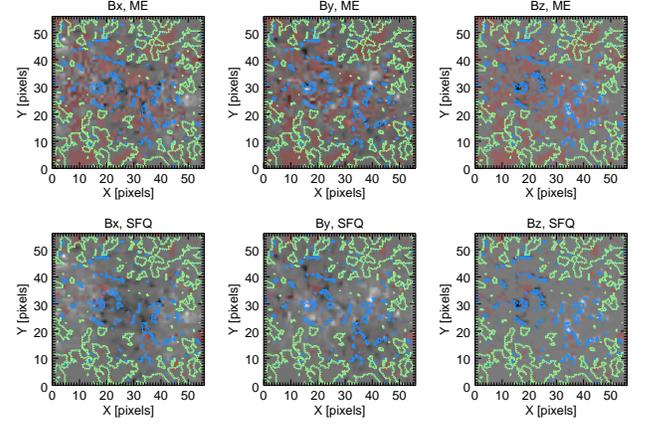}
	\caption{\label{fig:disambig_image_nom}
		Comparison of the disambiguation results of the ME (top row) and SFQ (bottom row) codes for the BIFROST field with the binning factor 9 at the level of nominal photosphere. Pixels with the  disambiguation errors are indicated in the semitransparent  red. Blue contours enclose regions with  almost vertical magnetic field (inclination angle $<$ 15 degrees). Green contours enclose weak field regions where  $B < 50$ G. Ticks on the contours show the direction towards the lower values.
	}
\end{figure}

For the nominal photosphere (see Table~\ref{table_disambig_nom_phot}), all tested methods fail to provide acceptable result. Both flux and pixel errors are of the  order of 20 -- 30\%, with only exception of large bin-factors and SFQ method, where the errors are about 10\%; see Figure~\ref{fig:disambig_image_nom}. In most cases (but not always), both methods work better in areas of stronger magnetic field but fail in areas of weak small-scale field. This failure of all these methods originates from the fact that none of the methods is capable of correctly processing the very small magnetic elements with the size of the order of one pixel  (salt and pepper patterns). {This might have important implications for $\pi$-disambiguation of the magnetic field at the quiet sun areas, which is, in particular, a substantial part of the full-disk disambiguation in the} SDO/HMI pipeline.

\begin{figure}\centering
	\includegraphics[width=0.99\columnwidth]{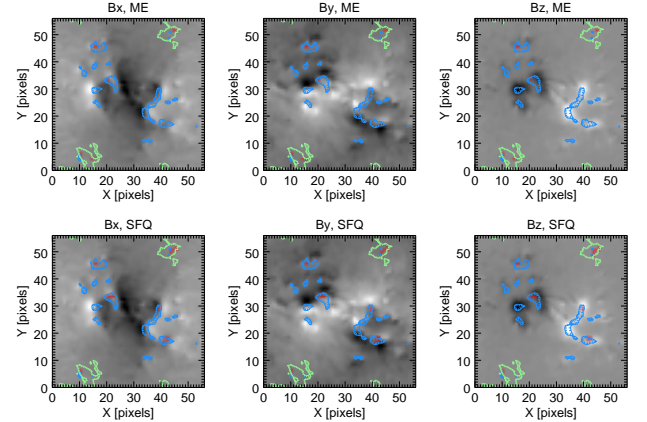}
	\caption{\label{fig:disambig_image_beta}
		Comparison of the disambiguation results of the ME (top row) and SFQ (bottom row) codes for the BIFROST field with the binning factor 9 at the level of correct $\beta$ photosphere. Pixels with the disambiguation errors are indicated in the semitransparent  red. Blue contours enclose regions with almost vertical magnetic field (inclination angle $<$ 15 degrees). Green contours enclose weak field regions where  $B < 50$ G. Ticks on the contours show the direction towards the lower values.
	}
\end{figure}

Table \ref{table_disambig_beta_phot} shows the test results for the level of typical AR $\beta$ photosphere. At this level, the magnetic field becomes noticeably smoother than that at the nominal photosphere, and all methods improve their recovery success rate by at least an order of magnitude. AA pixel errors exceed 5\% and  the flux errors exceed 2\%. SFQ and ME results are much better. On the finest grid the best quality is provided by SFQ, while the output of ME method contains an extended area where the transverse flux is inverted.  For the grids with lower resolutions (bin 6 -- 9) this artefact disappears and ME outperforms SFQ. However, both SFQ (all bins) and ME (bins 3 -- 9) give flux error of 1\% or less and, thus, their results can be used as boundary conditions for NLFFF extrapolations.  In principle, it is tempting to use the $\pi$-disambiguation mismatches between the different disambiguation methods to reprocess those pixels, where two methods give opposite results, to improve the solution. However, as clearly seen from Figure~\ref{fig:disambig_image_beta}, there are a number of pixels, where \textit{both methods fail} to recover the field azimuth, even though the total number of those `bad' pixels is very small. Figure~\ref{fig:disambig_image_beta} demonstrates that the failures are not random, but occur in the areas, where the field is either weak or almost vertical or both.

At the chromospheric level (Table~\ref{table_disambig_chromo}), the  magnetic field is more horizontal and smooth enough, so it is disambiguated perfectly by ME (there are errors in several pixels only for bins 4 and 6). The SFQ method also works excellent and fails only in several tens of pixels giving flux error from 0.0001\% to 0.01\%, depending on the binning. AA also perform better on the smooth chromospheric (than the photospheric) field with the flux error about 1.5\% for all bins.

These tests validate both ME and SFQ disambiguation methods at the $\beta$-photosphere and the chromosphere. This permits us to adopt that the $\pi$-ambiguity has been resolved perfectly, while performing further tests on the field preprocessing and extrapolation.
Note, that in the present article we do not study  dependence of the disambiguation accuracy on  position of the data cube at the solar disk. Most of the disambiguation approaches tend to generate more errors towards the solar limb. Here, the SFQ method has an advantage.   According to tests on real magnetograms  observed close to the disk center and artificially rotated towards the limb, it maintains high accuracy even there  \citep{2014SoPh..289.1499R}, while  other methods often produce artifacts in the transverse field components.	

\section{Preprocessing Tests}
\label{S_prepro}


Even the photospheric magnetic vector data with a perfectly resolved $\pi$-ambiguity represent a substantial challenge for the coronal magnetic field reconstruction. The problem is that the coronal modeling relies on the magnetic field force-freeness in the corona, which is dominated by the magnetic pressure (the plasma $\beta$ is less than 1), while the photospheric magnetic field is not force-free given that $\beta\gg1$ there. To overcome this mismatch it has been proposed \citep{2006SoPh..233..215W, 2007A&A...476..349F, 2011A&A...526A..70F, 2014SoPh..289...63J} to modify the measured photospheric magnetogram in such a way that a new, `preprocessed,' magnetogram is representative of a higher, chromospheric, height and, thus, more force-free than the original photospheric one. An appropriately performed preprocessing can facilitate the coronal field reconstruction by providing a boundary condition that is more suitable for a NLFFF extrapolation. The tests performed using available chromospheric diagnostics \citep{2010ApJ...719L..56J} confirm that the preprocessing is doing reasonably well; here we perform the tests of various available approaches to the preprocessing using the same modeling data cubes as above.

\subsection{ Preprocessing approaches}

\citet{2006SoPh..233..215W} proposed to preprocess the photospheric vector magnetogram data to drive the observed non-force-free photospheric data towards the suitable boundary condition in the chromosphere by minimizing a functional $L_{prep}$:
\begin{equation}
\label{Eq_prepr_TW}
L_{prep}=\mu_1L_1+\mu_2L_2+\mu_3L_3+\mu_4L_4\ ,
\end{equation}

\noindent where 

$L_1=(\Sigma_p \tilde{B}_x\tilde{B}_z)^2+(\Sigma_p\tilde{B}_y\tilde{B}_z)^2+ \left(\Sigma_p (\tilde{B}_z^2-\tilde{B}_x^2-\tilde{B}_y^2)\right)^2$\
,

$L_2=(\Sigma_px(\tilde{B}_z^2-\tilde{B}_x^2-\tilde{B}_y^2))^2+(\Sigma_py(\tilde{B}_z^2-\tilde{B}_x^2-\tilde{B}_y^2))^2+\left(\Sigma_p(y\tilde{B}_x\tilde{B}_z-x\tilde{B}_y\tilde{B}_z)\right)^2$\
,

$L_3=\Sigma_p(\tilde{B}_x-B_{x})^2+\Sigma_p(\tilde{B}_y-B_{y})^2 +
\Sigma_p(\tilde{B}_z-B_{z})^2$\ ,

$L_4=\Sigma_p(\Delta \tilde{B}_x)^2+\Sigma_p(\Delta \tilde{B}_y)^2+\Sigma_p(\Delta
\tilde{B}_z)^2$\ .

\noindent Here, minimizing $L_1$ and $L_2$ terms ensures that the final magnetogram corresponds, as close as possible, to the force-free and torque-free conditions, respectively. $L_3$ term is responsible for the similarity of the preprocessed data  to the original magnetogram, while $L_4$ term is responsible for the smoothing. Summations $\Sigma_p$ represent the surface integrals over all available grid nodes $p$ at the bottom boundary; $\Delta$ stands for the two-dimensional (2D) Laplace operator. The idea is to minimize $L_{prep}$ at once, so that all terms $L_{n}$ are getting small simultaneously. The weights $\mu_{n}$ are unknown \textit{a priori}; \citet{2006SoPh..233..215W} tested various choices for $\mu_{n}$ and came up with a strategy on how to choose these weights; the default set of the weights is $\mu=[1,~1,~10^{-3},~10^{-2}]$. Minimization of the functional  is performed  iteratively using Newton-Raphson scheme\footnote{The iteration step $s$ is being optimized itself during the minimization given that $s$ can be considered as one of the arguments of the functional; the optimum step is selected by the standard \lq\lq golden section\rq\rq\ method \citep{Press:2007:NRE:1403886} at each iteration step.} \citep{Press:2007:NRE:1403886}. Hereafter, we  refer to our  implementation of the preprocessing method developed by \citet{2006SoPh..233..215W} as TW preprocessing.

\citet{2007A&A...476..349F} proposed another form of  functional (\ref{Eq_prepr_TW}) to be minimized. The distinctions are in the form of $L_4$ term (the smoothing was performed using a median filter instead of the differential Laplas operator in \citet{2006SoPh..233..215W}), in varying the field only inside a given range defined by a field  threshold value instead of considering $L_3$, and in the minimization method selected---annealing simulation.  \citet{2011A&A...526A..70F} have shown that the \citet{2006SoPh..233..215W} method gives  smoother preprocessed field. Since both methods have a similar nature and show comparable final results, we only consider the  \citet{2006SoPh..233..215W} method for further testing.


\citet{2014SoPh..289...63J} proposed a slightly different approach to the preprocessing that consists of two distinct steps. The first step produces a potential extrapolation starting from the photospheric vertical magnetic field component $B_z$. Then, at a certain level, typically one pixel above the photosphere, the $B_z$ component is taken from the potential extrapolation, while a functional similar to Eq.~(\ref{Eq_prepr_TW}) is formed for the transverse field components only, and is being minimized  at the second step of the method.  An apparent advantage of this approach is the ability to control the height level, to which the preprocessed magnetogram must correspond. For our tests we used the preprocessing code provided by the authors \citep{2014SoPh..289...63J} with the default set of  the weights $\mu=[1,~1,~10^{-3},~1]$ and with a few cosmetic modifications needed to ease the code running.   In what follows, we  refer to the preprocessing method of \citet{2014SoPh..289...63J}  as JF preprocessing.


\subsection{ Performance metrics for the preprocessing codes}

Preprocessing methods are supposed to  modify the magnetic field vector measured at the photosphere to make it more compatible with the force-freeness of the magnetic field model in the corona. The corresponding  modification of the photospheric field could either result in removing the force  component from the magnetic field distribution without any noticeable change of the field strength, or, in addition to this removal, may yield the corresponding decrease of the field strength emulating an effective \lq\lq elevation\rq\rq\ of the field distribution up to a higher chromospheric level where the magnetic field does become a force free one.

Thus, we assess the preprocessing methods in the three following respects:
\begin{enumerate}
	\item Has the magnetic field become significantly more  force free after the preprocessing?
	\item How close is the preprocessed field to the magnetic field in the BIFROST model?
    \item Does the preprocessing cause an effective \lq\lq elevation\rq\rq\ of the field distribution up to a higher level?
\end{enumerate}

The force-freeness of the preprocessed magnetic field  is assessed  using the $L_1$ and $L_2$ terms from (\ref{Eq_prepr_TW}). In this work, we quantitatively measure the preprocessing efficiency by calculating the logarithmic ratio of $L_1$ and $L_2$ parameters before and after the preprocessing: $\log_{10}\frac{L_1^0}{L_1^p}$ and $\log_{10}\frac{L_2^0}{L_2^p}$. The higher values of these parameters indicate more efficient removal of the force-carrying magnetic field component.

 The consistency of the preprocessed magnetic field with the model one at a given level of the data cube can straightforwardly be assessed using  the  $L_3$ metrics in Eq.~(\ref{Eq_prepr_TW}). To assess if any  effective elevation has happened, we  calculate this metrics using the Bifrost data from the same level and a few higher levels. We also use a few more useful metrics defined below:
\begin{itemize}
	\item Normalized  discrepancy between the preprocessed and observed field components
	\begin{equation}\label{eq:L3_prep_metric}
	\tilde{L_3} = \sqrt{\frac{L_3}{\sum_p B^2}},
	\end{equation}
	\item Normalized  discrepancy between the absolute value of the preprocessed and observed fields
	\begin{equation}\label{eq:EB_prep_metric}
	E_B = \sqrt{\frac{\sum_p(\tilde{B} - B)^2}{\sum_p B^2}},
	\end{equation}
	\item Average angle between the preprocessed and observed field
	\begin{equation}\label{eq:angle_prep_metric}
	A = \arccos \left(\sum_p \frac{\tilde{\vec{B}} \cdot \vec{B}}{\tilde{B} B} / N \right),
	\end{equation}
	\item Slope of the regression dependence $\tilde{B} =SB + b$
	\begin{equation}\label{eq:slope_prep_metric}
	S = \frac{\sum_p \tilde{B} B - \sum_p \tilde{B} \sum_p B /N}{\sum_p B ^2 - (\sum_p B )^2 /N}
	\end{equation}
	\item Correlation coefficient
	\begin{equation}\label{eq:correlation_prep_metric}
		R = \frac{\sum_p \tilde{B}B - \sum_p \tilde{B} \sum_p B /N}{\sqrt{\sum_p \tilde{B}^2 - (\sum_p \tilde{B})^2 /N} \sqrt{\sum_p B^2 - (\sum_p B)^2 /N} }
	\end{equation}
\end{itemize}
Better preprocessing method is expected to give lower values of $	\tilde{L}_3 $, while the preferable values of the slope $S$ and the correlation coefficient $R$ are those closer to unity.
	If the preprocessed and the original fields are substantially different, $	\tilde{L}_3 $ will have a larger value. For instance, if $\tilde{\vec{B}} \equiv \vec{0}$ the $\tilde{L}_3$ metric will be equal to unity for any $\vec{B}$. Therefore we will interpret the preprocessing results with $\tilde{L}_3 \ge 0.5$ as unacceptable.
Additionally, we calculate metrics (\ref{eq:L3_prep_metric})---(\ref{eq:correlation_prep_metric}) for the model magnetic field taken at the height of one voxel to nail down a possible effective elevation that can be a side or intended effect of the preprocessing.

\subsection{Comparison of the alternative approaches.}

Tables \ref{table_prep_photo}, \ref{table_prep_beta}, and \ref{table_prep_chromo} show the metrics describing how force-free are the boundary conditions produced by the two codes.
For the nominal photosphere, all results have $	\tilde{L}_3 > 0.5$ meaning that the preprocessed field is strongly different from the original magnetogram. Therefore, we interpret the preprocessing results obtained for the nominal photosphere as incorrect and do not analyse them in what follows.
 For all levels and binning factors our implementation of the TW preprocessing \citep{2006SoPh..233..215W}  outperforms the JF preprocessing code \citep{2014SoPh..289...63J} providing significantly more force-free results. However, the $L_3$ metrics are nearly the same for both codes, meaning that both solutions are comparably close to the input magnetic field.

 \begin{figure}\centering
	\includegraphics[width=0.49\columnwidth]{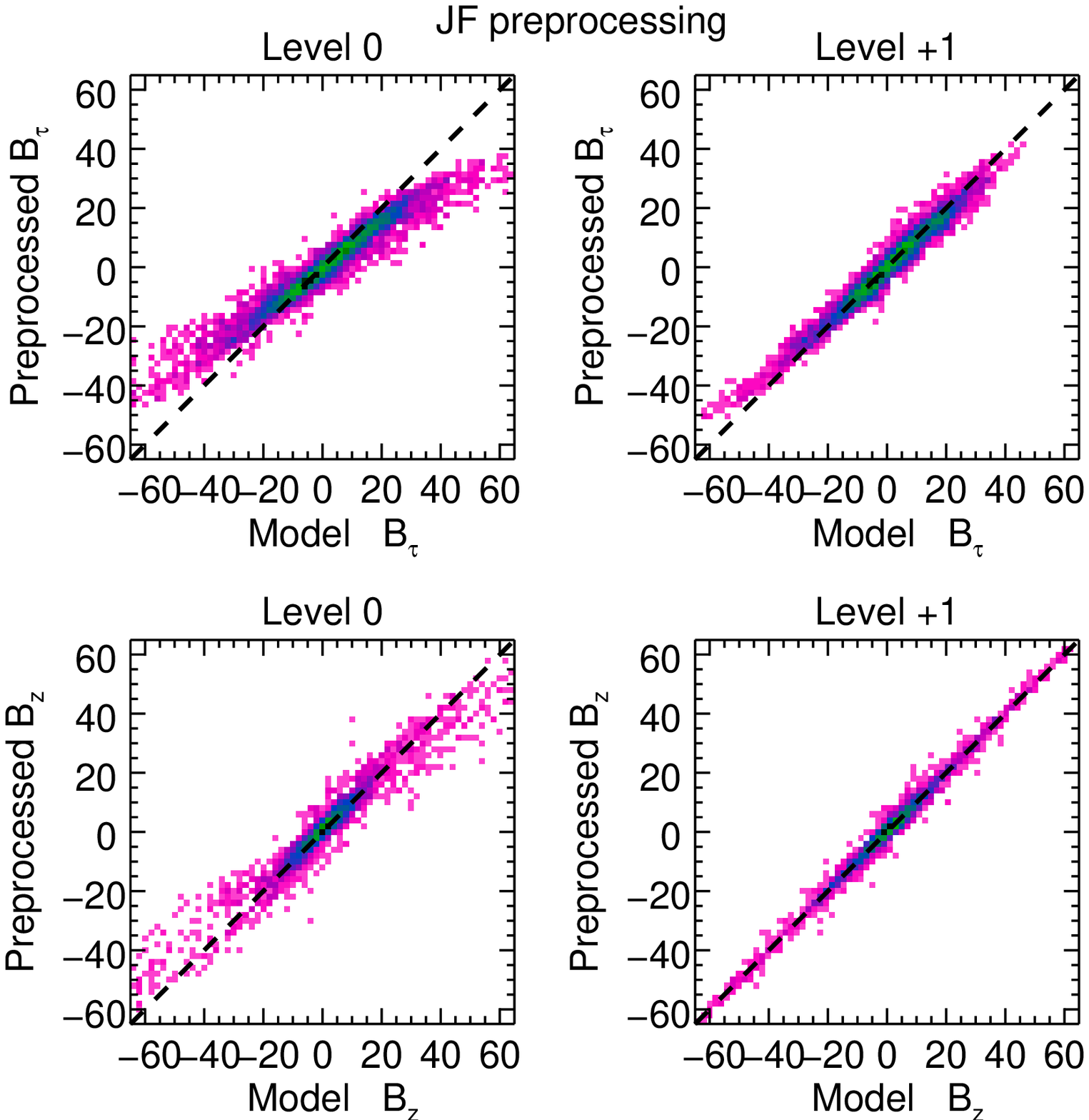}
	\includegraphics[width=0.49\columnwidth]{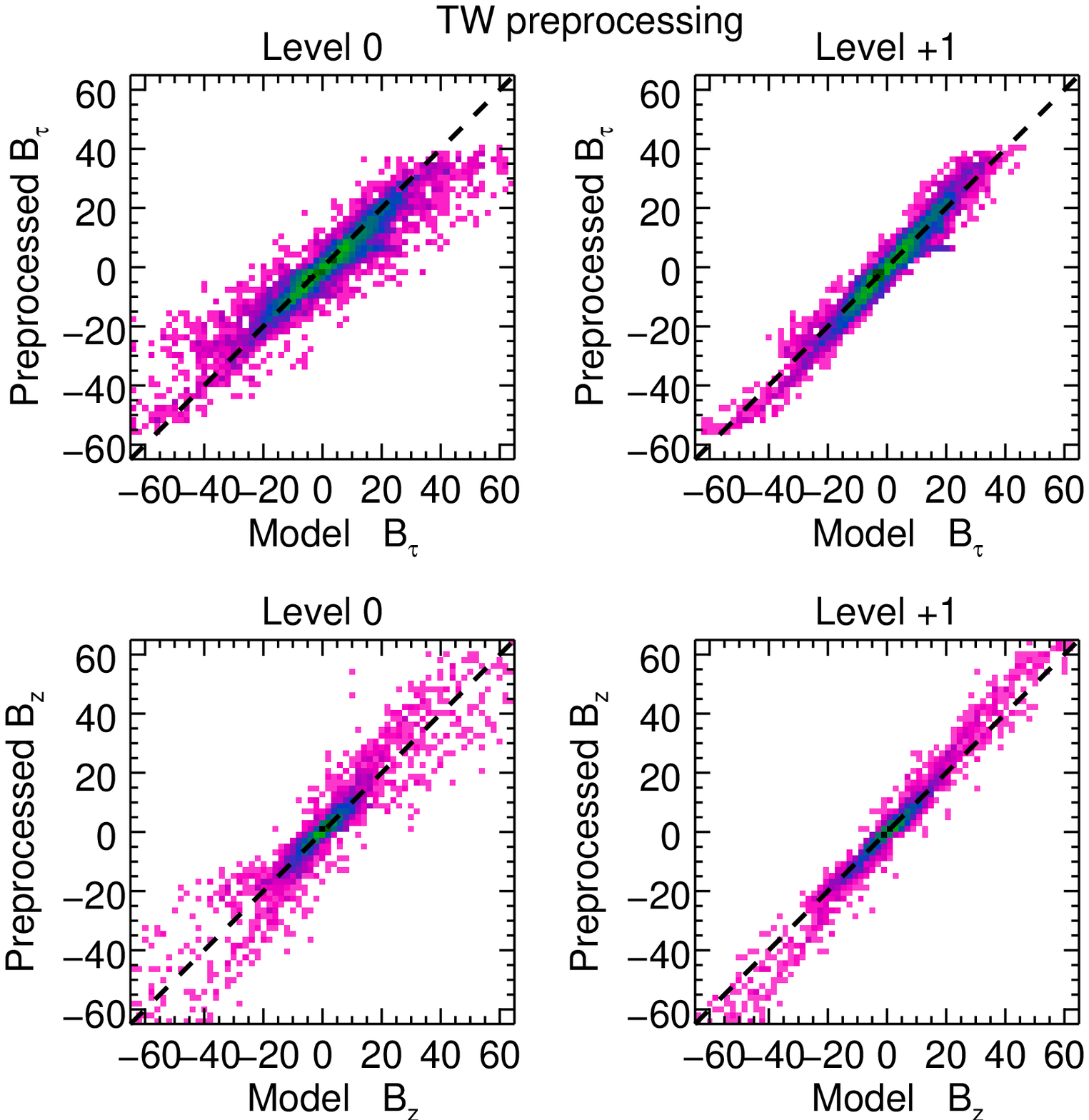}
	\caption{\label{fig:prep_JF_test}
		Dependence of the  magnetic field preprocessed with the JF (left panel) and TW (right panel) preprocessing codes   on the initial field on the  $\beta$-photosphere for bin=9.
	}
\end{figure}

The detailed quantitative comparison of the preprocessing results and the magnetic field in the Bifrost model is given in Tables \ref{table_prep_extrapol_photo} and \ref{table_prep_extrapol_chromo}.
We assess separately the full vector and the longitudinal and transverse components at two levels: the same level as the initial magnetogram and one level higher.
$E_B$  metric (Eq. \ref{eq:EB_prep_metric}) indicates the discrepancy  between the absolute values of the preprocessed and  modeled magnetic field, while the difference in the directions of the fields is shown by $A$ metric (Eq. \ref{eq:angle_prep_metric}).
{This metric} shows that there is an effective field elevation  caused by any of the JF (all bins) and TW (bins 4--9) preprocessing methods.
The effect is more pronounced for the JF preprocessing especially at low spatial resolution (bin 9).
However, the mean angle between the preprocessed and Bifrost field shows different behavior.
One level elevation does not improve the mean angle metric for the JF preprocessing,  while the TW method demonstrates slight improvement of it at bins 7 and 9.  Being applied to the chromospheric level (even though this might not be needed in practice), JF preprocessing again demonstrates effective elevation in $E_B$, while TW method shows no improvement at higher level  in either absolute value discrepancy  or in angle.

JF and TW methods preprocess longitudinal and transverse magnetic field components differently.
Therefore we assess the field components separately.
For  both longitudinal and transverse components we calculate the slope (Eq. \ref{eq:slope_prep_metric}) and correlation (Eq. \ref{eq:correlation_prep_metric}) metrics.
The former characterizes the absolute value of the field while the latter is to assess the spatial structuring of the field.

\begin{figure}\centering
	\includegraphics[width=0.99\columnwidth]{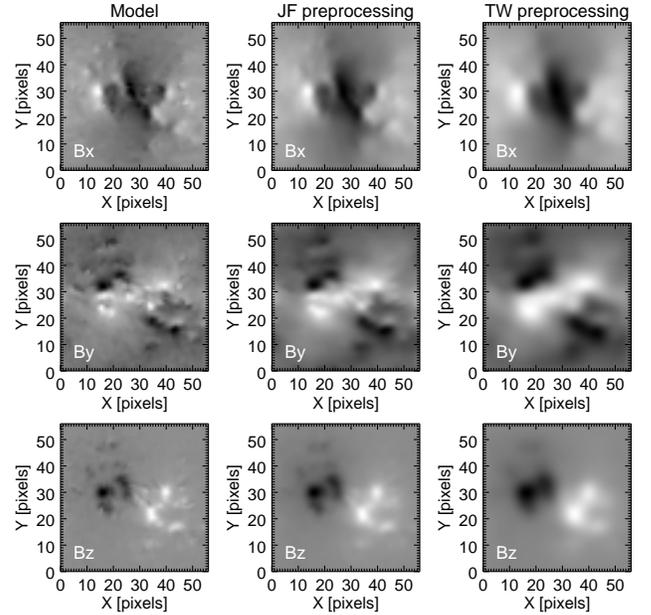}
	\caption{\label{fig:prep_compare}
		Comparison of the original (left column) Bifrost field at the $\beta$-photosphere level  with the results of JF (middle column) and TW (right column) preprocessing for bin=9.
	}
\end{figure}

Both slope and correlation metrics show the presence of an effective elevation in the longitudinal component preprocessed using the JF method.
Indeed, such a behavior is rather expected because the JF preprocessing fixes the longitudinal component to be equal to the potential field extrapolation results and does not  optimize it. The transverse component is, however, elevated by a height exceeding the one voxel size (but less than the two voxel sizes). Therefore, there is a mismatch in the heights to where the longitudinal and transverse components are elevated in the JF method. This mismatch is primarily responsible for the mismatch in the angle between the preprocessed and the (one level up) model field mentioned above.

		\begin{deluxetable}{c|cc|cc|cc}
			\tablecolumns{9}
			\tablewidth{0pc}
			\tabletypesize{\footnotesize}
			\tablecaption{   Results of the preprocessing test at the nominal photosphere} \label{table_prep_photo}
			\tablehead{\colhead{} &  \multicolumn{2}{c}{$\log_{10}{\frac{L_1^{o}}{L_1^{p}}}$} & \multicolumn{2}{c}{$\log_{10}{\frac{L_2^{o}}{L_2^{p}}}$}& \multicolumn{2}{c}{$L_3$}\\
				\colhead{Bin } & \colhead{JF} & \colhead{TW} & \colhead{JF} & \colhead{TW}  & \colhead{JF} & \colhead{TW}
			}
			\startdata
			2  &4.55	&11.26  &5.42	&11.28	&0.60   &0.72\\
			3  &4.50	&10.68	&5.42	&10.80	&0.67	&0.79\\
			4  &4.35	&10.42	&5.32	&10.89	&0.70	&0.83\\
			6  &4.20	&9.80	&4.97	&10.16	&0.74	&0.85\\
			7  & 4.11	&9.59   &4.72	&9.96	&0.74   &0.85\\
			9  & 3.99	&9.24   &4.29	&9.58	&0.74   &0.85\\
			\enddata
			
		\end{deluxetable}

After the TW preprocessing, the value of the longitudinal field becomes underestimated for the original level ($S<1$) but overestimated for the +1 voxel level ($S>1$), meaning that some effective elevation is present, but the elevation height is somewhere between 0 and 1 voxel. In contrast, the transverse component displays an elevation very close to the one voxel height; thus, the TW preprocessing also results in a mismatch in the elevation heights for the longitudinal and transverse components, likewise the JF method.

The JF preprocessing demonstrates  better cross-correlation metric $R$ for the longitude component for all levels and binning factors, both at the photospheric and chromospheric heights.  The cross-correlation coefficient for the transverse field is worth for both methods at all levels and heights. At high resolutions (bins 2-4) JF and TW show very similar results, while  at low resolution (bins $> 4$), JF demonstrates higher correlation with the modeled field.
This behavior is also demonstrated in Figure \ref{fig:prep_JF_test} in the form of two-dimensional histograms showing the scatter of the preprocessed field components versus the modeled ones.

	Our tests show that both methods have their own advantages and disadvantages. TW preprocessing improves force-freeness metrics by many orders of magnitude, but  significantly changes the field structure, especially for lower spatial resolutions, Figure~\ref{fig:prep_compare}. It also effectively elevates the field, but the elevation height seams to differ from one voxel and differs for the longitudinal and transverse components. The JF preprocessing better saves the field structure for low resolution grids and more distinctly elevates the longitudinal field to the height of one voxel, but the output is not as force-free as for the TW method;  in addition, the transverse component is effectively elevated by a height larger than the size of one voxel. These mismatches in the elevation heights for the field components revealed in both methods, will then have negative impact on the fidelity of the NLFFF extrapolations.

		
				\begin{deluxetable}{c|cc|cc|cc}
					\tablecolumns{9}
					\tablewidth{0pc}
					\tabletypesize{\footnotesize}
					\tablecaption{  Results of the preprocessing tests at the $\beta$-photosphere} \label{table_prep_beta}
					\tablehead{\colhead{} &  \multicolumn{2}{c}{$\log_{10}{\frac{L_1^{o}}{L_1^{p}}}$} & \multicolumn{2}{c}{$\log_{10}{\frac{L_2^{o}}{L_2^{p}}}$}& \multicolumn{2}{c}{$L_3$}\\
						\colhead{Bin } & \colhead{JF} & \colhead{TW} & \colhead{JF} & \colhead{TW}  & \colhead{JF} & \colhead{TW}
					}
					\startdata
					2  &2.64	&11.02	&2.65	&10.28	&0.24	&0.19\\
					3  &2.07	&10.79	&2.09	&9.59	&0.24	&0.20\\
					4  &2.38	&10.33	&2.41	&8.94	&0.31	&0.29\\
					6  &1.79	&9.61	&1.82	&8.35	&0.30	&0.30\\
					7  &1.30	&9.35	&1.29	&8.51	&0.27	&0.27\\
					9  &1.60	&9.56	&1.60	&7.81	&0.34   &0.37\\
					\enddata
					
				\end{deluxetable}

				
				\begin{deluxetable}{c|cc|cc|cc}
					\tablecolumns{9}
					\tablewidth{0pc}
					\tabletypesize{\footnotesize}
					\tablecaption{ Results of the preprocessing tests at the chromosphere}  \label{table_prep_chromo}
					\tablehead{\colhead{} &  \multicolumn{2}{c}{$\log_{10}{\frac{L_1^{o}}{L_1^{p}}}$} & \multicolumn{2}{c}{$\log_{10}{\frac{L_2^{o}}{L_2^{p}}}$}& \multicolumn{2}{c}{$L_3$}\\
						\colhead{Bin } & \colhead{JF} & \colhead{TW} & \colhead{JF} & \colhead{TW}  & \colhead{JF} & \colhead{TW}
					}
					\startdata
					2  &0.67	&9.37	&0.57	&9.22	&0.08	&0.039\\
					3  &0.67	&10.25	&0.57	&11.04	&0.10	&0.060\\
					4  &0.64	&10.31	&0.55	&10.53	&0.12	&0.088\\
					6  &0.66	&9.67	&0.57	&9.81	&0.16	&0.12\\
					7  &0.66	&9.47	&0.56	&9.55	&0.17	&0.13\\
					9  &0.62	&9.24	&0.53	&9.11	&0.20	&0.15\\
					\enddata
					
				\end{deluxetable}

				\begin{deluxetable*}{c|c|cc|cc|cc|cc|cc|cc}
				%
				\tablecolumns{15}
				\tablewidth{0pc}
				\tabletypesize{\footnotesize}
				\tablecaption{Photosphere effective elevation test \label{table_prep_extrapol_photo}}
				\tablehead{
					\multicolumn{2}{c|}{} & \multicolumn{4}{c|}{Full vector}& \multicolumn{4}{c|}{Longitudinal component} & \multicolumn{4}{c}{Transversal components}\\
					\multicolumn{2}{c|}{} &  \multicolumn{2}{c}{$E_B$}&  \multicolumn{2}{c|}{$A$}& \multicolumn{2}{c}{S}& \multicolumn{2}{c|}{R} & \multicolumn{2}{c}{S}& \multicolumn{2}{c}{R}\\
					\colhead{Bin } & \multicolumn{1}{c|}{level}& \colhead{JF} & \colhead{TW} & \colhead{JF} & \multicolumn{1}{c|}{TW}  & \colhead{JF} & \colhead{TW} & \colhead{JF} & \multicolumn{1}{c|}{TW} & \colhead{JF} & \colhead{TW} & \colhead{JF} & \colhead{TW}
				}

	\startdata
	\multirow{2}{*}{2}  &+0	&0.25	&0.16	&11.55	&10.85	&0.882	&0.960	&0.990	&0.9897	&0.709	&0.816	&0.980	&0.9821\\
						&+1	&0.14	&0.13	&22.62	&22.44	&0.982	&1.065	&0.994	&0.9895	&0.812	&0.936	&0.961	&0.9646\\
	\hline
    \multirow{2}{*}{3}  &+0	&0.23	&0.15	&12.32	&12.79	&0.868	&0.946	&0.988	&0.9851	&0.723	&0.842	&0.978	&0.9752\\
				  		&+1	&0.10	&0.11	&19.68	&19.43	&0.984	&1.070	&0.994	&0.9891	&0.859	&1.003	&0.975	&0.9740\\
	\hline
    \multirow{2}{*}{4}  &+0	&0.33	&0.24	&15.21	&16.18	&0.814	&0.890	&0.978	&0.9667	&0.645	&0.754	&0.962	&0.9472\\
				  		&+1	&0.13	&0.13	&23.62	&23.22	&0.980	&1.075	&0.992	&0.9841	&0.827	&0.983	&0.955	&0.9562\\
	\hline
    \multirow{2}{*}{6}  &+0	&0.32	&0.25	&14.12	&16.11	&0.799	&0.867	&0.976	&0.9586	&0.645	&0.748	&0.960	&0.9266\\
				  		&+1	&0.09	&0.12	&17.95	&17.02	&0.985	&1.081	&0.994	&0.9877	&0.863	&1.023	&0.975	&0.9637\\
	\hline
    \multirow{2}{*}{7}  &+0	&0.28	&0.21	&11.92	&13.91	&0.813	&0.891	&0.981	&0.9664	&0.679	&0.797	&0.968	&0.9327\\
				  		&+1	&0.07	&0.12	&12.29	&11.69	&0.985	&1.090	&0.996	&0.9903	&0.880	&1.053	&0.987	&0.9685\\
	\hline
    \multirow{2}{*}{9}  &+0	&0.39	&0.33	&14.02	&16.69	&0.754	&0.816	&0.969	&0.9324	&0.596	&0.685	&0.949	&0.8774\\
				  		&+1	&0.08	&0.15	&14.53	&13.28	&0.984	&1.095	&0.995	&0.9841	&0.862	&1.040	&0.980	&0.9516\\
	\hline
			\enddata
	\end{deluxetable*}

	\begin{deluxetable*}{c|c|cc|cc|cc|cc|cc|cc}
				%
				\tablecolumns{15}
				\tablewidth{0pc}
				\tabletypesize{\footnotesize}
				\tablecaption{Chromosphere effective elevation test \label{table_prep_extrapol_chromo}}
				\tablehead{
					\multicolumn{2}{c|}{} & \multicolumn{4}{c|}{Full vector}& \multicolumn{4}{c|}{Longitudinal component} & \multicolumn{4}{c}{Transversal components}\\
					\multicolumn{2}{c|}{} &  \multicolumn{2}{c}{$E_B$}&  \multicolumn{2}{c|}{$A$}& \multicolumn{2}{c}{S}& \multicolumn{2}{c|}{R} & \multicolumn{2}{c}{S}& \multicolumn{2}{c}{R}\\
					\colhead{Bin } & \multicolumn{1}{c|}{level}& \colhead{JF} & \colhead{TW} & \colhead{JF} & \multicolumn{1}{c|}{TW}  & \colhead{JF} & \colhead{TW} & \colhead{JF} & \multicolumn{1}{c|}{TW} & \colhead{JF} & \colhead{TW} & \colhead{JF} & \colhead{TW}
				}

	\startdata
	\multirow{2}{*}{2}  &+0	&0.08	&0.03	&2.93	&1.23	&0.963	&1.000	&0.9994	&0.9999	&0.888	&0.957	&0.9990	&0.9985\\
				  		&+1	&0.05	&0.03	&4.12	&3.48	&0.996	&1.033	&0.9995	&0.9995	&0.928	&0.998	&0.9995	&0.9979\\
	\hline
    \multirow{2}{*}{3}  &+0	&0.10	&0.05	&3.73	&2.60	&0.946	&0.998	&0.9989	&0.9995	&0.866	&0.960	&0.9977	&0.9934\\
				  		&+1	&0.05	&0.06	&5.02	&4.38	&0.994	&1.048	&0.9991	&0.9989	&0.924	&1.021	&0.9987	&0.9922\\
	\hline
    \multirow{2}{*}{4}  &+0	&0.12	&0.07	&4.20	&3.83	&0.931	&0.994	&0.9985	&0.9985	&0.847	&0.961	&0.9965	&0.9854\\
				  		&+1	&0.05	&0.08	&5.48	&4.93	&0.993	&1.059	&0.9988	&0.9981	&0.920	&1.041	&0.9976	&0.9838\\
	\hline
    \multirow{2}{*}{6}  &+0	&0.16	&0.09	&5.33	&5.39	&0.899	&0.986	&0.9971	&0.9966	&0.806	&0.956	&0.9933	&0.9762\\
				  		&+1	&0.05	&0.10	&6.65	&6.16	&0.990	&1.086	&0.9982	&0.9968	&0.913	&1.079	&0.9952	&0.9752\\
	\hline
    \multirow{2}{*}{7}  &+0	&0.18	&0.10	&5.72	&6.03	&0.884	&0.982	&0.9965	&0.9955	&0.788	&0.953	&0.9918	&0.9721\\
				  		&+1	&0.05	&0.12	&7.03	&6.64	&0.989	&1.098	&0.9979	&0.9962	&0.910	&1.096	&0.9942	&0.9706\\
	\hline
    \multirow{2}{*}{9}  &+0	&0.21	&0.11	&5.56	&6.86	&0.860	&0.976	&0.9959	&0.9940	&0.758	&0.952	&0.9893	&0.9622\\
    					&+1	&0.06	&0.14	&6.24	&7.38	&0.988	&1.121	&0.9975	&0.9953	&0.904	&1.129	&0.9914	&0.9594\\
	\hline
			\enddata
	\end{deluxetable*}

\vspace{0.51cm}
\section{NLFFF Tests}
\label{S_nlfff}


Different approaches and algorithms of the NLFFF reconstruction proposed so far include the vertical integration method, the boundary integral method, the Euler potential method, the Grad-Rubin methods, and various kind of evolutionary methods, such as magnetofrictional and optimization methods \citep[see, e.g., brief overview in][Section 5.3.3]{2005psci.book.....A}. Correspondingly, a variety of computing codes employing one or another version of these methods have been implemented. Although it is certainly interesting to cast all or most of the available NLFFF reconstruction methods vs realistic MHD models, such a study would be highly excessive for a single paper. Here we focus on two different implementations of the optimization method\footnote{We do not consider modifications proposed by \citet{2010A&A...516A.107W, 2012SoPh..281...37W} to address imperfection of real data  because our paper attempts to evaluate the best achievable performance of the NLFFF extrapolation codes themselves, i.e., not affected by possible negative influence of the measurement errors or lacking data.} \citep{2000ApJ...540.1150W} performed by our team members Alexey Stupishin \citep[hereafter AS, following][]{2004SoPh..219...87W} and Ivan Myshyakov \citep[hereafter IM, following][]{2009SoPh..257..287R}. The IM code was used for magnetic modeling to address various problems by \citet{2015Ge&Ae..55.1124K, 2016ARep...60..939L}, while AS code by \citet{2012ARep...56..790K, 2012SoPh..276...61B, 2016SoPh..291.2037Y}.

\begin{deluxetable*}{ c c | c c c c | c c c c | c c c c | c c c c }
	
	\tabletypesize{ \small }
	\tablewidth{ 0pt }
	\tablecaption{ Performance of the magnetic field reconstruction methods \label{IM_T02} }
	\tablehead{
		& & \multicolumn{8}{c}{} & \multicolumn{4}{c}{ $\beta$-photosphere, } & \multicolumn{4}{c}{ $\beta$-photosphere, } \vspace{ -0.45cm } \\
		\colhead{ Bin } & \colhead{ Impl } & \multicolumn{4}{c}{ Chromosphere } & \multicolumn{4}{c}{ $\beta$-photosphere } & \multicolumn{8}{c}{} \vspace{ -0.45cm } \\
		& & \multicolumn{8}{c}{} & \multicolumn{4}{c}{ JF preprocessed } & \multicolumn{4}{c}{ TW preprocessed } \vspace{ -0.15cm } \\
		& & \colhead{ $\theta^{\circ}$ } & \colhead{ $\theta _{j}^{\circ}$ } & \colhead{ $\theta _{m}^{\circ}$ } & \colhead{ $\theta _{mj}^{\circ}$ } & \colhead{ $\theta^{\circ}$ } & \colhead{ $\theta _{j}^{\circ}$ } & \colhead{ $\theta _{m}^{\circ}$ } & \colhead{ $\theta _{mj}^{\circ}$ } & \colhead{ $\theta^{\circ}$ } & \colhead{ $\theta _{j}^{\circ}$ } & \colhead{ $\theta _{m}^{\circ}$ } & \colhead{ $\theta _{mj}^{\circ}$ } & \colhead{ $\theta^{\circ}$ } & \colhead{ $\theta _{j}^{\circ}$ } & \colhead{ $\theta _{m}^{\circ}$ } & \colhead{ $\theta _{mj}^{\circ}$ }
	}
	
	\startdata
    \multirow{2}{*}{ 3 } & IM &  9.3 &  4.5 & 17.8 & 11.4 & 16.4 & 11.1 & 30.5 & 20.9 & 11.5 & 13.7 & 28.2 & 19.5 & 12.4 & 13.7 & 26.7 & 18.8 \\
                         & AS & 24.4 &  8.7 & 22.2 & 12.7 & 33.8 & 14.9 & 24.5 & 20.6 & 35.9 & 19.0 & 24.5 & 19.5 & 33.8 & 18.2 & 24.3 & 20.1 \\
    \\
    \multirow{2}{*}{ 4 } & IM & 10.7 &  5.4 & 18.1 & 11.5 & 18.9 & 13.4 & 34.3 & 22.3 & 13.0 & 16.5 & 28.5 & 21.2 & 13.7 & 16.9 & 27.1 & 20.3 \\
                         & AS & 24.9 &  9.4 & 21.9 & 12.4 & 34.5 & 16.3 & 24.5 & 20.7 & 36.6 & 21.3 & 24.3 & 20.0 & 34.5 & 20.8 & 24.1 & 20.5 \\
    \\
    \multirow{2}{*}{ 6 } & IM & 10.6 &  5.9 & 16.0 & 10.0 & 18.2 & 13.4 & 27.2 & 17.4 & 13.7 & 16.9 & 23.4 & 15.8 & 15.1 & 18.9 & 22.6 & 16.4 \\
                         & AS & 26.8 & 10.7 & 21.6 & 12.0 & 34.3 & 18.1 & 23.4 & 18.2 & 39.8 & 25.5 & 23.4 & 15.4 & 34.2 & 24.2 & 23.0 & 16.9 \\
    \\
    \multirow{2}{*}{ 7 } & IM & 11.0 &  6.3 & 15.8 &  9.8 & 14.3 & 10.6 & 21.3 & 13.2 & 12.5 & 14.2 & 20.2 & 13.5 & 14.0 & 17.9 & 19.4 & 14.2 \\
                         & AS & 28.3 & 11.5 & 21.6 & 11.8 & 39.1 & 21.1 & 23.0 & 17.1 & 42.0 & 25.9 & 23.0 & 14.4 & 38.8 & 27.7 & 22.4 & 15.1 \\
    \\
    \multirow{2}{*}{ 9 } & IM & 11.8 &  7.0 & 15.1 &  9.4 & 17.7 & 14.7 & 24.2 & 15.2 & 16.2 & 20.2 & 20.7 & 13.9 & 19.4 & 26.2 & 20.0 & 14.6 \\
                         & AS & 30.1 & 13.3 & 22.1 & 12.2 & 43.0 & 26.1 & 24.0 & 18.9 & 46.1 & 32.6 & 24.1 & 14.9 & 42.8 & 35.0 & 23.2 & 15.8 \\
	\enddata
	
	\tablecomments{ Bin -- is binnging factor. Impl -- is implementation of the optimization method. Four subsequent large columns contain numerical characteristics of reconstructed field. Column's title provide information of the starting layer and whether preprocessing was applied. }
	
\end{deluxetable*}

\subsection{Description of the NLFFF optimization methods}

In both implementations used in this paper the NLFFF reconstructions are performed following the optimization method \citep{2000ApJ...540.1150W}. The main idea of the optimization method is to transform some trial configuration of the magnetic field (usually a potential extrapolation from the bottom boundary) to a final force-free field configuration. This is achieved by minimization of the following, positively defined, functional:

\begin{equation}\label{Eq_nlfff_func}
L=\int\limits_{V} \left[B^{-2}\left[ [\nabla\times\mathbf{B}]\times\mathbf{B}\right]^2 + |\nabla\cdot\mathbf{B}|^2\right] w(x,y,z) dV,
\end{equation}
where $w(x,y,z)$ is a 'weight' function defined below.

It is obvious that if the magnetic field $\mathbf{B}$ has a force-free configuration everywhere in the volume of interest $V$ then $L$ must be zero. In practice, the solution with $L=0$ is hardly achievable, so minimization of the functional leads to an approximate force-free solution for the magnetic configuration. Solving the minimization problem (see Appendixes in \citet{2000ApJ...540.1150W,2004SoPh..219...87W}, for the detailed math) yields two equations, which optimize the magnetic field in the inner volume and on the boundaries of the computational domain.

The two implementations of the optimization method used in our study are different in how the boundary zone is treated. IM implementation takes into account both optimization equations as detailed by \citet{2009SoPh..257..287R} such as the magnetic field is allowed to reconfigure everywhere in the volume of interest including its top and side boundaries; $w\equiv 1$ in this implementation. The magnetic field at the bottom boundary, which represents the input magnetogram, remains fixed during the optimization. Having variable magnetic field at the top and side boundaries is potentially helpful, because, given that the NLFFF reconstruction is initiated with a potential configuration, fixing the potential field at these boundaries may have negative impact on the force-freeness of the reconstructed magnetic field.

The alternative, AS implementation follows \citet{2004SoPh..219...87W} approach  in selecting the weight function: $w(x,y,z) = w_x(x) w_y(y) w_z(z)$, where the factors $w_x(x)$ and $w_y(y)$ are both equal to 1 in the internal area of the simulation cube $(0.1-0.9) \cdot L_{x,y}$. In the buffer zone outside this internal area the factors $w_x(x)$ and $w_y(y)$ decrease towards zero as the $\cos$-functions. The other factor, $w_z(z)=1$ for the heights  $(0-0.9) \cdot L_{z}$ and then goes to zero as the $\cos$-function. The magnetic field at the top and side boundaries is frozen to be the potential one; the same as has been used to initialize the optimization. After each successful iteration the step is increased by a factor of 1.03, while after each unsuccessful iteration the step is reduced by the factor of 0.8. The optimization ends when the step has become less than 0.01 of the initial step.

One modification of the AS implementation relative to the original method is the selection of initial approximation for the magnetic field  \citep[that follows a multigrid extension proposed by][]{2008SoPh..247..269M}: for the most sparse grid (bin 9 in our tests) the initial approximation is the potential field, while for each next, denser grid the initial field is taken as an appropriate interpolation of the final (NLFFF) state of the previous grid. The mentioned modifications optimize the simulation time but have almost no effect on the final metrics of the reconstructed NLFFF cube.

	Alternative implementations has different time performances. Both codes spend comparable time for a single iteration. For example, in case of bin 9, IM code does $\approx 770$ iterations per second, while the speed of AS code is $\sim 200$ iterations per second (calculations were performed on 4 core 3.4~GHz processor). However, AS code has to make only about  $2,500$ iterations in total to get the final result for bin=9 due to dynamically changed time step;  the number of required iterations is getting significantly smaller (several hundreds) for the higher-resolution grids as they use interpolated solution of the scarcer grid as the initial condition, which is a much better approximation to the reality than the potential extrapolation used for bin=9 case. IM code has a fixed time step, makes $10^{5}$ iterations in total and even more for lower bin factors. Therefore AS implementation requires a considerably shorter computational time.


\subsection{Metrics for evaluation of the NLFFF method performance}

There are two questions we want to answer about the performance of our NLFFF extrapolation codes: (1) how close the final data cubes are to the targeted force-free state and (2) how well they reproduce the original field in the entire 3D domain. To address the first question we use the same metrics as we have used to evaluate the force-freeness of the original model field itself, Eqns~(\ref{IM_E01})--(\ref{IM_E02}).

To assess how close the NLFFF extrapolated data cube is to the corresponding model data cube, we use the ``angular'' metrics similar to Eqn~(\ref{IM_E01}):
\begin{equation}\label{IM_E03}
    \theta_{m}
    =
    arccos
    \left(
        \frac{
            \sum_{i}^{N}
            \tau_{i}
        }
        {
            N
        }
    \right)
    ,
    \quad
    \theta_{mj}
    =
    arccos
    \left(
        \frac{
            \sum_{i}^{N}
            \left|
                \textit{\textbf{j}}
            \right|_{i}
            \tau_{i}
        }
        {
            \sum_{i}^{N}
            \left|
                \textit{\textbf{j}}
            \right|_{i}
        }
    \right)
    ,
    \quad $$$$
    \tau_{i}
    =
    \frac{
        \textit{\textbf{B}}_{{\rm NLFFF},~i}
        \cdot
        \textit{\textbf{B}}_{i}
    }
    {
        \left|
            \textit{\textbf{B}}_{{\rm NLFFF}}
        \right|_{i}
        \left|
            \textit{\textbf{B}}
        \right|_{i}
    },
\end{equation}
where  $\textbf{\textit{j}}$ is the electric current density, computed for reconstructed field, the summation is performed over the voxels of the analyzed volume subdomain, 
$\tau_i$ is the cosine of the angle between the restored and model magnetic field at the $i$-th node of the computational grid; $\theta_m$ is the angle averaged over all nodes, in the ideal case of the force-free field it must be zero; $\theta_{mj}$ is a similar metrics but weighted with the restored electric current that ensures that the contribution form nodes with strong electric current dominates this metrics.
For a voxel-to-voxel inspection we compute the local error (residual) $ \Delta_{\alpha}[j]$, the local relative error $\delta_{\alpha}[j]$, and local normalized residual $\chi_{\alpha}^2[j]$ as 

\begin{equation}\label{Eq_NLFFF_err_loc_def}
   \Delta_{\alpha}[j]= B_{\rm NLFFF, \alpha}[j]-\overline{B}_{\alpha}[j], \qquad $$$$
   \delta_{\alpha}[j]= \frac{B_{\rm NLFFF, \alpha}[j]-\overline{B}_{\alpha}[j]}{ \langle\overline{B}_{\alpha}[j]\rangle}, \qquad \alpha=x,~y,~{\rm or}~z
   ,
\end{equation}
\begin{equation}\label{Eq_NLFFF_chi2_loc_def}
   \chi_{\alpha}^2[j]= \frac{(B_{\rm NLFFF, \alpha}[j]-\overline{B}_{\alpha}[j])^2}{\delta B_{\alpha}^2[j]}, \qquad \alpha=x,~y,~{\rm or}~z
   ,
\end{equation}
where $j$ is the number of a given voxel, $\langle\overline{B}_{\alpha}\rangle=\sqrt{\overline{B}_{\alpha}^2 + \delta B_{\alpha}^2}$, $\overline{B}_{\alpha}[j]$ and $\delta B_{\alpha}^2$ are defined for each binning factor by Equations~(\ref{Eq_B_mean_def}) and (\ref{Eq_B_var_def}), while $B_{\rm NLFFF, \alpha}$ is the corresponding component of the magnetic field obtained from the extrapolation.
Here, to compute the relative error we take into account that after the cube rebinning the magnetic field in each voxel is only known to the accuracy of $\overline{B}_{\alpha}\pm \delta B_{\alpha}$. Thus, in the denominators of $\delta_{\alpha}[j]$ in Eq.~(\ref{Eq_NLFFF_err_loc_def}) we use $\langle\overline{B}_{\alpha}\rangle$ rather than $\overline{B}_{\alpha}$; otherwise, in `singular' points, where $\overline{B}_{\alpha}$ in the denominator is very close to zero, such a metrics would artificially underestimate the accuracy. However, to compute a similar metrics for the absolute value of the magnetic field vector, we do not add any $\delta B$ because the absolute value is never too close to zero in the analyzed volume.

To characterize the extrapolation performance in a given subdomain, which can be, for example, a given layer or the entire data cube, we use the normalized rms residual $\Delta_{\rm rms}$, the normalized rms error $\delta_{\rm rms}$, and 'effective $\chi^2$' metrics
defined as

\begin{equation}\label{Eq_NLFFF_NormRes_def}
   \Delta_{\rm rms, \alpha}=\sqrt{\frac{\sum\limits_{j=1}^{N_{\rm vox}}  \Delta_{\alpha}^2[j]}{\sum\limits_{j=1}^{N_{\rm vox}} \overline{B}_{\alpha}^2[j]}}
   , \qquad \alpha=x,~y,~{\rm or}~z
   ,
\end{equation}

\begin{equation}\label{Eq_NLFFF_err_def}
   \delta_{\rm rms, \alpha}=\sqrt{\frac{1}{N_{\rm vox}}\sum\limits_{j=1}^{N_{\rm vox}}  \delta_{\alpha}^2[j]}
   , \qquad \alpha=x,~y,~{\rm or}~z
   ,
\end{equation}

\begin{equation}\label{Eq_NLFFF_chi2_def}
   \chi_{\rm eff, \alpha}^2=\frac{1}{N_{\rm vox}}\sum\limits_{j=1}^{N_{\rm vox}} \chi_{\alpha}^2[j] 
   , \qquad \alpha=x,~y,~{\rm or}~z
   ,
\end{equation}
where the summation is performed over the subdomain of the data cube\footnote{All reconstructed data cubes obtained in our study are available at our project web-site: \url{http://www.ioffe.ru/LEA/SF_AR/files/Magnetic_data_cubes/Extrapolations-Bifrost/index.html}.} used for the analysis; $N_{\rm vox}$ is the total number of voxels in the selected subdomain. 
The normalized rms residual, Eq.~(\ref{Eq_NLFFF_NormRes_def}), is similar to metrics (\ref{eq:EB_prep_metric}) used to evaluate the preprocessing performance; this metric gives more weight to the voxels, where the magnetic field is strong. In contrast, metric (\ref{Eq_NLFFF_err_def}) gives equal weight to any voxel, with either strong or weak field. The $\chi^2$ metrics weights the voxel in accordance with the uncertainty to which the magnetic field is known in the given voxel.

\subsubsection{NLFFF extrapolations from the chromospheric boundary}


We expect that any extrapolation approach will perform best if the bottom boundary condition is close to being force-free. Thus, testing the extrapolation performance from the nearly force-free chromospheric layer will characterize the true potential of the given extrapolation code itself. For this reason we begin our tests from the NLFFF data cubes extrapolated from the chromospheric level.

Table~\ref{IM_T02} presents angular metrics that characterize the force-freeness of the reconstructed field (the angles $\theta$ and $\theta_j$) and its closeness to the original model field (the angles $\theta_m$ and $\theta_{mj}$) computed for the entire 3D volume above the bottom boundary and excluding the top and side buffer zones. It is interesting that the IM code provides a more force-free magnetic field than the one in the model: the corresponding values $\theta$ and $\theta_j$ in Table~\ref{IM_T02} are systematically lower than those in Table~\ref{IM_T01}. This implies that the dynamics and the finite gas pressure in the coronal volume described by the MHD model produce a measurable deviation of the coronal magnetic field from the force-free state, while the NLFFF optimization drives the magnetic data cube towards another, more force-free, solution. Nevertheless, the extrapolated field is reasonably close to the model one; see $\theta_m$ and $\theta_{mj}$ metrics. In contrast, the field restored with the AS code is less force-free than the model one, which is likely a negative effect of the fixed top and side boundary conditions and the buffer zone employed in this method.

\begin{figure}
 \textbf{(a) IM chr } \ \ \textbf{(b) AS chr}  \ \ \textbf{(c) IM ph}  \ \  \textbf{(d) AS ph} \\
\centering
\includegraphics[width=0.243\columnwidth]{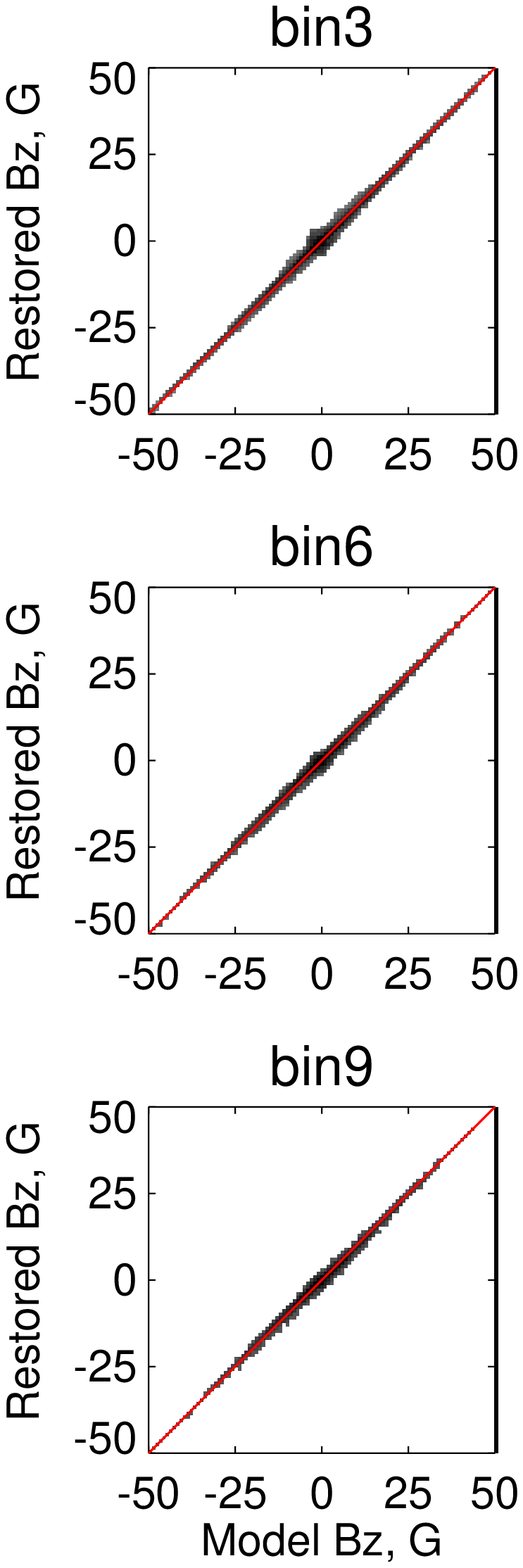}
\includegraphics[width=0.215\columnwidth, clip]{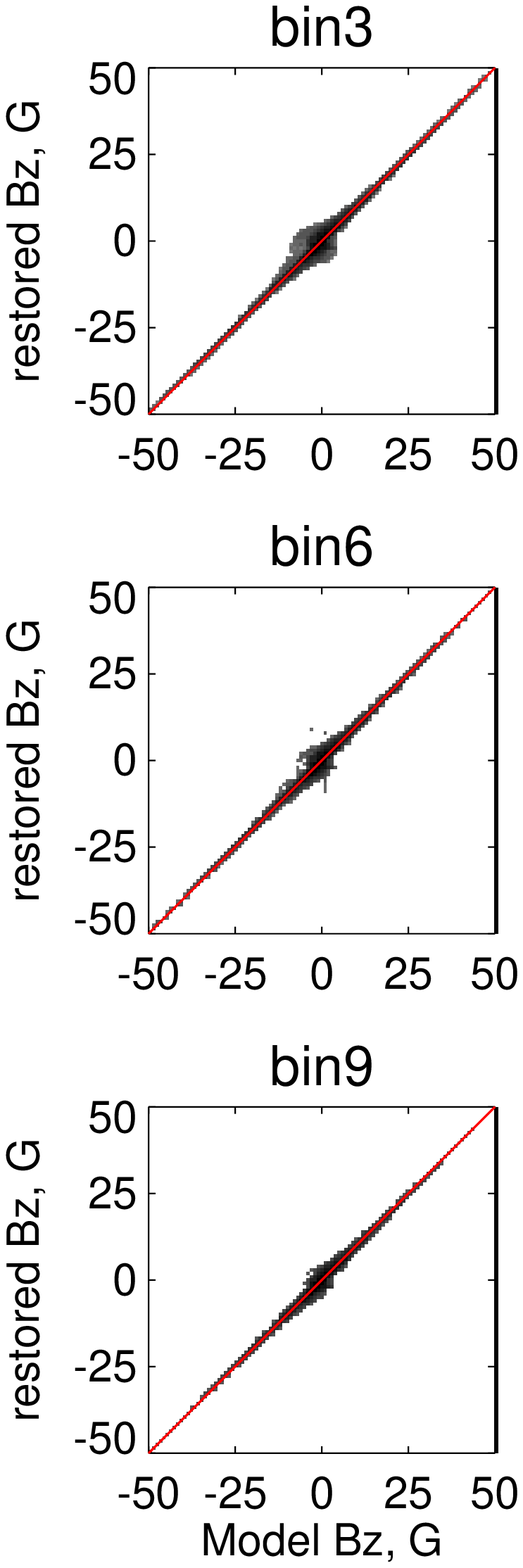}
\includegraphics[width=0.23\columnwidth, clip]{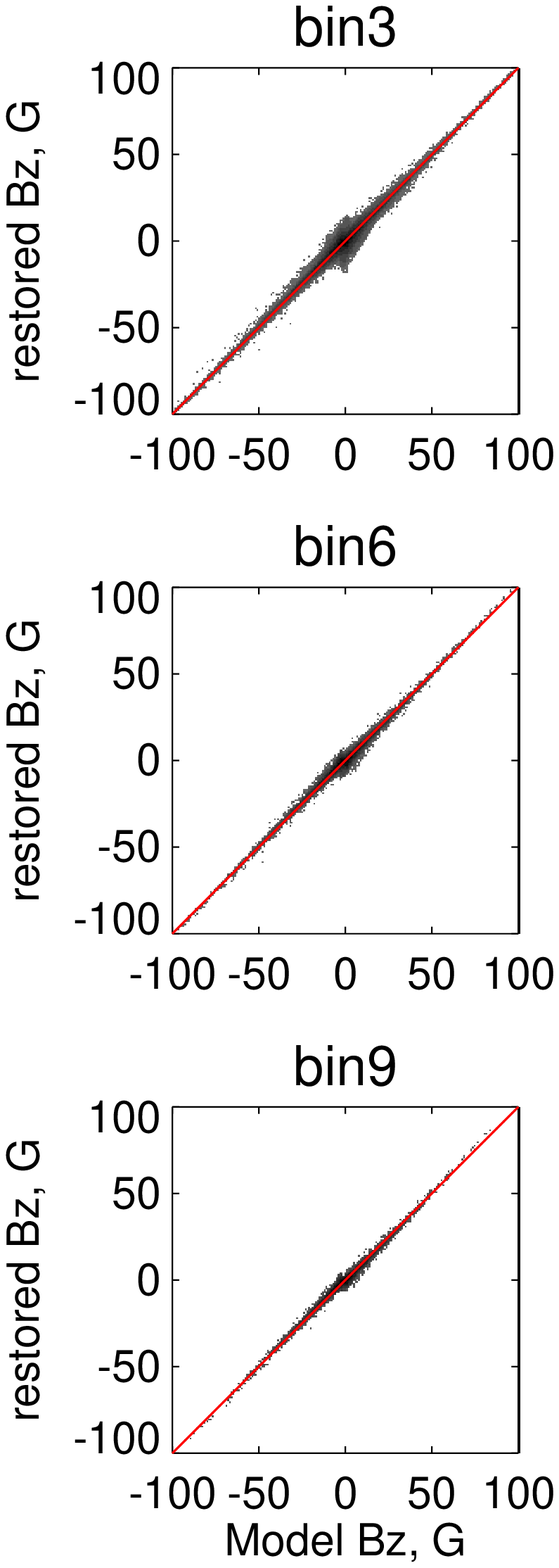}
\includegraphics[width=0.23\columnwidth, clip]{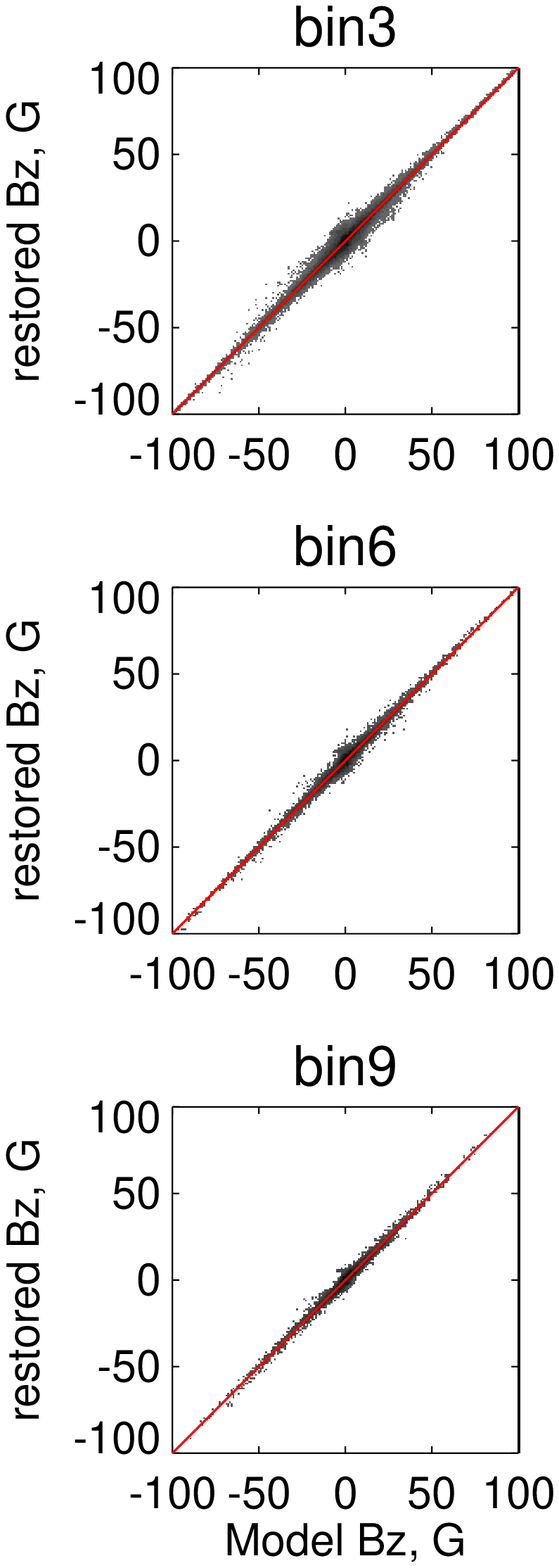}\\
\caption{\label{f_bifrost_385_NLFFF_2dh_Bz}
 2D histograms of $B_z$ reconstruction obtained using two methods, IM \& AS, from the chromosphere and the {$\beta$-}photosphere without preprocessing---in 3D volume. The buffer zone is discarded everywhere.
 }
\end{figure}

\begin{figure}
 \textbf{(a) IM chr } \ \ \textbf{(b) AS chr}  \ \ \textbf{(c) IM ph}  \ \  \textbf{(d) AS ph} \\
\centering
\includegraphics[width=0.243\columnwidth]{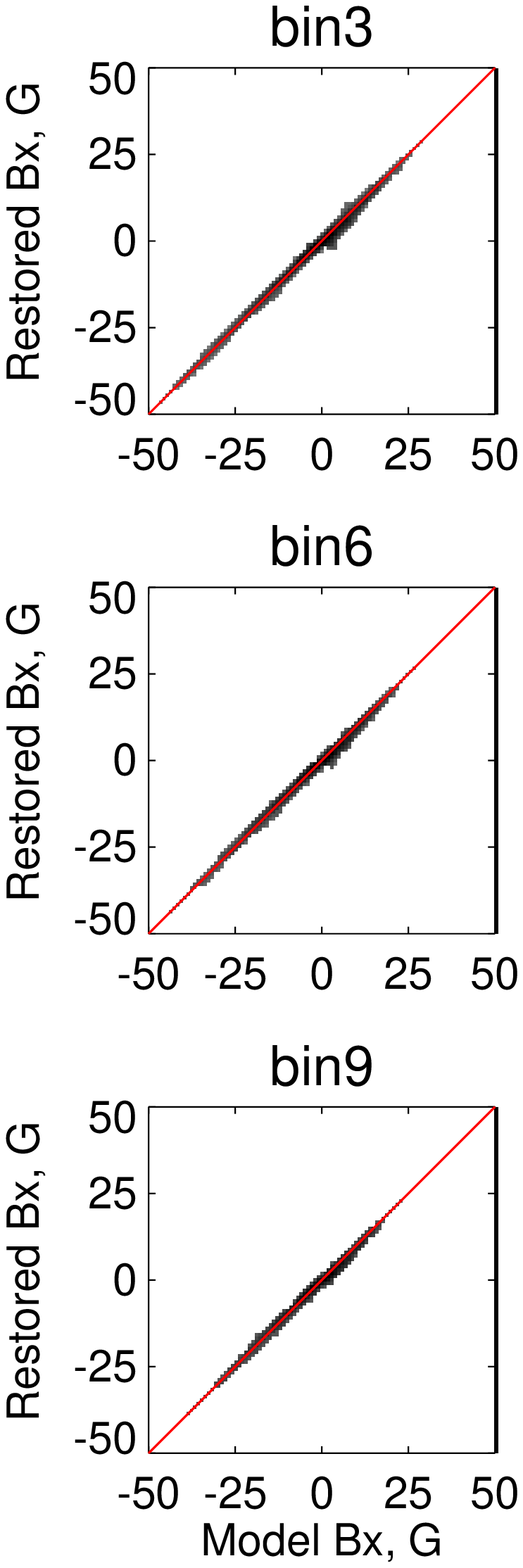}
\includegraphics[width=0.215\columnwidth, clip]{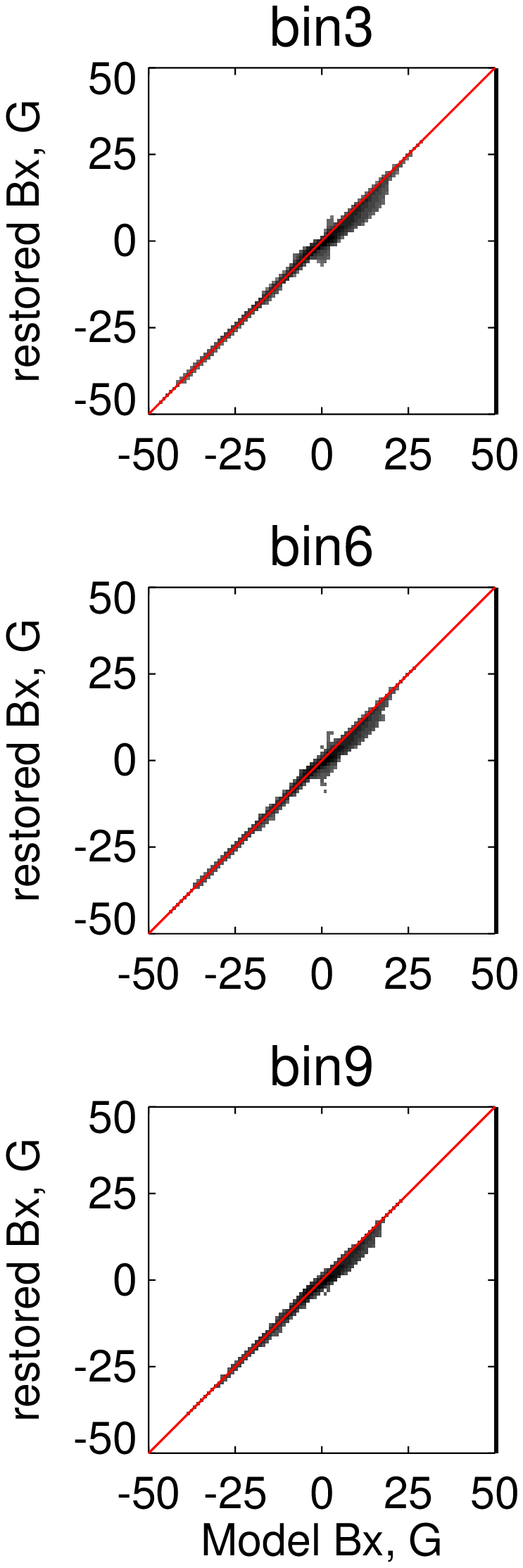}
\includegraphics[width=0.23\columnwidth, clip]{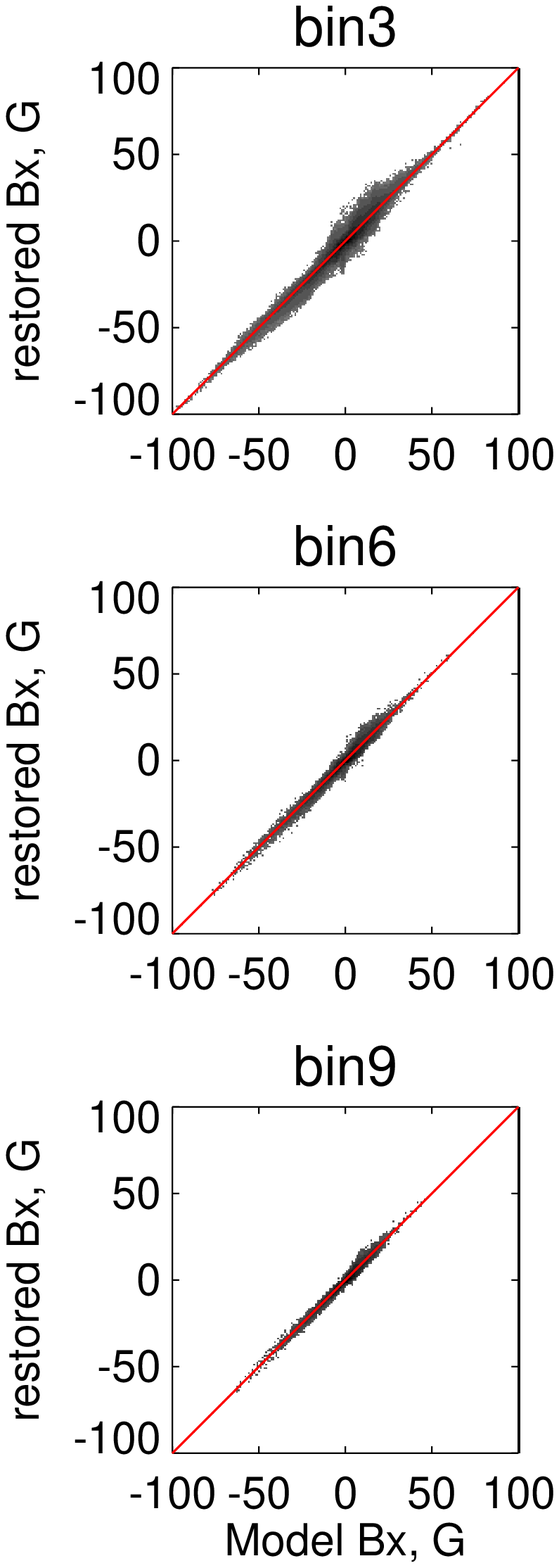}
\includegraphics[width=0.23\columnwidth, clip]{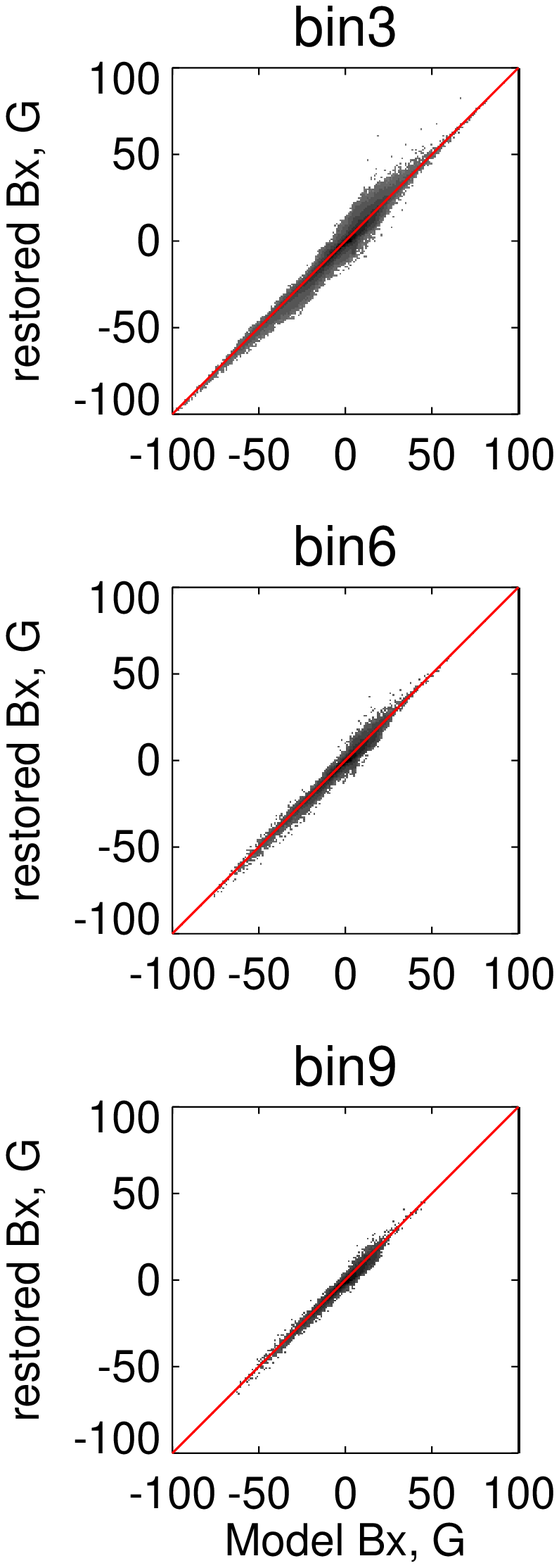}\\
\caption{\label{f_bifrost_385_NLFFF_2dh_Bx}
2D histograms of $B_x$ reconstruction obtained using two methods, IM \& AS, from the chromosphere and the {$\beta$}-photosphere without preprocessing---in 3D volume. The buffer zone is discarded everywhere.
 }
\end{figure}

\begin{figure}
 \textbf{(a) IM chr } \ \ \textbf{(b) AS chr}  \ \ \textbf{(c) IM ph}  \ \  \textbf{(d) AS ph} \\
\centering
\includegraphics[width=0.243\columnwidth]{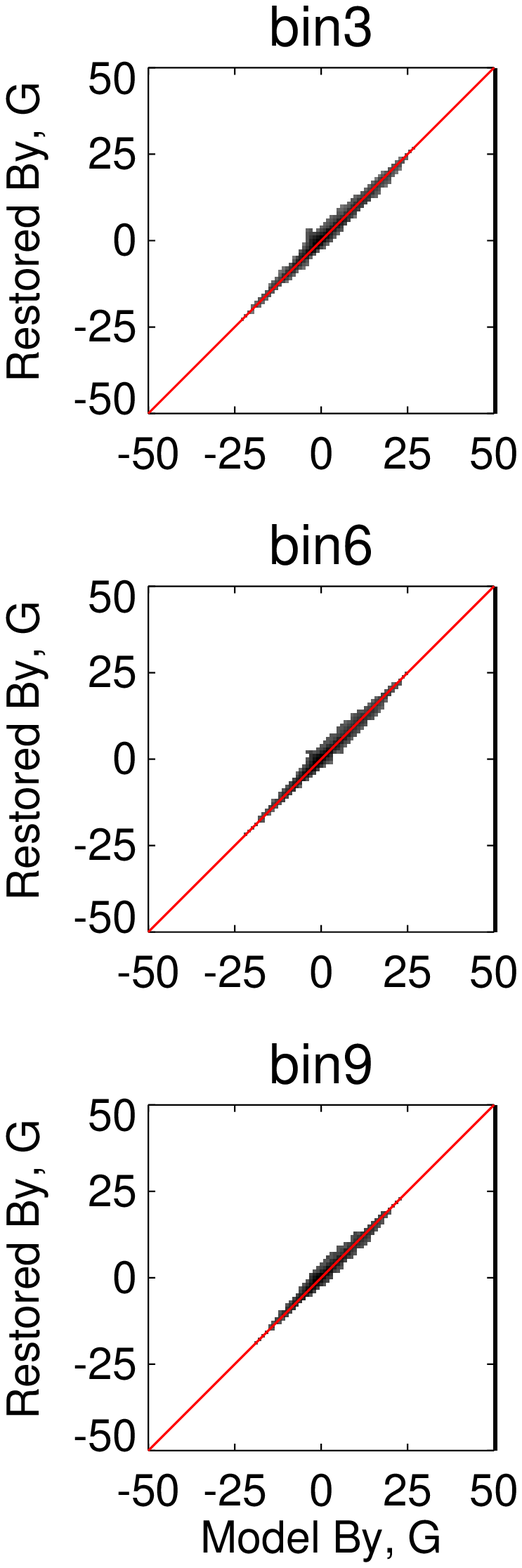}
\includegraphics[width=0.215\columnwidth, clip]{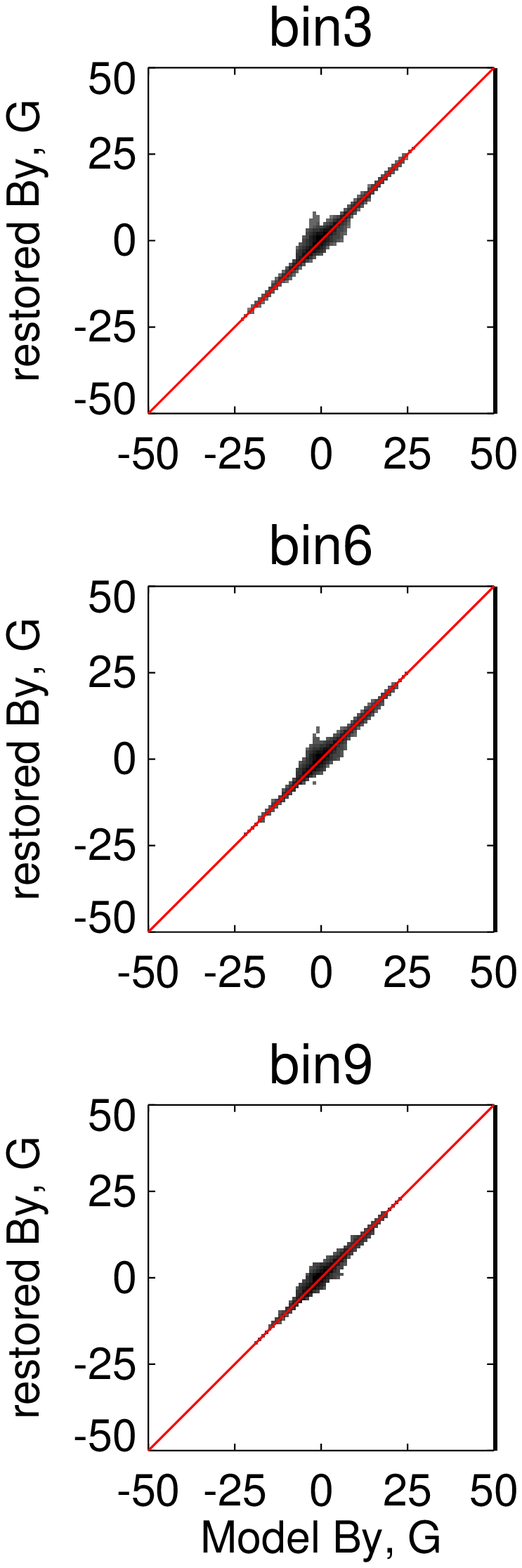}
\includegraphics[width=0.23\columnwidth, clip]{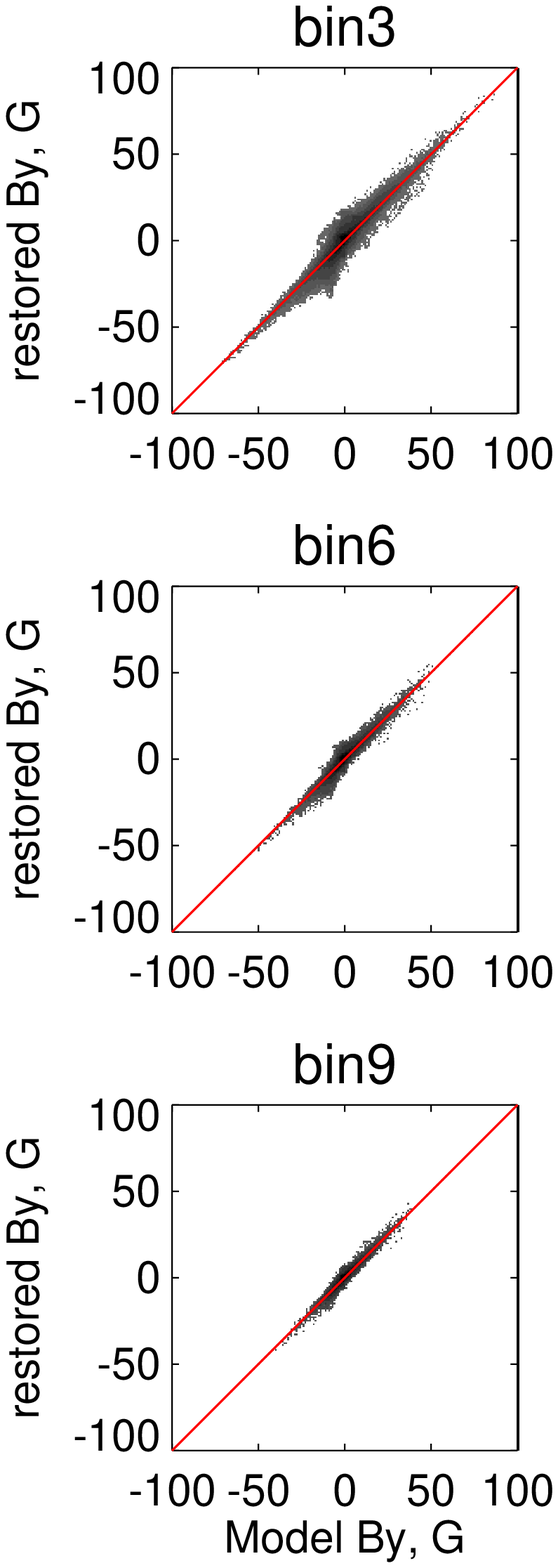}
\includegraphics[width=0.23\columnwidth, clip]{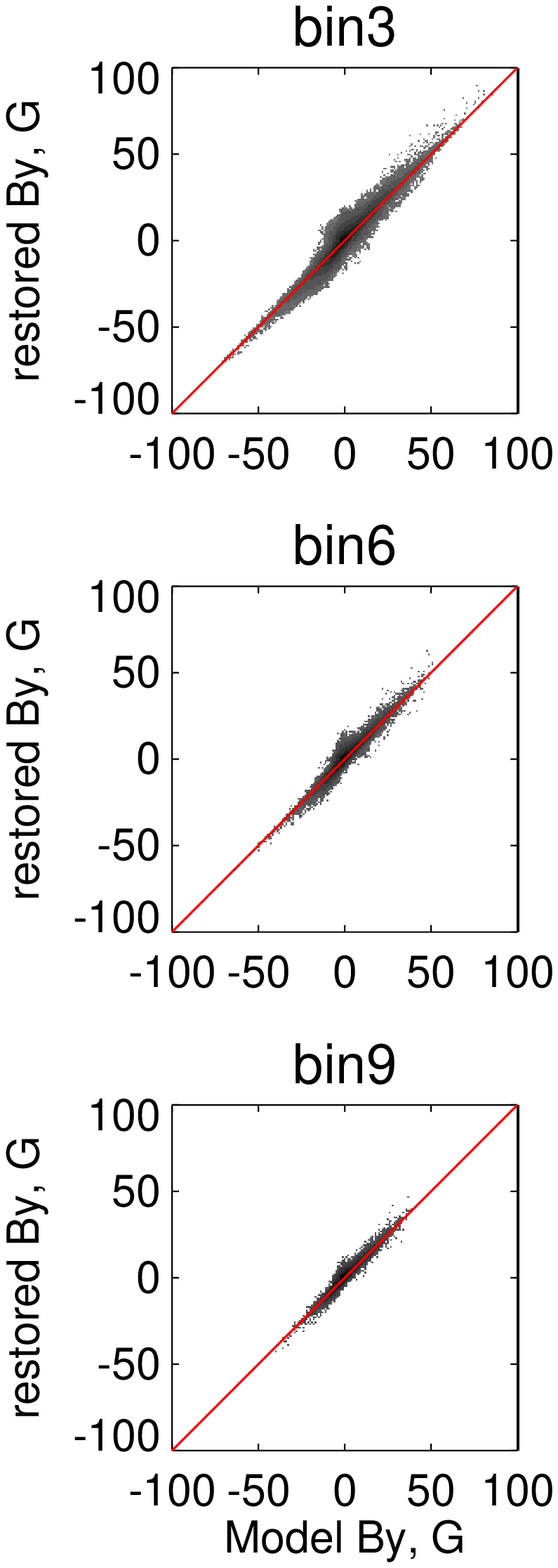}\\
\caption{\label{f_bifrost_385_NLFFF_2dh_By}
2D histograms of $B_y$ reconstruction obtained using two methods, IM \& AS, from the chromosphere and the {$\beta$-}photosphere without preprocessing---in 3D volume. The buffer zone is discarded everywhere.
 }
\end{figure}

\begin{figure}\centering
\includegraphics[width=0.57\columnwidth]{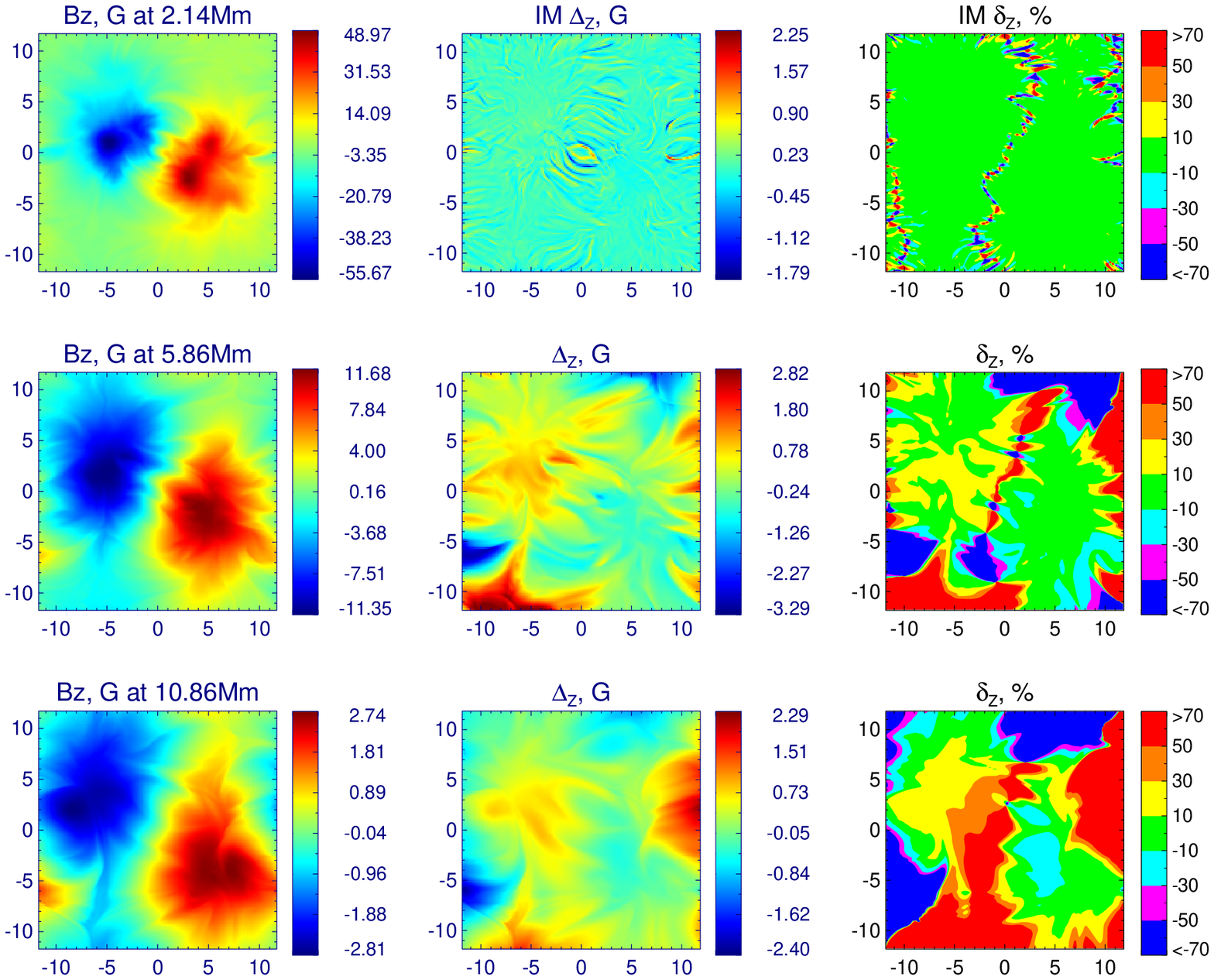} \ \  
\includegraphics[width=0.37\columnwidth,clip]{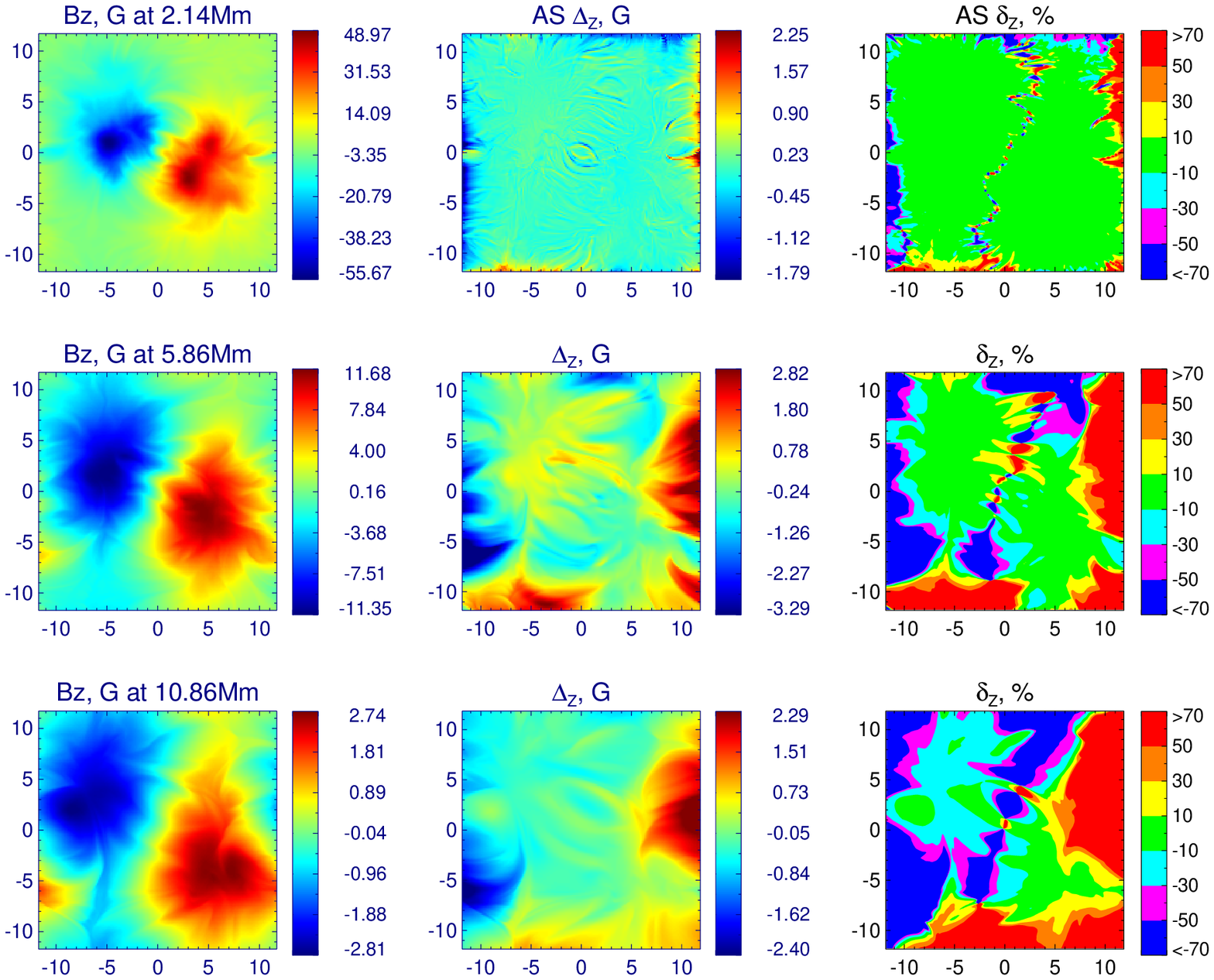}\\
\caption{\label{f_bifrost_385_rebin4z_chrNLFFF_Bz} The model $B_z$ field distributions (left column) and performance of the IM (next two columns) and AS (two right columns) NLFFF extrapolations ($B_z$ component) from the chromospheric level (bin=3) at three levels (their height are shown at the panel titles); {second and fourth columns: residual between the extrapolated and the model field; third and fifth column: relative error. The residual is within 2--3~G at all levels, while the relative error increases with height, because the field strength decreases with the height. The relative error is also bigger along the `neutral lines,' where the field is close to zero. The results for other components and other binning factors are similar to those shown in this figure. {The animated version of this figure shows the same information but for all layers of the reconstructed $B_z$ data cubes separately for the IM and AS extrapolations. Each frame of the animations shows four panels at a given layer: (a) the model field, (b) the restored field, (c) the residual, and (d) the relative error.}}
 }
\end{figure}

\begin{figure}\centering
\includegraphics[width=0.57\columnwidth]{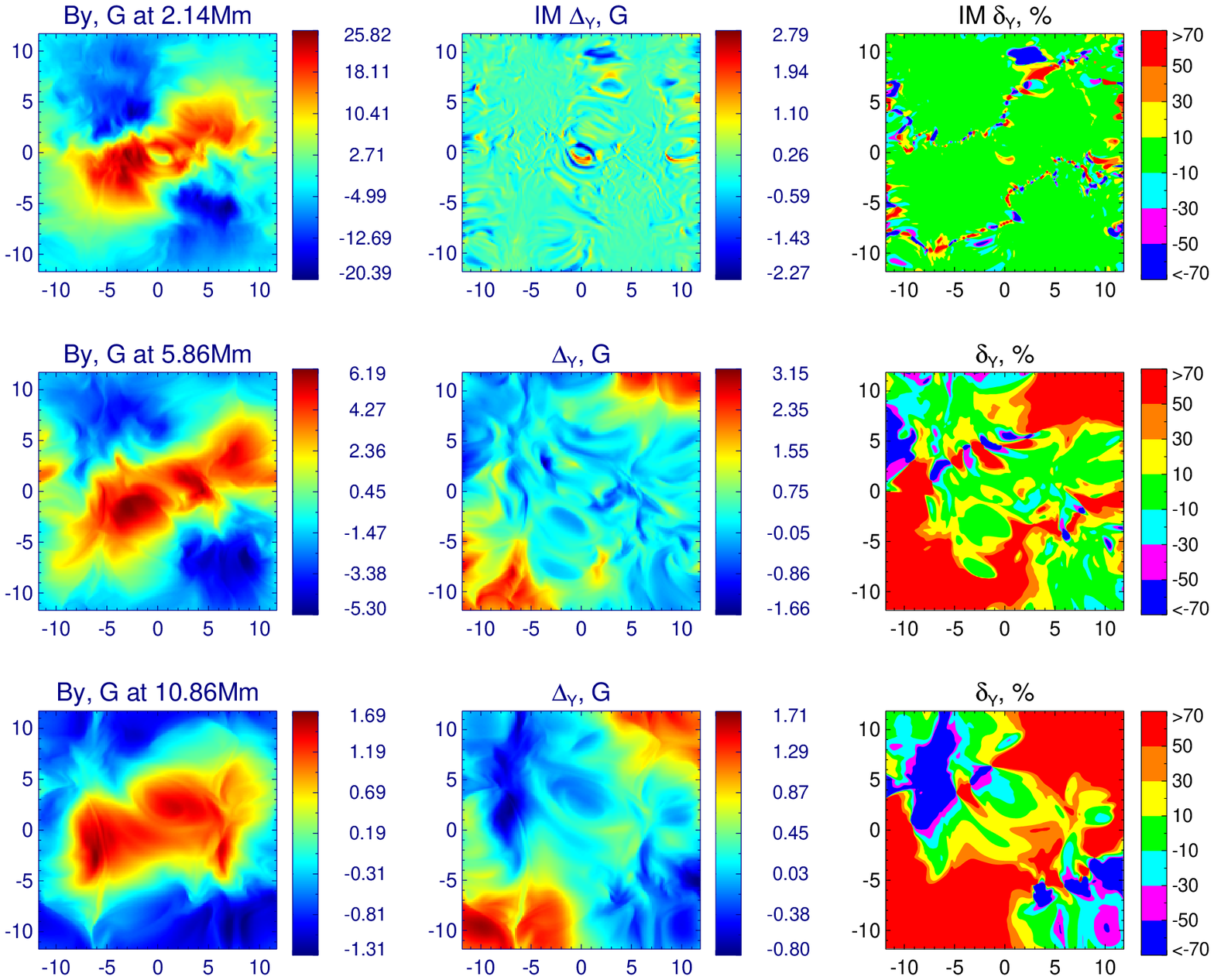} \ \ 
\includegraphics[width=0.37\columnwidth, clip]{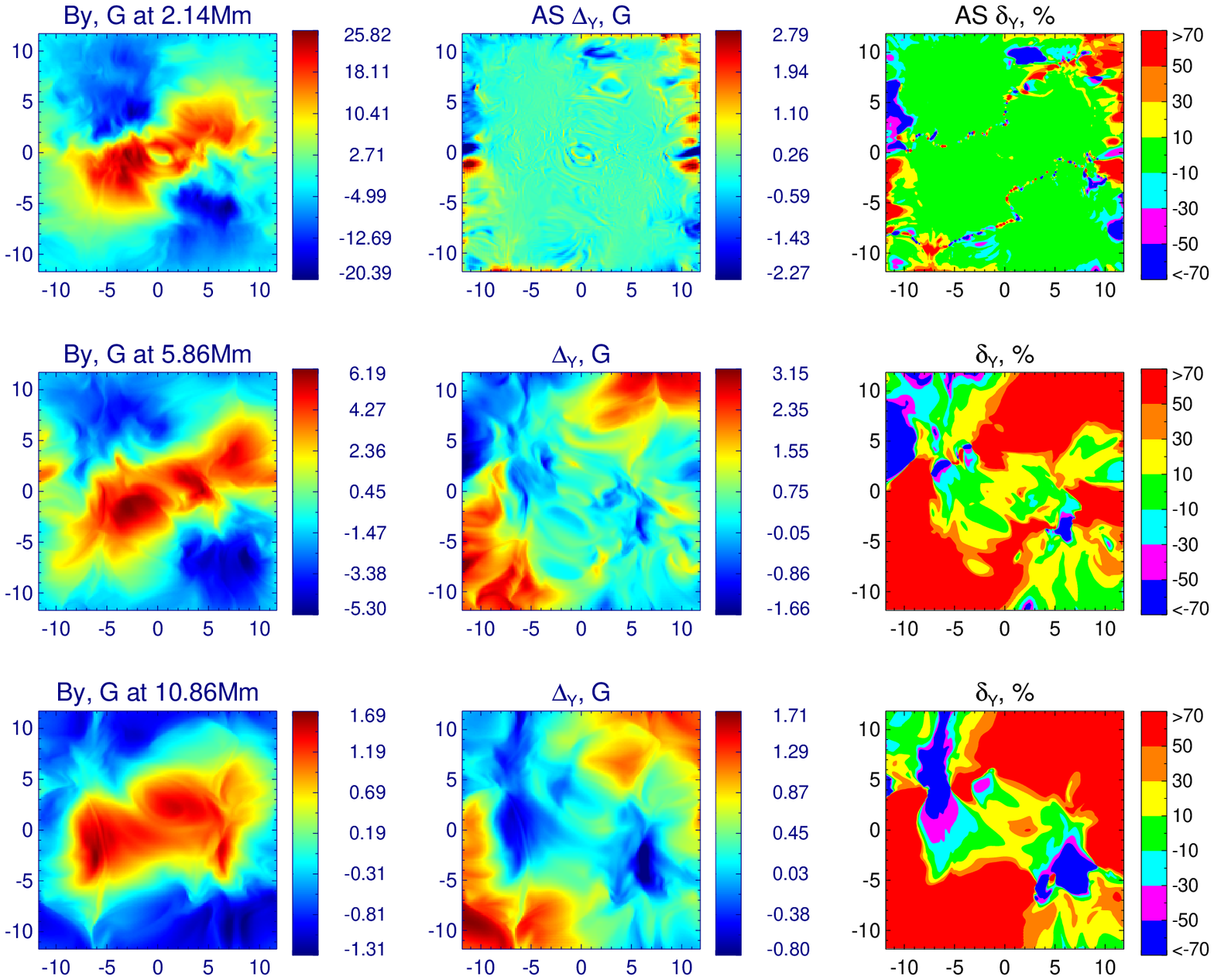}\\
\caption{\label{f_bifrost_385_rebin4z_chrNLFFF_By} The model $B_y$ field distributions (left column) and performance of the IM (next two columns) and AS (two right columns) NLFFF extrapolations ($B_y$ component) from the chromospheric level (bin=3) at three levels (their height are shown at the panel titles); {second and fourth columns: residual between the extrapolated and the model field; third and fifth column: relative error. The residual is within 2--3~G at all levels, while the relative error increases with height, because the field strength decreases with the height. The relative error is also bigger along the `neutral lines,' where the field is close to zero. The results for other components and other binning factors are similar to those shown in this figure. {The animated version of this figure shows the same information but for all layers of the reconstructed $B_y$ data cubes separately for the IM and AS extrapolations. Each frame of the animations shows four panels at a given layer: (a) the model field, (b) the restored field, (c) the residual, and (d) the relative error.}}
 }
\end{figure}

\newpage
Likewise in the preprocessing tests, we are looking if the restored values of the magnetic field components (or the absolute values) correlate with the model ones and what is the scatter around the cross-correlation curves. Figures~\ref{f_bifrost_385_NLFFF_2dh_Bz}---\ref{f_bifrost_385_NLFFF_2dh_By}, two left columns, display these cross-correlation plots in the form of 2D histograms superimposed on the $y=x$ diagonal line. These plots suggest that the performance of the extrapolations improves for higher binning factors; in particular the `clouds` in the area of poorer restored weak field values (around zero) are bigger for the smaller binning factors. Comparison between the two methods suggests that the IM approach works better for small magnetic field values, while for large magnetic field both methods perform comparably or, sometimes, the AS method performs marginally better. We will return to this comparison later.

Figures~\ref{f_bifrost_385_rebin4z_chrNLFFF_Bz},~\ref{f_bifrost_385_rebin4z_chrNLFFF_By} give a clear visual idea of the optimization code performance at three layers: one close to the bottom of the cube, another one---in a middle height, and the last one---close to the top buffer zone in the AS code (for fair comparison of the methods we exclude the same buffer zone in both AS and IM cases). The error (2nd and 4th columns) of the magnetic field reconstruction slightly increases with height: although the scale of the error variation is almost the same at all of the three presented levels, $-2~G\lesssim  \Delta_{z} \lesssim 2~G$, the areas occupied by blue or red colors (indicating bigger errors) increase with the height.  As a result, the relative error (3rd and 5th columns) increases with the height noticeably: the green area where the relative error is within $\pm10\%$ is getting smaller with height.  The relative error is getting large at  neutral lines (i.e., where the magnetic field is about zero) at any layer, even at the one closest to the bottom.

At the low heights the two competing methods perform comparably well: although the AS methods shows smaller error in the middle of the plot, where the magnetic field is reasonably large, it also produces some artifacts close to the boundaries, which is not surprising given that the method employs a buffer zone at the boundary regions. It is, however, interesting, that at the intermediate heights the AS method provides a bigger green area in the plot, where the relative error of the field reconstruction is within 10~\%, than the IM method, although the IM method outperforms the AS one at higher heights. A qualitatively similar picture is observed for two other components of the magnetic field: $B_y$ (Fig.~\ref{f_bifrost_385_rebin4z_chrNLFFF_By}) and $B_x$ (not shown).

\begin{deluxetable*}{c|cccc|cccc|cccc|cccc}

\rotate
				
\tablecolumns{15}

\tablewidth{0pc}

\tabletypesize{\footnotesize}

\tablecaption{Normalized rms error at a given level for bin-factor 9. \label{table_rms_error}}

\tablehead{
\multicolumn{1}{c|}{} & \multicolumn{4}{c|}{$\delta_{rms}(B)$}& \multicolumn{4}{c|}{$\delta_{rms}(Bx)$} & \multicolumn{4}{c|}{$\delta_{rms}(By)$} & \multicolumn{4}{c}{$\delta_{rms}(Bz)$}\\
\multicolumn{1}{c|}{} &  \multicolumn{2}{c}{chromo}&  \multicolumn{2}{c|}{{$\beta$-}photo}& \multicolumn{2}{c}{chromo}& \multicolumn{2}{c|}{{$\beta$-}photo} & \multicolumn{2}{c}{chromo} & \multicolumn{2}{c|}{{$\beta$-}photo} & \multicolumn{2}{c}{chromo} & \multicolumn{2}{c}{{$\beta$-}photo}\\
				\multicolumn{1}{c|}{Level, Mm} & \colhead{IM} & \multicolumn{1}{c}{AS} & \colhead{IM} & \multicolumn{1}{c|}{AS}  & \colhead{IM} & \multicolumn{1}{c}{AS} & \colhead{IM} & \multicolumn{1}{c|}{AS} & \colhead{IM} & \multicolumn{1}{c}{AS} & \colhead{IM} & \multicolumn{1}{c|}{AS} & \colhead{IM} & \multicolumn{1}{c}{AS} & \colhead{IM} & \multicolumn{1}{c}{AS}
				}

\startdata
     0.86 & -  & -  &     0.00 &     0.00 & -  & -  &     0.00 &     0.00 & -  & -  &     0.00 &     0.00 & -  & -  &     0.00 &     0.00 \\
     1.29 & -  & -  &    13.22 &    15.80 & -  & -  &    41.50 &    51.53 & -  & -  &    48.41 &    64.89 & -  & -  &    28.98 &    32.92 \\
     1.71 & -  & -  &    11.83 &    19.44 & -  & -  &    54.58 &    77.71 & -  & -  &    68.40 &   122.66 & -  & -  &    30.94 &    55.48 \\
     2.14 &     0.00 &     0.00 &    11.28 &    17.82 &     0.00 &     0.00 &    52.36 &    83.10 &     0.00 &     0.00 &    76.79 &   150.87 &     0.00 &     0.00 &    34.28 &    37.58 \\
     2.57 &     1.43 &     2.21 &    11.27 &    16.83 &    12.60 &    13.75 &    45.84 &    73.84 &    18.60 &    45.44 &    98.33 &   177.45 &    11.36 &    11.21 &    32.92 &    35.95 \\
     3.00 &     2.34 &     4.38 &    11.26 &    17.54 &    19.70 &    27.23 &    46.90 &    79.76 &    31.38 &   144.01 &   147.28 &   259.00 &    22.30 &    20.17 &    33.03 &    42.88 \\
     3.43 &     3.25 &     6.40 &    11.43 &    16.50 &    24.16 &    33.32 &    46.40 &    79.76 &    48.00 &   184.75 &   190.45 &   280.19 &    32.75 &    28.91 &    43.18 &    56.51 \\
     3.86 &     4.31 &     7.09 &    12.03 &    17.02 &    25.46 &    38.36 &    47.57 &    91.95 &    67.50 &   157.94 &   172.51 &   230.89 &    33.39 &    32.08 &    49.91 &    65.25 \\
     4.29 &     5.33 &     8.55 &    12.84 &    16.35 &    29.48 &    48.58 &    55.01 &   102.57 &    71.85 &   125.21 &   147.31 &   178.25 &    35.67 &    50.82 &    57.52 &    76.18 \\
     4.71 &     6.25 &     8.98 &    13.75 &    16.65 &    33.59 &    57.48 &    65.65 &   113.74 &    75.95 &   129.38 &   156.30 &   184.49 &    39.86 &    53.36 &    72.54 &    83.33 \\
     5.14 &     7.08 &     9.97 &    14.79 &    16.05 &    33.71 &    68.16 &    62.19 &   125.21 &    85.61 &   169.44 &   185.57 &   221.02 &    45.60 &    63.24 &    82.46 &    84.68 \\
     5.57 &     7.93 &    10.28 &    16.08 &    16.36 &    35.51 &    84.52 &    61.14 &   138.04 &   114.93 &   203.77 &   233.19 &   252.55 &    53.54 &    70.00 &    94.02 &    91.71 \\
     6.00 &     8.72 &    11.19 &    17.46 &    16.22 &    37.03 &    81.69 &    60.85 &   139.94 &   138.18 &   226.76 &   273.98 &   261.13 &    55.65 &    74.83 &    98.34 &    92.08 \\
     6.43 &     9.63 &    11.73 &    18.86 &    16.85 &    40.70 &    93.39 &    68.74 &   153.62 &   160.61 &   257.38 &   319.05 &   290.81 &    55.29 &    77.39 &    95.58 &    94.24 \\
     6.86 &    10.47 &    12.36 &    19.81 &    16.87 &    44.69 &   106.27 &    77.61 &   165.73 &   138.80 &   221.37 &   285.29 &   243.29 &    56.07 &    86.16 &    89.36 &   101.62 \\
     7.29 &    11.34 &    13.21 &    21.22 &    18.02 &    46.95 &   119.11 &    84.25 &   177.37 &   148.52 &   228.13 &   309.66 &   251.31 &    58.98 &    95.69 &    85.24 &   111.87 \\
     7.71 &    12.30 &    14.66 &    22.99 &    19.37 &    49.63 &   132.08 &    92.43 &   189.03 &   170.13 &   246.65 &   344.42 &   266.38 &    64.64 &   113.54 &    83.16 &   129.34 \\
     8.14 &    13.37 &    16.07 &    24.90 &    21.19 &    53.43 &   145.35 &   100.12 &   198.74 &   189.50 &   260.63 &   370.92 &   282.46 &    74.22 &   137.59 &    82.77 &   154.39 \\
     8.57 &    14.41 &    17.74 &    27.06 &    22.95 &    57.27 &   158.31 &   109.08 &   210.81 &   200.99 &   267.05 &   392.14 &   287.85 &    83.27 &   156.08 &    85.29 &   173.00 \\
     9.00 &    15.50 &    19.49 &    29.60 &    25.21 &    61.65 &   171.14 &   119.76 &   220.42 &   217.77 &   287.73 &   431.59 &   311.45 &    84.57 &   155.05 &    82.42 &   171.96 \\
     9.43 &    16.92 &    21.85 &    32.90 &    27.76 &    66.06 &   181.68 &   127.53 &   228.34 &   243.57 &   325.13 &   497.96 &   348.70 &    90.25 &   159.59 &    84.62 &   176.12 \\
     9.86 &    18.70 &    24.49 &    37.03 &    30.87 &    68.89 &   189.89 &   136.12 &   231.95 &   258.39 &   339.71 &   539.04 &   364.20 &   107.26 &   177.86 &    96.88 &   195.19 \\
    10.29 &    20.69 &    27.21 &    41.73 &    33.76 &    72.72 &   197.35 &   147.00 &   236.25 &   258.66 &   316.44 &   528.81 &   336.67 &   129.45 &   197.34 &   108.42 &   215.54 \\
    10.71 &    23.21 &    30.01 &    47.38 &    36.83 &    79.75 &   210.62 &   162.30 &   245.62 &   282.73 &   328.77 &   564.24 &   348.96 &   160.20 &   220.26 &   130.77 &   238.96 \\
    11.14 &    26.26 &    32.92 &    54.07 &    39.79 &    87.27 &   225.24 &   179.17 &   255.57 &   315.55 &   345.40 &   591.32 &   368.48 &   198.86 &   246.45 &   159.97 &   265.59 \\
          	\\
			\enddata
		\end{deluxetable*}

\begin{deluxetable*}{c|cccc|cccc|cccc|cccc}

\rotate
				
\tablecolumns{15}

\tablewidth{0pc}

\tabletypesize{\footnotesize}

\tablecaption{Normalized rms residual at a given level for bin-factor 9. \label{table_delta_error}}

\tablehead{
\multicolumn{1}{c|}{} & \multicolumn{4}{c|}{$\Delta_{rms}(B)$}& \multicolumn{4}{c|}{$\Delta_{rms}(Bx)$} & \multicolumn{4}{c|}{$\Delta_{rms}(By)$} & \multicolumn{4}{c}{$\Delta_{rms}(Bz)$}\\
\multicolumn{1}{c|}{} &  \multicolumn{2}{c}{chromo}&  \multicolumn{2}{c|}{{$\beta$-}photo}& \multicolumn{2}{c}{chromo}& \multicolumn{2}{c|}{{ $\beta$-}photo} & \multicolumn{2}{c}{chromo} & \multicolumn{2}{c|}{{$\beta$-}photo} & \multicolumn{2}{c}{chromo} & \multicolumn{2}{c}{{$\beta$-}photo}\\
				\multicolumn{1}{c|}{Level, Mm} & \colhead{IM} & \multicolumn{1}{c}{AS} & \colhead{IM} & \multicolumn{1}{c|}{AS}  & \colhead{IM} & \multicolumn{1}{c}{AS} & \colhead{IM} & \multicolumn{1}{c|}{AS} & \colhead{IM} & \multicolumn{1}{c}{AS} & \colhead{IM} & \multicolumn{1}{c|}{AS} & \colhead{IM} & \multicolumn{1}{c}{AS} & \colhead{IM} & \multicolumn{1}{c}{AS}
				}

\startdata
     0.86 & -  & -  &     0.00 &     0.00 & - & -  &     0.00 &     0.00
& -  & -  & 0.00 &     0.00 & -  & -  &     0.00 & 0.00 \\
      1.29 & -  & -  &     5.16 &     5.95 & - & -  &    10.47 &
11.99 & -  & -  & 14.83 &    16.32 & -  & -  &     5.72 & 6.52 \\
      1.71 & -  & -  &     4.34 &     6.09 & - & -  &    10.26 &
14.07 & -  & -  & 16.27 &    20.13 & -  & -  &     6.43 & 9.65 \\
      2.14 &     0.00 &     0.00 &     4.02 & 4.95 &     0.00 &     0.00
&    10.00 &    14.95 &     0.00 &     0.00 &    16.45 &    21.28 &
0.00 &     0.00 &     7.03 &     8.04 \\
      2.57 &     0.82 &     0.82 &     4.25 & 4.64 &     3.41 &     3.25
&    10.35 &    15.24 &     6.29 &     5.87 &    17.80 &    22.35 &
2.37 &     2.01 &     8.12 &     8.26 \\
      3.00 &     1.61 &     1.59 &     4.85 & 4.72 &     5.22 &     5.91
&    11.21 &    16.83 &     8.48 &     9.47 &    19.20 &    23.70 &
4.17 &     3.28 &     9.42 &     8.91 \\
      3.43 &     2.43 &     2.17 &     5.66 & 4.74 &     6.62 &     8.71
&    12.19 &    18.68 &    11.19 &    12.73 &    20.84 &    25.71 &
5.48 &     4.39 &    10.95 &     9.69 \\
      3.86 &     3.27 &     2.74 &     6.62 & 5.08 &     7.75 &    10.45
&    13.32 &    20.55 &    14.39 &    15.33 &    23.85 &    28.95 &
6.27 &     5.63 &    12.33 &    10.69 \\
      4.29 &     4.12 &     3.32 &     7.70 & 5.31 &     8.58 &    13.20
&    14.85 &    23.71 &    14.30 &    18.39 &    25.44 &    30.09 &
7.28 &     6.95 &    13.96 &    11.56 \\
      4.71 &     5.01 &     3.89 &     8.87 & 5.87 &     9.74 &    16.56
&    17.05 &    27.22 &    15.72 &    21.65 &    28.52 &    32.70 &
8.46 &     8.28 &    15.83 &    12.86 \\
      5.14 &     5.92 &     4.52 &    10.10 & 6.27 &    11.25 &    20.00
&    19.54 &    31.35 &    17.90 &    27.42 &    33.43 &    37.25 &
9.52 &     9.54 &    17.73 &    13.89 \\
      5.57 &     6.84 &     5.13 &    11.34 & 6.94 &    12.75 &    23.50
&    21.99 &    35.14 &    19.98 &    32.22 &    39.22 &    41.67 &
10.51 &    10.86 &    19.57 &    15.18 \\
      6.00 &     7.79 &     5.72 &    12.54 & 7.35 &    14.38 &    26.84
&    24.59 &    39.05 &    22.80 &    37.90 &    46.14 &    46.25 &
11.52 &    12.36 &    21.36 &    16.41 \\
      6.43 &     8.74 &     6.24 &    13.68 & 7.93 &    16.09 &    30.25
&    27.22 &    42.38 &    26.35 &    42.63 &    53.99 &    50.05 &
12.59 &    13.90 &    23.07 &    17.87 \\
      6.86 &     9.65 &     6.74 &    14.79 & 8.34 &    17.65 &    33.24
&    29.66 &    45.49 &    30.74 &    47.13 &    62.82 &    53.38 &
13.64 &    15.46 &    24.67 &    19.30 \\
      7.29 &    10.47 &     7.18 &    15.90 & 9.06 &    18.88 &    36.20
&    31.71 &    48.31 &    36.20 &    51.63 &    71.56 &    57.49 &
14.59 &    17.02 &    26.12 &    20.77 \\
      7.71 &    11.22 &     7.66 &    17.03 & 9.71 &    20.02 &    39.07
&    33.60 &    51.25 &    39.73 &    54.41 &    77.35 &    59.64 &
15.51 &    18.43 &    27.53 &    22.09 \\
      8.14 &    11.92 &     8.22 &    18.20 & 10.73 &    21.30 &
42.53 &    35.82 &    54.56 &    41.81 &    55.22 &    81.15 &    60.63
&    16.33 &    19.82 &    28.93 &    23.35 \\
      8.57 &    12.57 &     8.90 &    19.49 & 11.74 &    22.77 &
45.91 &    38.20 &    57.75 &    44.67 &    57.16 &    85.70 &    62.17
&    16.97 &    20.92 &    30.34 &    24.39 \\
      9.00 &    13.19 &     9.73 &    21.00 & 13.11 &    23.99 &
48.98 &    40.15 &    60.23 &    47.68 &    59.17 &    91.25 &    64.35
&    17.88 &    22.33 &    32.16 &    25.74 \\
      9.43 &    13.93 &    10.83 &    22.86 & 14.53 &    25.10 &
51.61 &    41.95 &    62.20 &    50.31 &    60.94 &    96.05 &    65.71
&    19.40 &    24.04 &    34.69 &    27.57 \\
      9.86 &    14.82 &    12.18 &    25.18 & 16.29 &    26.07 &
54.19 &    43.69 &    63.82 &    52.47 &    61.42 &   100.18 &    66.14
&    21.48 &    26.32 &    38.13 &    29.97 \\
     10.29 &    15.86 &    13.80 &    28.03 & 18.12 &    26.99 &
56.54 &    45.50 &    65.12 &    54.51 &    61.29 &   104.38 &    65.47
&    24.13 &    29.03 &    42.77 &    33.04 \\
     10.71 &    17.17 &    15.63 &    31.58 & 20.18 &    27.87 &
59.03 &    47.44 &    66.23 &    56.75 &    60.18 &   108.74 &    64.02
&    27.54 &    32.55 &    49.08 &    36.91 \\
     11.14 &    18.96 &    17.74 &    35.95 & 22.37 &    29.07 &
61.01 &    49.48 &    66.74 &    59.80 &    58.38 &   113.66 &    61.87
&    32.25 &    37.10 &    57.94 &    41.99 \\
 	\\
			\enddata
		\end{deluxetable*}

The fact that we created our model data cubes by rebinning the original data cube implies that the magnetic field value $\overline{B}_{\alpha}$ in a given voxel is only known to the accuracy $\overline{B}_{\alpha}\pm \delta B_{\alpha}$ defined by Eqns~(\ref{Eq_B_mean_def}),~(\ref{Eq_B_var_def}); thus, even the most precise field reconstruction would only recover the field to this accuracy. Stated another way, an ideally perfect reconstruction would have $\chi_{\alpha}^2[j] \sim 1$ and $\chi_{\rm eff, \alpha}^2 \sim 1$. 
We evaluated the $\chi_{\rm eff, \alpha}^2$ metrics for our extrapolation data cubes and found them to be much larger than 1 in all cases. Inspection of the inputs used to compute $\chi_{\rm eff, \alpha}^2$ metrics shows that the $ \delta B_{\alpha}$ from Eqn.~(\ref{Eq_B_var_def}) are very small implying that the magnetic field in the model volume is known with very high accuracy even after rebinning. No reconstruction can be performed with that high accuracy; so the formal $\chi^2$ test fails.

\begin{figure}
 \textbf{(a) IM chr } \ \ \textbf{(b) AS chr}  \ \ \textbf{(c) IM ph}  \ \  \textbf{(d) AS ph} \\
\centering
\includegraphics[width=0.23\columnwidth]{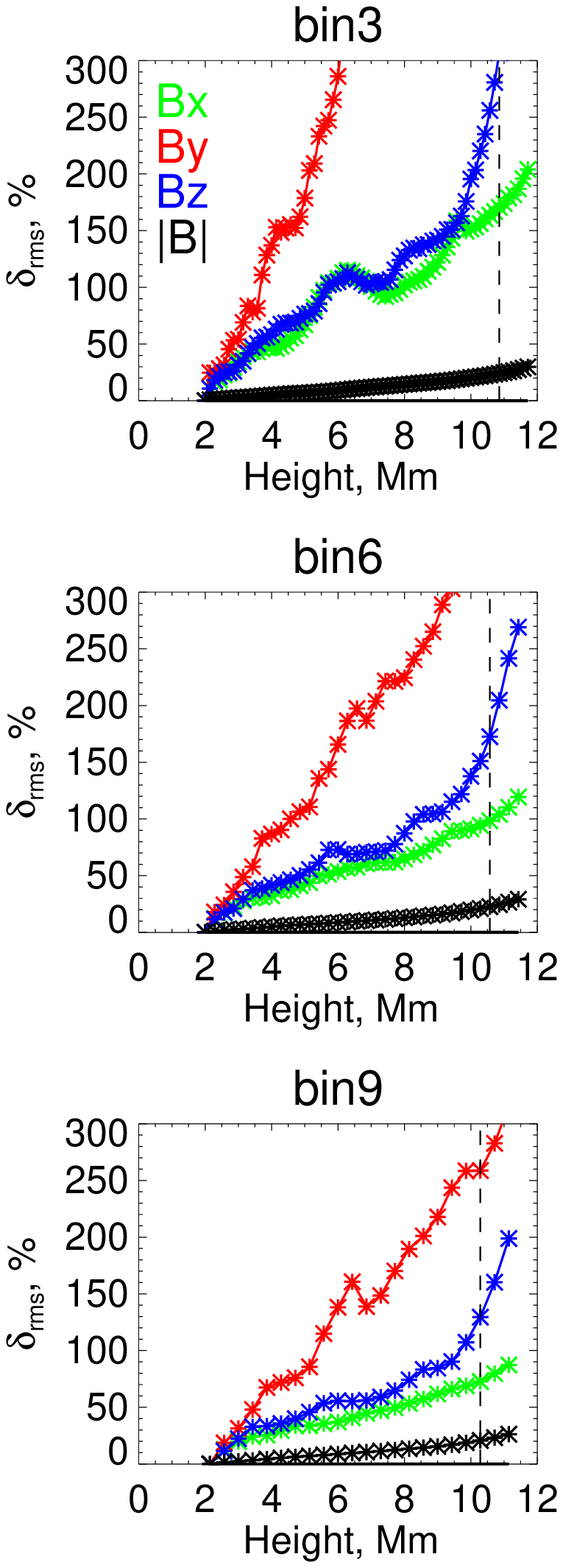}
\includegraphics[width=0.23\columnwidth]{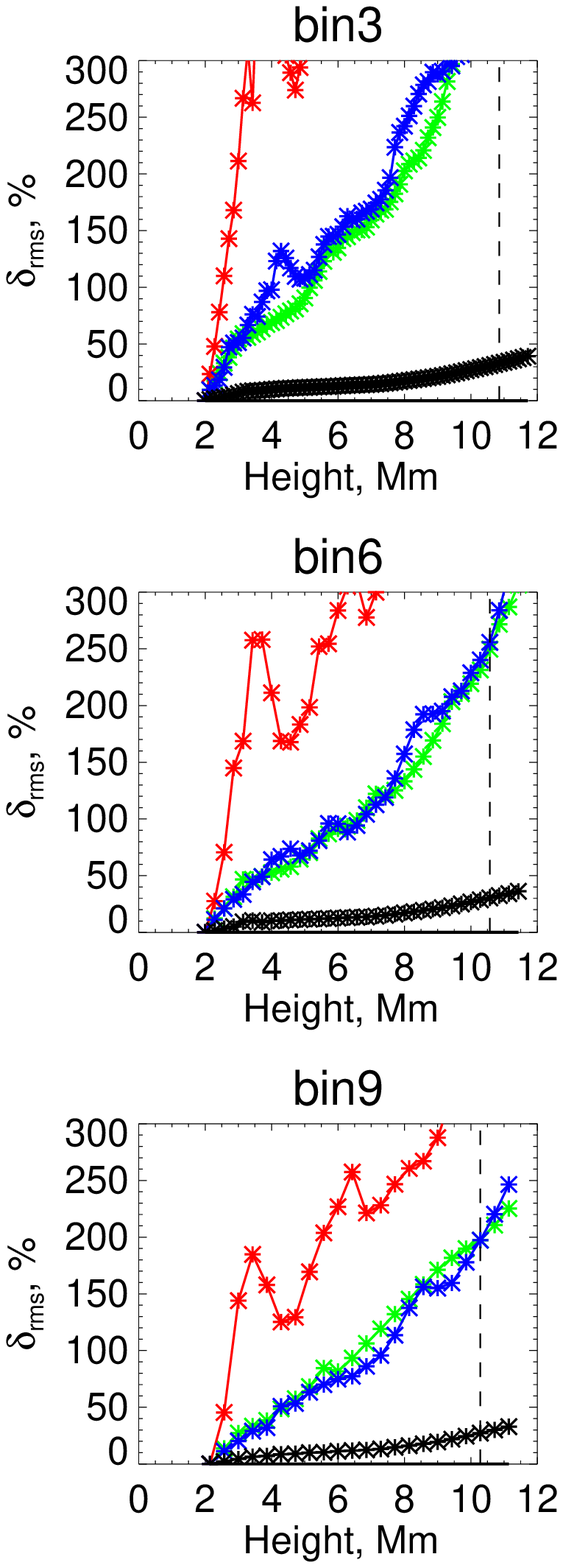}
\includegraphics[width=0.23\columnwidth]{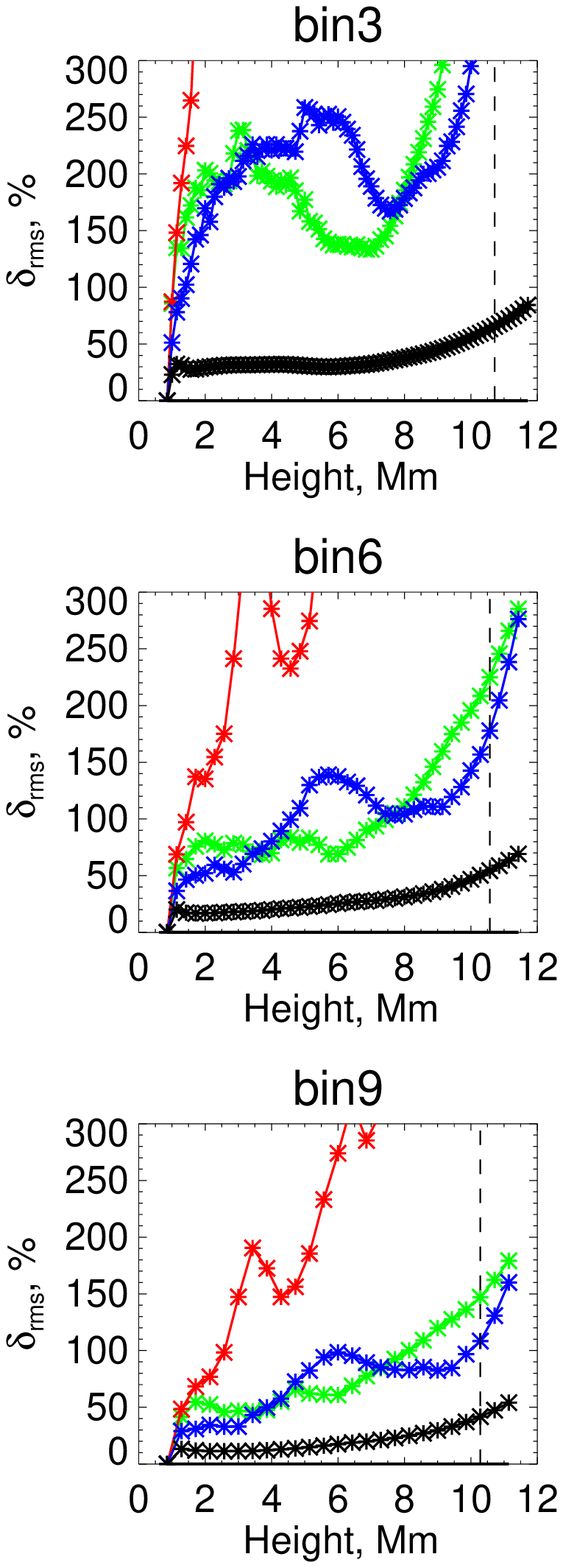}
\includegraphics[width=0.23\columnwidth]{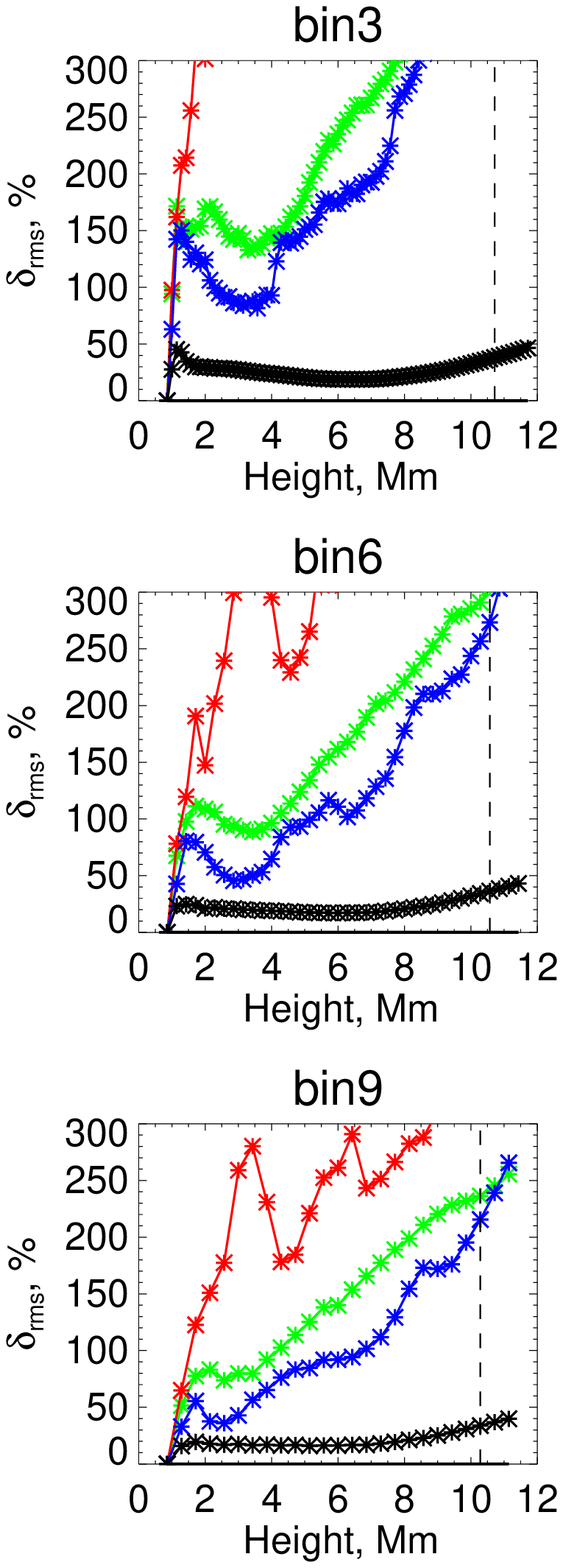}\\
\caption{\label{f_bifrost_385_NLFFF_sigmaVSheight}
Relative rms error in a layer as a function of height for the NLFFF reconstructions obtained using two methods, IM \& AS, from the chromosphere and the {$\beta$-}photosphere without preprocessing. The side buffer zones are discarded everywhere, while the height of the top buffer zone is shown by the dashed vertical line.
 }
\end{figure}

However, the reconstruction often provides the accuracy of 10---30\% (see Figures~\ref{f_bifrost_385_rebin4z_chrNLFFF_Bz},~\ref{f_bifrost_385_rebin4z_chrNLFFF_By}), which is fully acceptable for most of the practical applications, even though it is not that perfect as the accuracy of the original model field. To further quantify that,
Figure~\ref{f_bifrost_385_NLFFF_sigmaVSheight}a,b shows the height dependence for the normalized rms error of the absolute value and all components of the magnetic field for a representative set of three different binnings. 
A few observations can be made from this Figure. Firstly, the performance of both methods improves from small to large bin factors.   Given that Figures~\ref{f_bifrost_385_rebin4z_chrNLFFF_Bz} and \ref{f_bifrost_385_rebin4z_chrNLFFF_By} contain domains with the local error larger than 70\%, the rms error can be rather large, especially at higher heights.  Secondly, reconstruction of the $B_y$ component is not that good as for $B_x$ or $B_z$ components. This is an outcome of asymmetry of the original model, in which the $B_y$ values are systematically smaller than other components. Thirdly, the absolute value of the magnetic field is recovered much better than any single component of the field, which looks unexpected but as we show below this is an outcome of large errors coming from the weak field contributions. And finally, we find that the IM code works on average better than the AS code, while extrapolation is being made starting from the chromospheric magnetogram; see Table~\ref{table_rms_error} for the level-by-level comparison of the normalized rms error in the case of bin=9. We believe, this is a direct outcome of taking into account the equations for the force-free field boundaries of the modeling cube: having an almost force-free field bottom boundary conditions at the chromospheric level is better consistent with the force-free boundaries (IM) than with the buffer zone with the potential boundaries (AS).

\begin{figure}
\includegraphics[width=0.96\columnwidth]{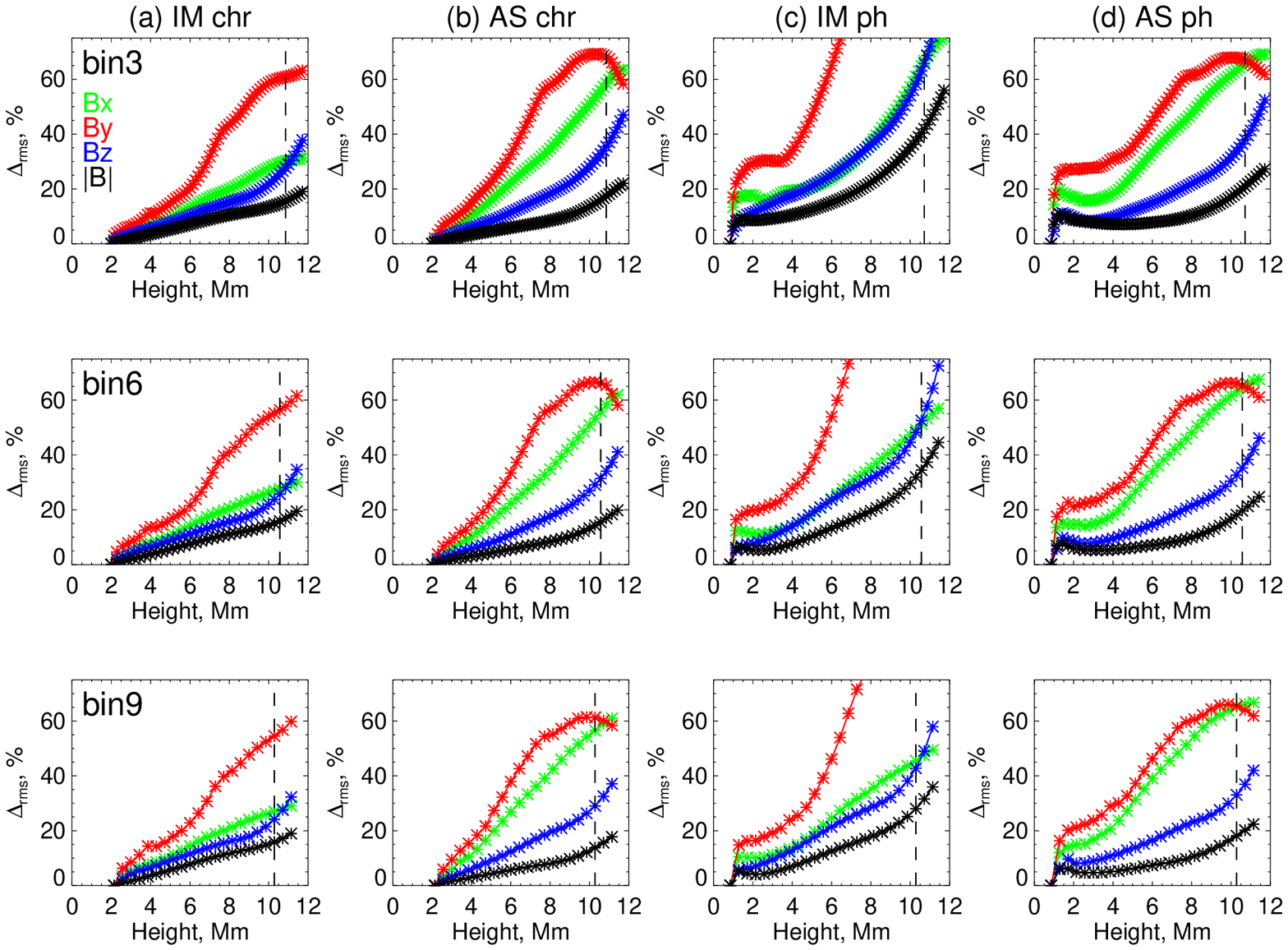}
\caption{\label{f_bifrost_385_NLFFF_deltaVSheight}
Relative rms residual in a layer as a function of height for the NLFFF reconstructions obtained using two methods, IM \& AS, from the chromosphere and the {$\beta$-}photosphere without preprocessing. The side buffer zones are discarded everywhere, while the height of the top buffer zone is shown by the dashed vertical line.
 }
\end{figure}

The absolute values of the normalized rms errors are rather large, which is the outcome of large errors in recovering small (close to zero) values of the field components. To explicitly demonstrate that, we turn to the normalized rms residual, which is weighted with the strong field contributions.
Figure~\ref{f_bifrost_385_NLFFF_deltaVSheight}a,b displays the height dependence of the normalized rms residuals for IM and AS codes respectively; see Table~\ref{table_delta_error} for the level-by-level comparison of the normalized rms residual in the case of bin=9. This plot and the Table confirm that the magnetic field is recovered more accurately in the voxels with the large magnetic field. In particular, now the absolute value of the field and its components are restored with a comparable accuracy (although the transverse components often display a bigger error than the longitudinal component). This metric only slightly depends on the grid resolution because the most numerous voxels with a weak field (which are more numerous for higher resolution data cubes) give only a weak contribution to this metric. The two methods recover the absolute value and $B_z$ component comparably well, but the IM code outperforms the AS one in recovering the transverse components (especially, $B_x$). In all cases the normalized rms residual for the absolute value $|B|$ and vertical component $B_z$ is within 20\% while for the transverse components---within 40\%.
Overall, we can conclude that these extrapolations from the force-free chromospheric level work rather well.

\subsubsection{NLFFF extrapolations from the {$\beta$-}photospheric boundary}

For the extrapolation from the $\beta$-photosphere boundary (without preprocessing) we computed the same metrics as for the chromospheric case. The angular metrics Table~\ref{IM_T02} show generally the same trends as for the chromosphere, but with larger values of the angles, which is expected. Again, the IM code often produces more force-free data cube than the model one. Figures~\ref{f_bifrost_385_NLFFF_2dh_Bz}---\ref{f_bifrost_385_NLFFF_2dh_By}c,d display the cross-correlation between the restored and model field in the volume above the $\beta$-photosphere (the buffer zone excluded). Although the scatter around the $y=x$ diagonal is larger than in the chromospheric case, the distribution follows the $y=x$ dependence remarkably well for all components of the magnetic field. This is highly important because this tells us that the NLFFF extrapolations, even when start from a non-force-free photospheric boundary, preserve the correct height scale, which is essential for all tasks that include the model-to-data comparison.

\begin{figure}[!t]\centering
\includegraphics[width=0.57\columnwidth]{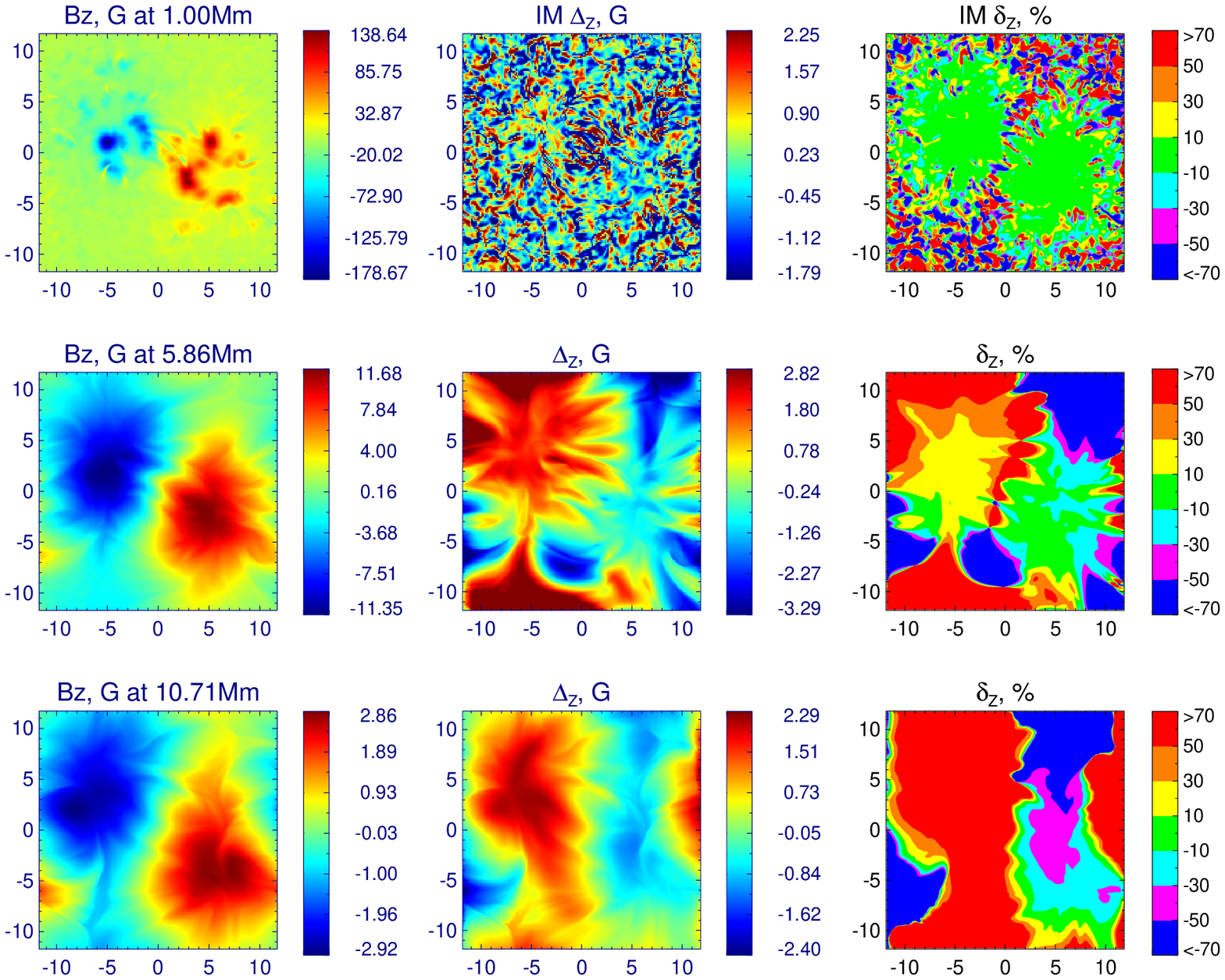} \  
\includegraphics[width=0.367\columnwidth, clip]{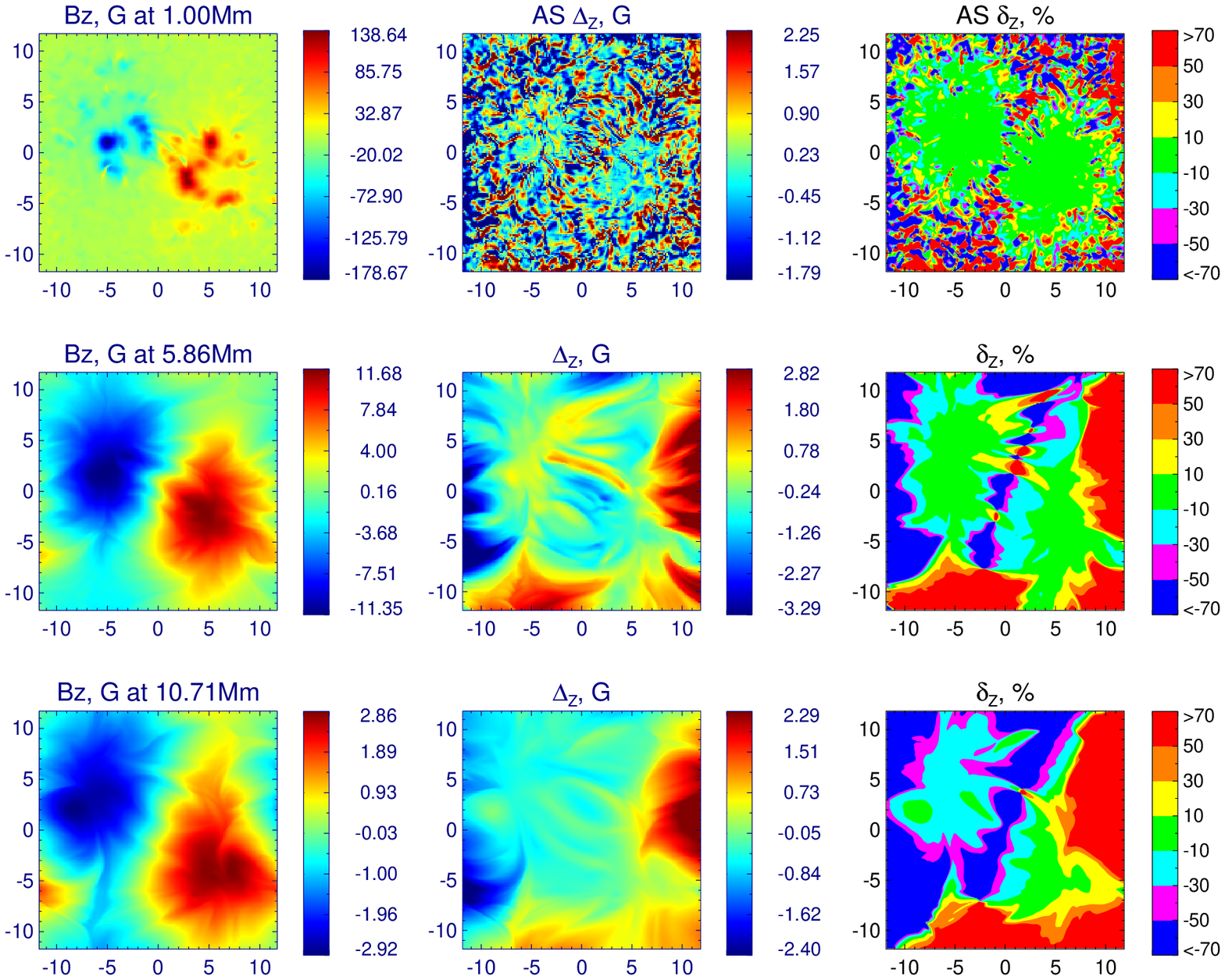}\\
\caption{\label{f_bifrost_385_rebin4z_phNLFFF_Bz} The model $B_z$ field distributions (left column) and performance of the IM (next two columns) and AS (two right columns) NLFFF extrapolations ($B_z$ component) from the {$\beta$-}photospheric level (bin=3) at three levels (their height are shown at the panel titles); {second and fourth columns: residual between the extrapolated and the model field; third and fifth column: relative error. The residual is within 2--3~G at all levels, while the relative error increases with height, because the field strength decreases with the height. The relative error is also bigger along the `neutral lines,' where the field is close to zero. The results for other components and other binning factors are similar to those shown in this figure. {The animated version of this figure shows the same information but for all layers of the reconstructed $B_z$ data cubes separately for the IM and AS extrapolations. Each frame of the animations shows four panels at a given layer: (a) the model field, (b) the restored field, (c) the residual, and (d) the relative error.}}
 }
\end{figure}

Not surprisingly, the normalized rms error and residual are bigger, while extrapolating from the photosphere compared with the chromospheric case. In particular, Figure~\ref{f_bifrost_385_NLFFF_deltaVSheight}c,d that the normalized rms residual metrics for the photospheric case are roughly by a factor of two-three larger that those for the chromospheric case. A qualitative difference between the former and the latter is that now we see a relatively big jump at the curves just above the bottom boundary. This jump is an immediate result from the conflict between the forced photospheric boundary, which is not allowed to change, and the implied force-free condition above this level: in fact, a few more layers above the $\beta$-photosphere are also not force-free; thus, the restored field noticeably deviates from the model one there.

Another interesting observation is that here the AS method works better than the IM one for the absolute value $|B|$ and the vertical component $B_z$, although not so well for the transverse components, especially---for $B_x$, while the $B_y$ component is recovered comparably imperfect by both methods. This mismatch in the accuracy of restoration of various components of the magnetic field is a likely cause of the large angle error in the AS extrapolation code. We conclude that the extrapolation from the ($\beta$-)photospheric level works acceptably well for many practical applications, even though measurably less perfect than the extrapolation from the chromospheric level;  see Tables~\ref{table_rms_error}~and~\ref{table_delta_error} for the level-by-level comparison of the normalized rms error and residual in the case of bin=9.

%
%
%
%
%
%

\subsubsection{NLFFF extrapolations from a preprocessed boundary}


Finally, we tested extrapolations from the photospheric level after its preprocessing; specifically, we used both preprocessed methods for both IM and AS extrapolation codes. Now, unlike the cases without preprocessing, we include the bottom (preprocessed) layer in the metrics computation given that it was modified compared to the original model field.  Table~\ref{IM_T02} shows that there is no coherent behavior of the metrics vs model grid resolution. Sometimes, but not always, the overall field is getting more force-free than without preprocessing (better $\theta$ metric), while less force-free in the sub-volume with a strong electric field (worse $\theta_j$ metric).

\begin{figure}
 \textbf{(a) IM chr } \ \ \textbf{(b) AS chr}  \ \ \textbf{(c) IM ph}  \ \  \textbf{(d) AS ph} \\
\centering
\includegraphics[width=0.258\columnwidth]{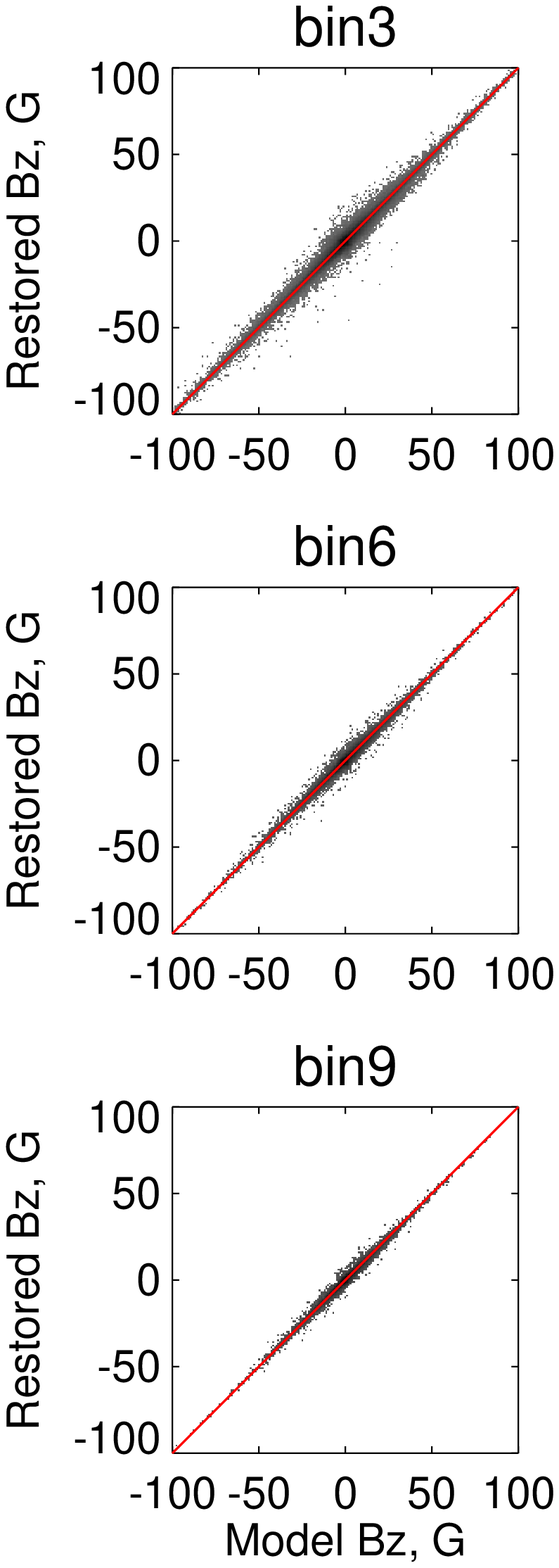}
\includegraphics[width=0.23\columnwidth, clip]{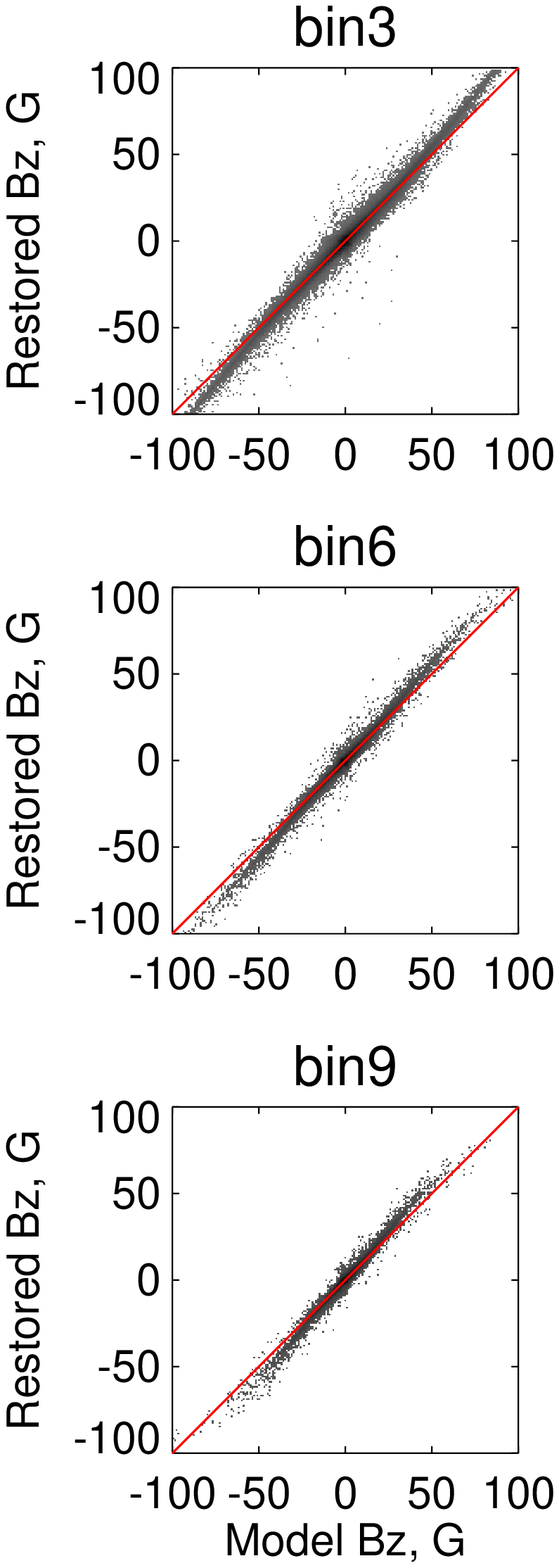}
\includegraphics[width=0.226\columnwidth, clip]{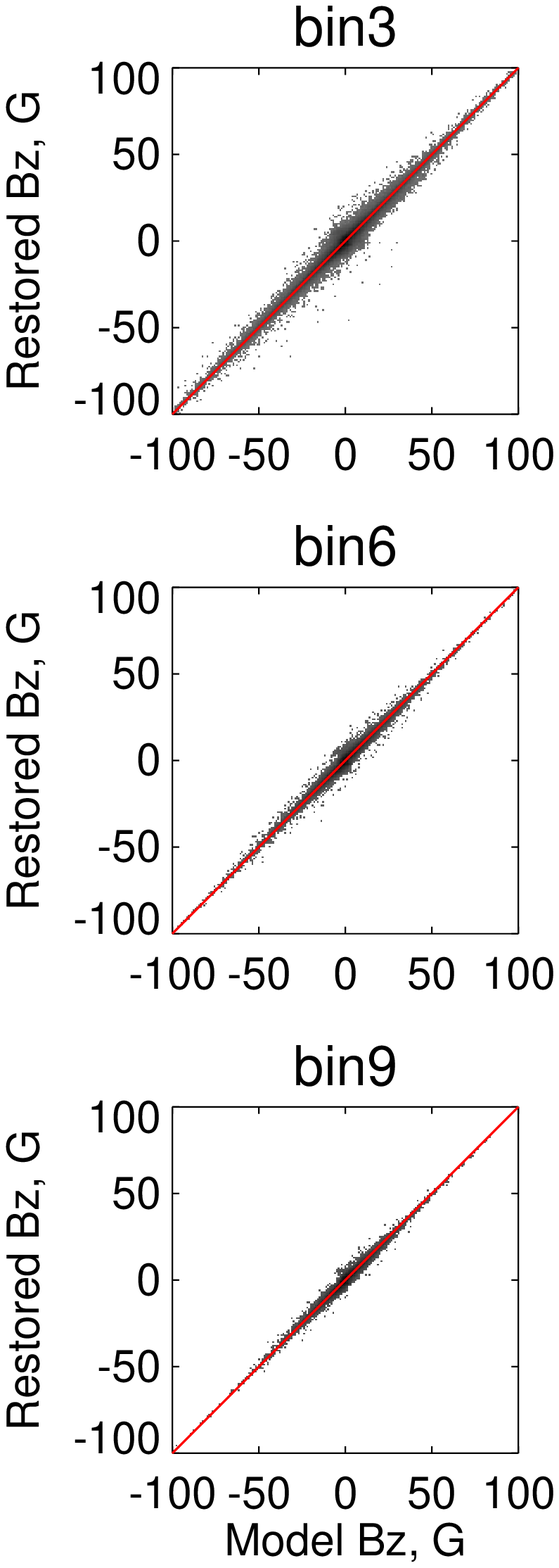}
\includegraphics[width=0.23\columnwidth, clip]{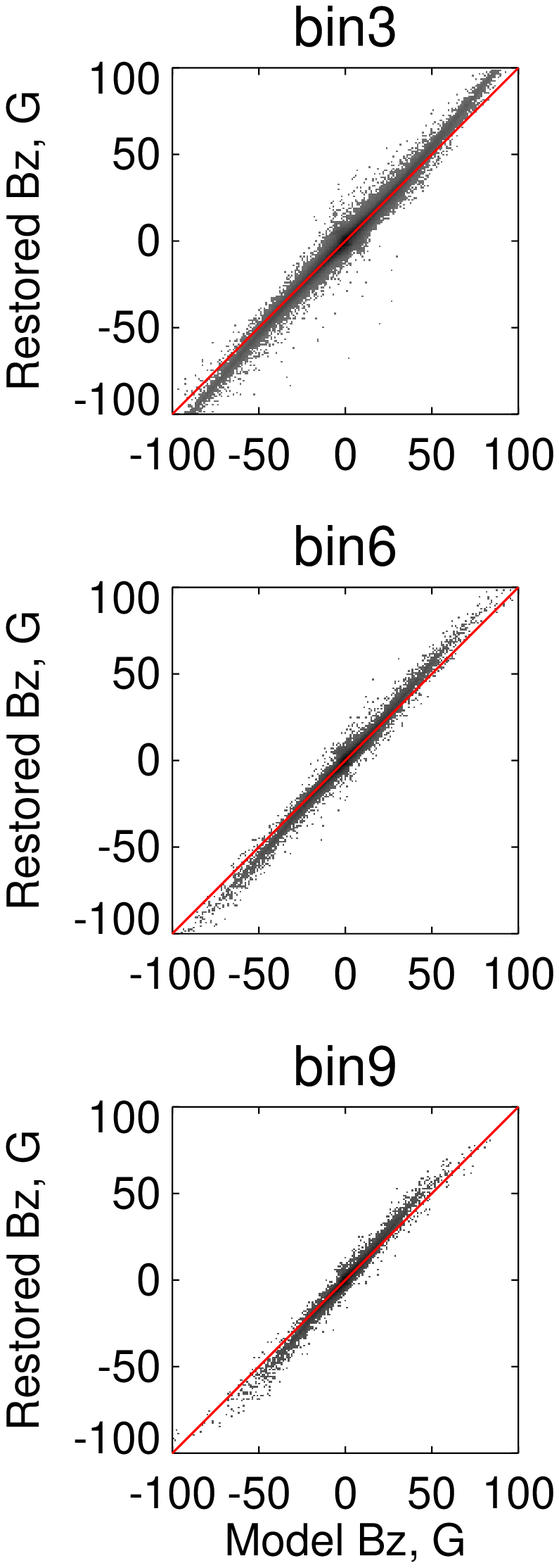}\\
\caption{\label{f_bifrost_385_NLFFF_2dh_prepr_Bz}
 2D histograms of $B_z$ reconstruction obtained using two methods, IM \& AS, from  the {$\beta$-}photosphere with a preprocessing---in 3D volume. The buffer zone is discarded everywhere.
 }
\end{figure}

\begin{figure}
 \textbf{(a) IM chr } \ \ \textbf{(b) AS chr}  \ \ \textbf{(c) IM ph}  \ \  \textbf{(d) AS ph} \\
\centering
\includegraphics[width=0.258\columnwidth]{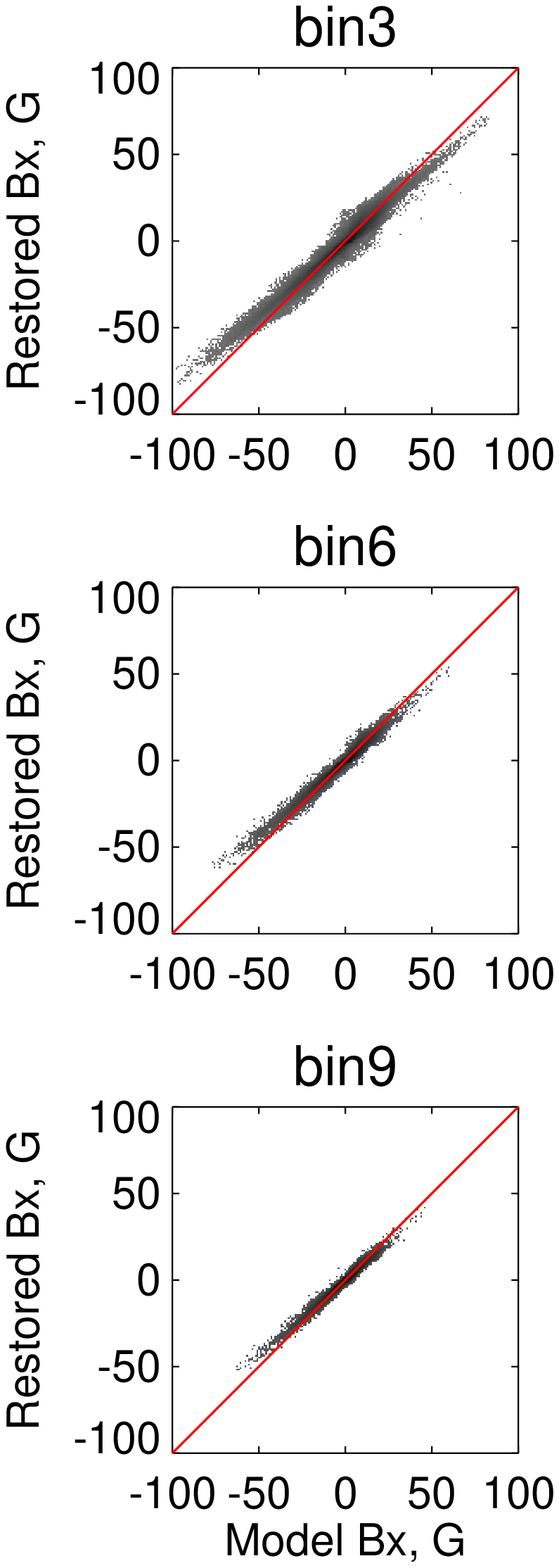}
\includegraphics[width=0.23\columnwidth, clip]{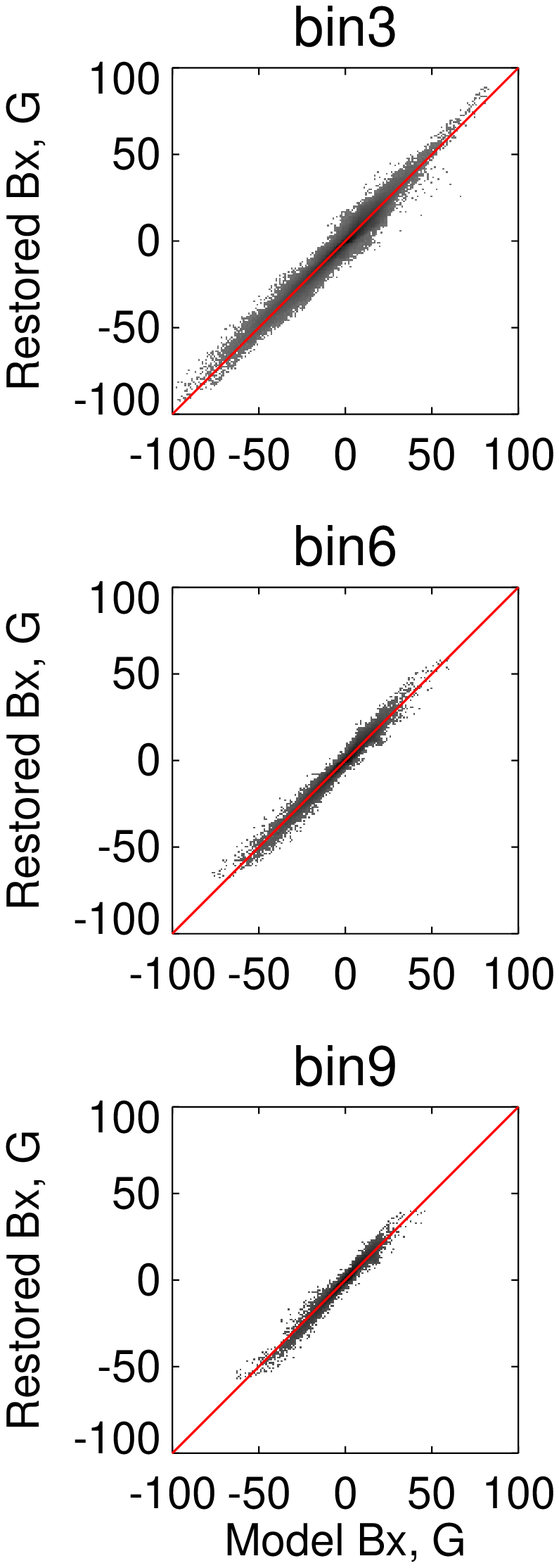}
\includegraphics[width=0.23\columnwidth, clip]{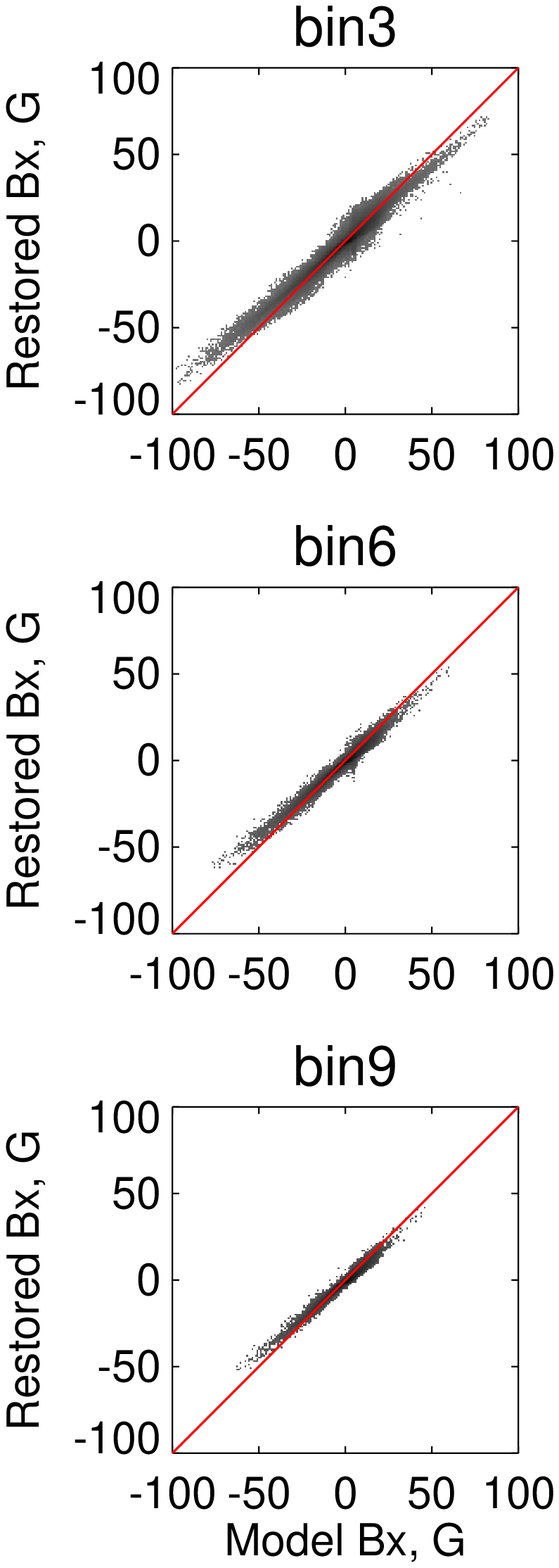}
\includegraphics[width=0.23\columnwidth, clip]{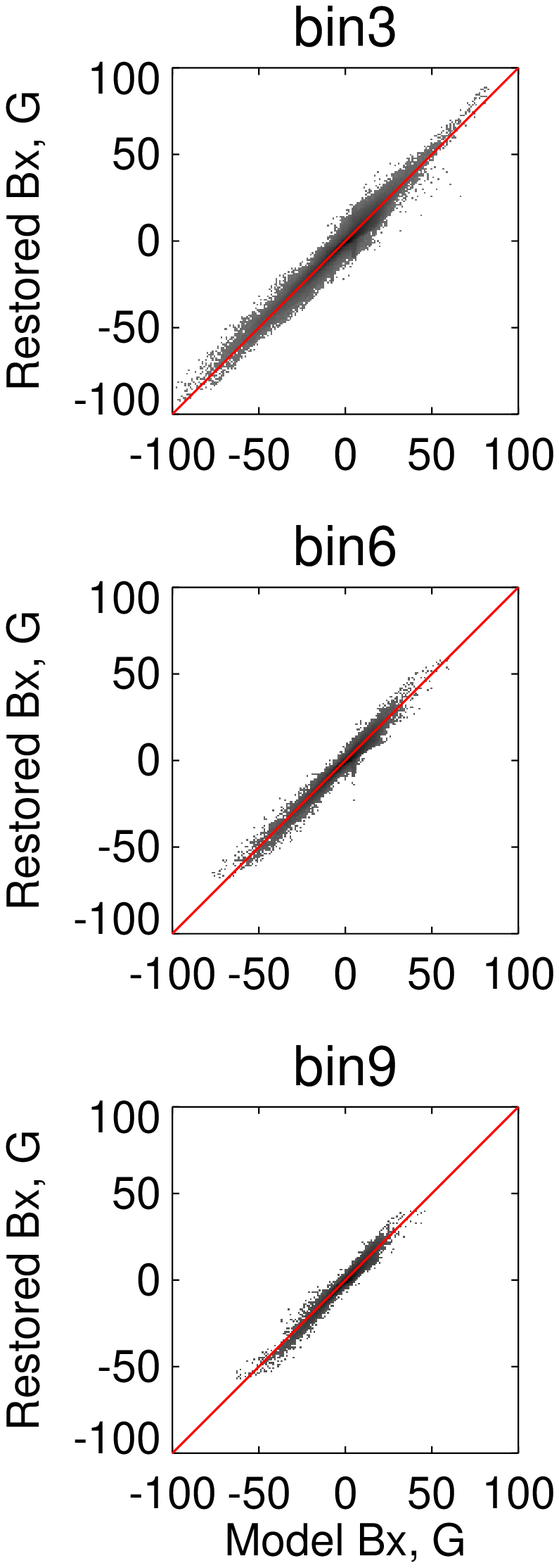}\\
\caption{\label{f_bifrost_385_NLFFF_2dh_prepr_Bx}
 2D histograms of $B_x$ reconstruction obtained using two methods, IM \& AS, from  the {$\beta$-}photosphere with a preprocessing---in 3D volume. The buffer zone is discarded everywhere.
 }
\end{figure}

\begin{figure}
  \textbf{(a) IM chr } \ \ \textbf{(b) AS chr}  \ \ \textbf{(c) IM ph}  \ \  \textbf{(d) AS ph} \\
\centering
\includegraphics[width=0.26\columnwidth]{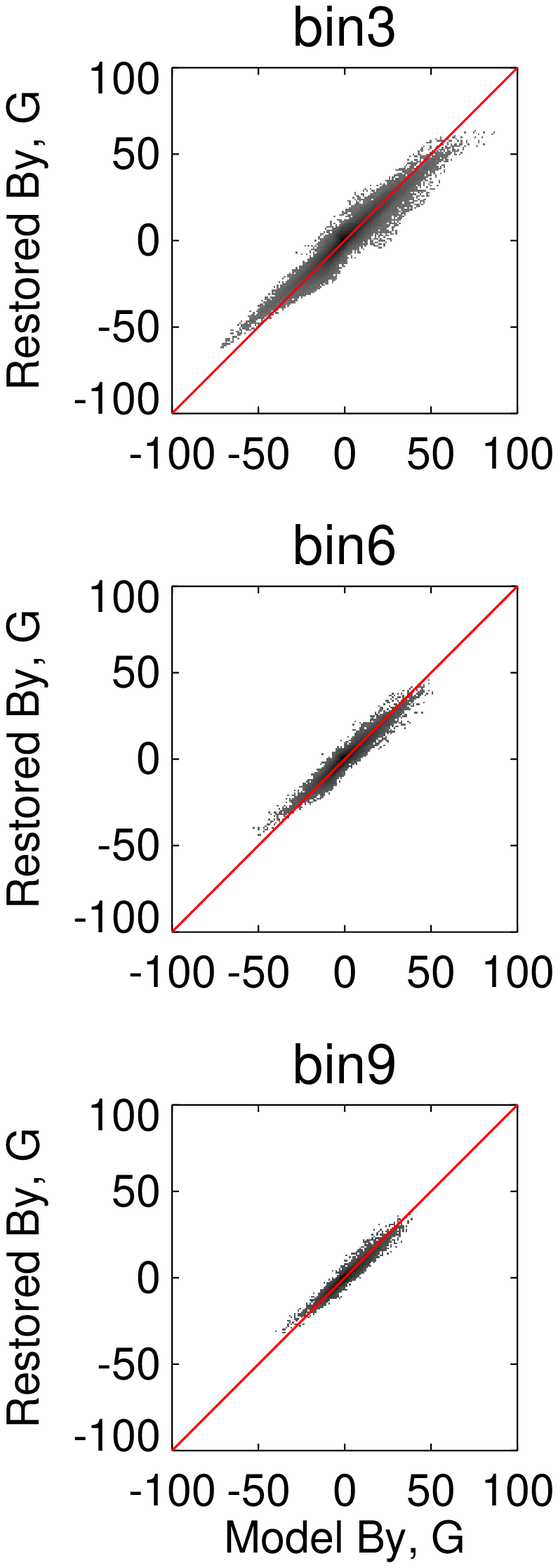}
\includegraphics[width=0.23\columnwidth, clip]{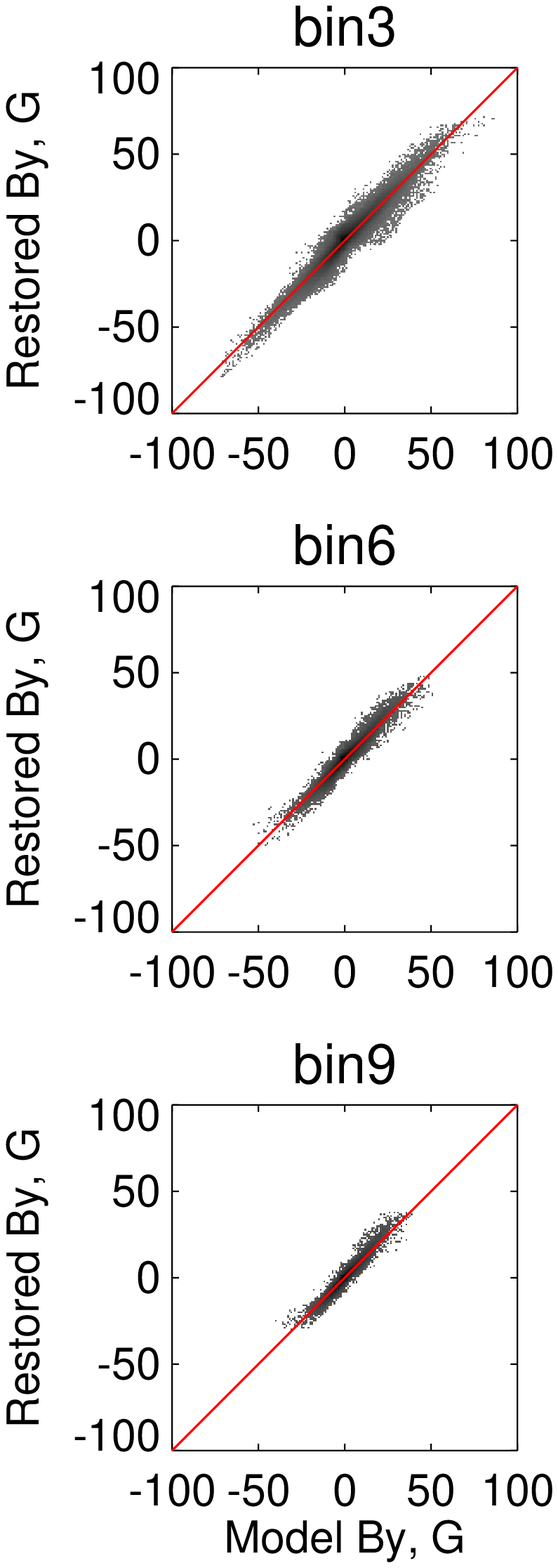}
\includegraphics[width=0.234\columnwidth, clip]{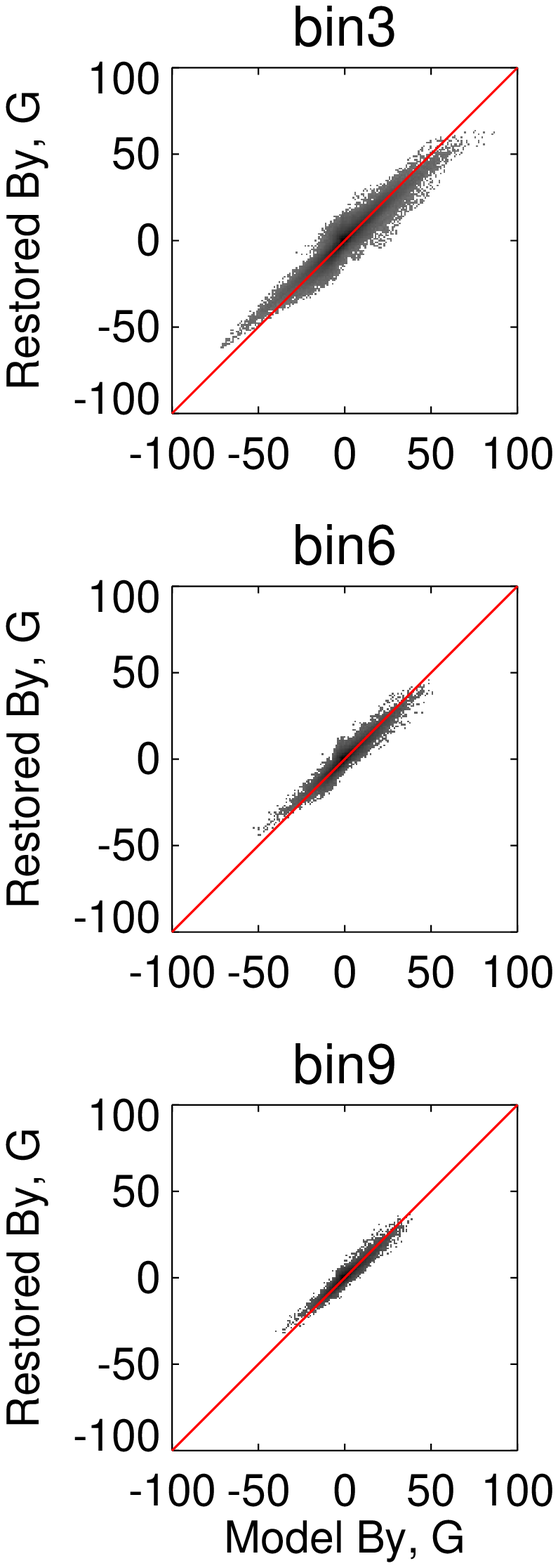}
\includegraphics[width=0.23\columnwidth, clip]{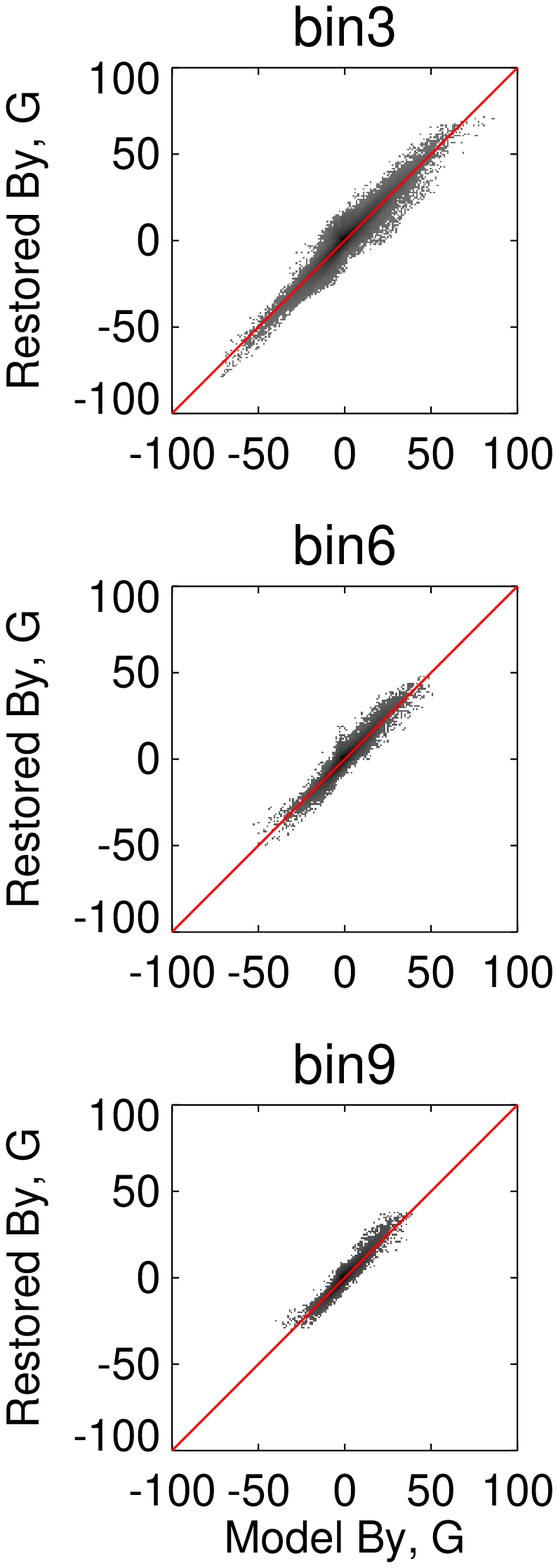}\\
\caption{\label{f_bifrost_385_NLFFF_2dh_prepr_By}
 2D histograms of $B_y$ reconstruction obtained using two methods, IM \& AS, from  the photosphere with a preprocessing---in 3D volume. The buffer zone is discarded everywhere.
 }
\end{figure}

\begin{figure}
\includegraphics[width=0.96\columnwidth]{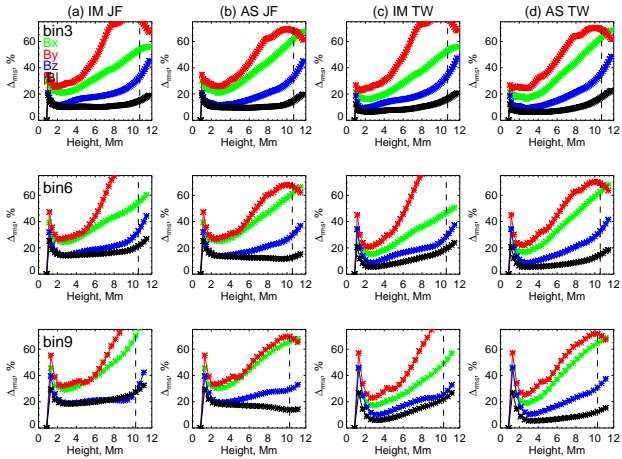}\\
\caption{\label{f_bifrost_385_NLFFF_deltaVSheight_prepr}
Relative rms residual in a layer as a function of height for the NLFFF reconstructions obtained using two methods, IM \& AS, from the photosphere with a preprocessing. The side buffer zones are discarded everywhere, while the height of the top buffer zone is shown by the dashed vertical line.
 }
\end{figure}

\begin{deluxetable*}{c|cccc|cccc|cccc|cccc}

\rotate
				
\tablecolumns{15}

\tablewidth{0pc}

\tabletypesize{\footnotesize}

\tablecaption{Normalized rms error after preprocessing at a given level for bin-factor 9. \label{table_rms_prepr_error}}

\tablehead{
\multicolumn{1}{c|}{} & \multicolumn{4}{c|}{$\Delta_{rms}(B)$}& \multicolumn{4}{c|}{$\Delta_{rms}(Bx)$} & \multicolumn{4}{c|}{$\Delta_{rms}(By)$} & \multicolumn{4}{c}{$\Delta_{rms}(Bz)$}\\
\multicolumn{1}{c|}{} &  \multicolumn{2}{c}{JF}&  \multicolumn{2}{c|}{TW}& \multicolumn{2}{c}{JF}& \multicolumn{2}{c|}{TW} & \multicolumn{2}{c}{JF} & \multicolumn{2}{c|}{TW} & \multicolumn{2}{c}{JF} & \multicolumn{2}{c}{TW}\\
				\multicolumn{1}{c|}{Level, Mm} & \colhead{IM} & \multicolumn{1}{c}{AS} & \colhead{IM} & \multicolumn{1}{c|}{AS}  & \colhead{IM} & \multicolumn{1}{c}{AS} & \colhead{IM} & \multicolumn{1}{c|}{AS} & \colhead{IM} & \multicolumn{1}{c}{AS} & \colhead{IM} & \multicolumn{1}{c|}{AS} & \colhead{IM} & \multicolumn{1}{c}{AS} & \colhead{IM} & \multicolumn{1}{c}{AS}
				}

\startdata
     1.29 &    11.74 &    11.74 &    18.52 &    18.52 &    49.39 &    49.39 &    59.91 &    59.91 &    53.99 &    53.99 &    62.80 &    62.80 &    43.53 &    43.53 &    56.00 &    56.00 \\
     1.71 &    10.20 &    10.62 &    15.88 &    16.19 &    57.55 &    60.68 &    63.15 &    67.46 &    58.04 &    75.16 &    67.31 &    81.91 &    42.61 &    40.57 &    49.65 &    47.24 \\
     2.14 &     9.26 &    11.65 &    14.66 &    17.41 &    57.71 &    63.22 &    66.38 &    72.56 &    63.11 &   112.86 &    71.07 &   128.83 &    40.51 &    42.27 &    47.20 &    53.71 \\
     2.57 &     8.86 &    12.26 &    13.99 &    18.46 &    55.86 &    62.28 &    68.84 &    71.48 &    66.53 &   157.49 &    76.49 &   174.11 &    37.28 &    42.30 &    43.73 &    52.01 \\
     3.00 &     8.69 &    12.49 &    13.30 &    18.31 &    59.76 &    64.16 &    73.87 &    67.36 &    86.77 &   219.79 &    99.08 &   244.61 &    36.86 &    43.26 &    45.01 &    45.72 \\
     3.43 &     8.79 &    12.83 &    12.71 &    18.49 &    63.75 &    67.47 &    79.79 &    71.74 &   115.00 &   264.09 &   123.13 &   295.16 &    42.79 &    58.30 &    54.18 &    57.00 \\
     3.86 &     9.24 &    12.91 &    12.32 &    18.34 &    68.76 &    76.11 &    87.46 &    74.34 &   119.52 &   209.52 &   125.65 &   238.25 &    44.34 &    58.69 &    57.04 &    56.78 \\
     4.29 &     9.81 &    13.12 &    11.95 &    18.28 &    77.17 &    86.33 &   101.38 &    84.83 &   119.39 &   166.58 &   125.54 &   192.68 &    48.39 &    78.24 &    59.27 &    78.25 \\
     4.71 &    10.40 &    13.17 &    11.42 &    18.37 &    92.92 &    98.20 &   125.24 &    95.05 &   120.45 &   166.99 &   128.14 &   193.75 &    51.69 &    75.26 &    57.69 &    71.12 \\
     5.14 &    11.06 &    13.13 &    10.92 &    18.17 &    86.43 &   105.57 &   119.66 &   101.92 &   136.51 &   201.81 &   147.75 &   237.17 &    58.54 &    82.58 &    58.61 &    84.14 \\
     5.57 &    11.84 &    13.23 &    10.69 &    18.36 &    82.65 &   117.65 &   116.60 &   112.03 &   173.13 &   229.95 &   178.85 &   268.30 &    70.34 &    86.36 &    64.69 &    89.18 \\
     6.00 &    12.71 &    13.46 &    10.75 &    18.53 &    76.01 &   120.13 &   109.49 &   115.10 &   206.64 &   241.49 &   205.23 &   285.48 &    76.77 &    88.02 &    69.53 &    92.58 \\
     6.43 &    13.74 &    13.87 &    11.15 &    19.07 &    78.13 &   133.96 &   106.72 &   129.77 &   243.42 &   268.11 &   245.12 &   312.53 &    77.84 &    89.20 &    71.81 &    95.07 \\
     6.86 &    14.67 &    14.01 &    11.54 &    19.34 &    84.33 &   146.48 &   104.80 &   141.80 &   210.74 &   223.94 &   204.64 &   263.12 &    78.33 &    97.22 &    74.96 &   103.99 \\
     7.29 &    15.73 &    14.64 &    12.25 &    20.48 &    87.35 &   158.28 &   100.91 &   154.88 &   225.33 &   229.43 &   217.68 &   266.84 &    79.19 &   105.43 &    74.85 &   113.87 \\
     7.71 &    16.96 &    15.73 &    13.23 &    21.98 &    92.34 &   169.14 &   108.47 &   166.54 &   254.39 &   243.45 &   251.71 &   284.09 &    86.63 &   122.39 &    77.93 &   132.37 \\
     8.14 &    18.36 &    16.89 &    14.46 &    23.76 &    95.86 &   179.78 &   111.08 &   179.02 &   274.37 &   257.53 &   280.08 &   299.48 &   101.12 &   144.44 &    84.03 &   157.78 \\
     8.57 &    19.88 &    18.29 &    15.87 &    25.53 &   100.29 &   190.13 &   113.97 &   189.64 &   287.52 &   262.93 &   300.73 &   305.69 &   115.77 &   162.48 &    90.98 &   177.24 \\
     9.00 &    21.59 &    19.83 &    17.45 &    27.68 &   106.53 &   200.27 &   121.84 &   200.94 &   306.10 &   283.20 &   335.13 &   327.94 &   116.86 &   160.11 &    89.20 &   175.69 \\
     9.43 &    23.76 &    21.92 &    19.41 &    30.11 &   112.49 &   206.55 &   129.09 &   208.11 &   328.16 &   318.05 &   374.39 &   367.12 &   123.33 &   163.78 &    90.70 &   179.46 \\
     9.86 &    26.38 &    24.33 &    21.73 &    33.08 &   119.47 &   211.15 &   135.16 &   215.00 &   325.05 &   331.21 &   380.54 &   379.92 &   143.46 &   180.35 &   104.73 &   198.94 \\
    10.29 &    29.27 &    26.70 &    24.18 &    35.67 &   127.75 &   215.23 &   143.72 &   220.97 &   303.74 &   306.72 &   356.55 &   350.54 &   168.32 &   198.96 &   123.37 &   219.53 \\
    10.71 &    32.69 &    29.16 &    27.12 &    38.61 &   136.22 &   224.39 &   164.62 &   233.03 &   321.87 &   317.83 &   380.88 &   361.37 &   204.36 &   219.73 &   157.59 &   243.95 \\
    11.14 &    36.57 &    31.54 &    30.78 &    41.03 &   150.26 &   233.82 &   196.81 &   246.27 &   330.97 &   336.39 &   402.63 &   380.18 &   248.93 &   243.80 &   202.33 &   270.82 \\
         	\\
			\enddata
		\end{deluxetable*}

\begin{deluxetable*}{c|cccc|cccc|cccc|cccc}

\rotate
				
\tablecolumns{15}

\tablewidth{0pc}

\tabletypesize{\footnotesize}

\tablecaption{Normalized rms residual after preprocessing at a given level for bin-factor 9. \label{table_delta_prepr_error}}

\tablehead{
\multicolumn{1}{c|}{} & \multicolumn{4}{c|}{$\Delta_{rms}(B)$}& \multicolumn{4}{c|}{$\Delta_{rms}(Bx)$} & \multicolumn{4}{c|}{$\Delta_{rms}(By)$} & \multicolumn{4}{c}{$\Delta_{rms}(Bz)$}\\
\multicolumn{1}{c|}{} &  \multicolumn{2}{c}{JF}&  \multicolumn{2}{c|}{TW}& \multicolumn{2}{c}{JF}& \multicolumn{2}{c|}{TW} & \multicolumn{2}{c}{JF} & \multicolumn{2}{c|}{TW} & \multicolumn{2}{c}{JF} & \multicolumn{2}{c}{TW}\\
				\multicolumn{1}{c|}{Level, Mm} & \colhead{IM} & \multicolumn{1}{c}{AS} & \colhead{IM} & \multicolumn{1}{c|}{AS}  & \colhead{IM} & \multicolumn{1}{c}{AS} & \colhead{IM} & \multicolumn{1}{c|}{AS} & \colhead{IM} & \multicolumn{1}{c}{AS} & \colhead{IM} & \multicolumn{1}{c|}{AS} & \colhead{IM} & \multicolumn{1}{c}{AS} & \colhead{IM} & \multicolumn{1}{c}{AS}
				}

\startdata
   0.86 &      0.00 &      0.00 &      0.00 & 0.00 &      0.00 &
0.00 &      0.00 & 0.00 &      0.00 &      0.00 &      0.00 & 0.00 &
0.00 &     0.00 &     0.00 &     0.00 \\
      1.29 &    29.29 &    29.29 &    26.56 & 26.56 &    45.63 &
45.63 &    45.25 &    45.25 &    55.17 &    55.17 &    56.91 &    56.91
&    39.50 &    39.50 &    45.92 &    45.92 \\
      1.71 &    21.60 &    22.08 &    14.34 & 14.73 &    32.32 &
33.14 &    25.63 &    26.07 &    38.78 &    39.05 &    33.85 &    34.23
&    25.75 &    26.57 &    24.67 &    25.13 \\
      2.14 &    19.45 &    19.37 &     8.92 & 8.78 &    31.00 &    31.60
&    20.03 &    20.51 &    32.92 &    33.22 &    25.70 &    26.96 &
21.46 &    21.44 &    15.89 &    15.80 \\
      2.57 &    18.64 &    18.57 &     6.59 & 6.49 &    29.58 &    30.49
&    17.55 &    18.96 &    31.95 &    33.25 &    22.89 &    25.30 &
19.50 &    19.95 &    11.61 &    11.93 \\
      3.00 &    18.40 &    18.08 &     5.91 & 5.60 &    29.15 &    30.36
&    17.13 &    19.01 &    32.66 &    34.19 &    23.21 &    26.27 &
18.94 &    19.39 &    10.23 &    10.45 \\
      3.43 &    18.38 &    17.94 &     5.92 & 5.54 &    29.74 &    31.01
&    17.96 &    20.47 &    33.52 &    35.64 &    24.69 &    28.71 &
18.89 &    19.62 &    10.14 &    10.72 \\
      3.86 &    18.50 &    17.72 &     6.26 & 5.54 &    30.72 &    32.31
&    19.16 &    22.23 &    34.65 &    37.27 &    27.35 &    31.74 &
19.15 &    19.89 &    10.83 &    11.20 \\
      4.29 &    18.68 &    17.64 &     6.78 & 5.72 &    32.15 &    33.68
&    20.42 &    23.73 &    35.77 &    39.20 &    30.40 &    35.03 &
19.44 &    20.50 &    11.71 &    12.29 \\
      4.71 &    18.88 &    17.42 &     7.40 & 5.86 &    33.71 &    36.13
&    21.42 &    26.79 &    34.55 &    38.45 &    29.45 &    35.50 &
19.88 &    21.01 &    12.88 &    13.22 \\
      5.14 &    19.16 &    17.31 &     8.22 & 6.16 &    35.22 &    38.30
&    22.60 &    29.64 &    35.13 &    39.57 &    30.17 &    37.11 &
20.43 &    21.86 &    14.23 &    14.61 \\
      5.57 &    19.46 &    17.13 &     9.16 & 6.45 &    37.10 &    41.45
&    24.16 &    33.71 &    37.05 &    42.99 &    32.86 &    41.66 &
20.84 &    22.48 &    15.50 &    15.61 \\
      6.00 &    19.77 &    16.99 &    10.16 & 6.78 &    38.85 &    43.93
&    25.62 &    36.87 &    40.38 &    46.37 &    36.50 &    45.82 &
21.09 &    23.30 &    16.69 &    16.84 \\
      6.43 &    20.16 &    16.82 &    11.28 & 7.12 &    40.78 &    46.74
&    27.34 &    40.45 &    44.86 &    51.05 &    41.18 &    51.15 &
21.23 &    24.01 &    17.90 &    17.94 \\
      6.86 &    20.66 &    16.73 &    12.50 & 7.51 &    43.30 &    49.32
&    29.58 &    43.44 &    49.93 &    53.92 &    46.11 &    54.96 &
21.29 &    24.96 &    19.12 &    19.35 \\
      7.29 &    21.18 &    16.53 &    13.69 & 7.86 &    46.06 &    52.16
&    32.05 &    46.76 &    55.65 &    56.38 &    51.60 &    58.07 &
21.27 &    25.80 &    20.30 &    20.73 \\
      7.71 &    21.71 &    16.31 &    14.82 & 8.27 &    48.53 &    54.39
&    34.25 &    49.37 &    62.31 &    59.58 &    58.46 &    61.74 &
21.14 &    26.66 &    21.31 &    22.17 \\
      8.14 &    22.28 &    15.96 &    15.92 & 8.68 &    50.98 &    56.81
&    36.37 &    52.40 &    67.37 &    61.75 &    63.85 &    63.70 &
21.04 &    27.27 &    22.19 &    23.48 \\
      8.57 &    22.83 &    15.56 &    16.93 & 9.17 &    53.37 &    58.99
&    38.39 &    55.26 &    72.32 &    63.43 &    68.75 &    65.06 &
20.92 &    27.76 &    22.75 &    24.70 \\
      9.00 &    23.43 &    14.99 &    17.90 & 9.71 &    56.15 &    61.07
&    40.59 &    58.11 &    78.44 &    66.49 &    74.81 &    67.90 &
20.79 &    27.76 &    22.88 &    25.63 \\
      9.43 &    24.26 &    14.50 &    18.96 & 10.40 &    59.99 &
62.86 &    43.22 &    60.53 &    83.43 &    68.34 &    79.91 &    70.10
&    21.38 &    27.94 &    23.14 &    26.82 \\
      9.86 &    25.53 &    14.05 &    20.28 & 11.35 &    64.48 &
64.39 &    45.93 &    62.70 &    87.81 &    69.48 &    84.18 &    71.38
&    23.46 &    28.34 &    24.00 &    28.56 \\
     10.29 &    27.35 &    13.90 &    21.94 & 12.46 &    69.77 &
65.58 &    48.90 &    64.60 &    91.45 &    69.47 &    87.28 &    71.56
&    27.39 &    29.31 &    25.68 &    30.80 \\
     10.71 &    29.65 &    13.84 &    23.96 & 13.88 &    76.15 &
66.75 &    52.43 &    66.50 &    93.72 &    67.87 &    88.76 &    69.95
&    33.51 &    30.68 &    28.42 &    33.85 \\
     11.14 &    32.36 &    14.19 &    26.68 & 15.25 &    83.72 &
67.72 &    56.88 &    68.36 &    95.79 &    65.31 &    89.83 &    67.26
&    42.24 &    32.93 &    32.76 &    37.38 \\
     	\\
			\enddata
		\end{deluxetable*}

For the IM code the use of any  preprocessing improves agreement between the restored and model field for smaller binning factors (higher grid resolution): the TW preprocessing works better for the bins 3 and 4, while the JF one works better for larger bin factors, where the extrapolation with the preprocessings shows only marginal improvement (if any) compared with the extrapolation without preprocessing.

However, the prize we pay for this marginally improved angular metrics is the corrupted height scale, which is vividly demonstrated by Figures~\ref{f_bifrost_385_NLFFF_2dh_prepr_Bz}---\ref{f_bifrost_385_NLFFF_2dh_prepr_By}. Specifically, the magnetic field extrapolated from the JF preprocessed boundary condition shows a nice $y=x$ pattern for the $B_z$ component, while deviates from that for both $B_x$ and $B_y$ components, which underestimate the magnetic field in the volume. In contrast, the use of the TW preprocessed boundary condition overestimates the magnetic field components, but not equally: the $B_z$ component is clearly stronger overestimated than the transverse ones, which almost follow the $y=x$ regression in some cases.

Figure~\ref{f_bifrost_385_NLFFF_deltaVSheight_prepr} reveals one more problem with the field reconstruction starting from a preprocessed boundary: the field restoration error is large starting from the very bottom layer with the normalized rms residual exceeding 10--15\% in many cases, especially for the case of TW preprocessing. We believe that this happens because the magnetic field becomes overly smooth due to the preprocessing (this smoothing is stronger for the TW preprocessing, see Section~\ref{S_prepro} and, specifically, Fig.~\ref{fig:prep_compare}{; the amount of smoothing can be reduced by lowering the $\mu_4$ parameter; however, it is well beyond our study to optimize the choice of the preprocessing parameters}). At somewhat higher layers, where the field in the model is smooth itself, the metric decreases and, thus, the field is more accurately reconstructed at the intermediate heights compared with the low height; see Tables~\ref{table_rms_prepr_error}~and~\ref{table_delta_prepr_error} for the quantitative level-by-level comparison of the normalized rms error and residual in the case of bin=9 reconstruction from the preprocessed boundaries.
Overall, the whole variety of the tests does not justify the use of preprocessing of the photospheric boundary given that a marginal or no improvement in the volumetric metrics is achieved at the expense of the height scale corruption.

%
%
%
%
%
%



\section{Discussion \& Conclusions}

Here we have demonstrated that a realistic MHD model, in the presented case---a \textit{en024048{\_}hion} simulation \citep{2016A&A...585A...4C} obtained with the Bifrost code \citep{2011A&A...531A.154G}, can be very efficiently used to cast various tools used for the coronal magnetic field reconstruction. In particular, we have evaluated the performance of the $\pi$-disambiguation codes, magnetogram preprocessing codes, and NLFFF extrapolation codes developed following the optimization method \citep{2000ApJ...540.1150W}.

We have found that the currently used $\pi$-disambiguation codes work pretty well at the AR photosphere and chromosphere, but often fail at the quiet sun photosphere. This can become important when the question of the magnetic field at the quiet sun is specifically addressed. Here we are primarily interested in the performance of the reconstruction tools in ARs; thus, we adopted that the $\pi$-ambiguity has been perfectly resolved.

Then, we have assessed the performance of two different preprocessing approaches aimed to improve the bottom boundary condition toward force-freeness. Although the tested preprocessing codes do produce a more force-free boundary, there is an unsolicited byproduct of these preprocessings---the poorly controlled elevation of the magnetic field components, which are different for the longitudinal and transverse field components. This mismatch results in a poorly controlled systematic error in the height scale in the extrapolated data cube.

On the other hand, comparison between the volumetric metrics of the magnetic data cubes extrapolated from the photospheric level either with or without preprocessing are not much different from each other, while extrapolation without preprocessing preserves the correct height scale. From this perspective, we conclude that the use of NLFFF extrapolation from the actual photospheric magnetogram (without any preprocessing{, at least, when noise in the data does not represent a problem}) is preferable. {We have to note, however, that this conclusion is based only on the tests performed with the `standard' parameters $\mu_i$, $i=1...4$, specified for the two alternative preprocessing method. It is well possible that there are combinations of the parameters $\mu_i$, which ensure a consistent elevation of the magnetic vector components by exactly one voxel (or another integer number of voxels) along with suppressing noise in the data, which might be helpful. But searching for such optimized combinations is clearly beyond the scope of this study.}

Finally, we compared to each other the results of NLFFF extrapolation using two different versions of the optimization method---one that uses the weighting function and the other one that employs the full set of NLFFF equations (including the one for the top and side boundaries). Although we found the metrics of the two codes to be comparable, still the codes that utilizes the full set of the NLFFF equations works systematically better than the one with the weighting function. We believe that this is because the use of the full set of equations is better consistent with the assumption of the field force-freeness than the presence of the boundary buffer zone with the potential boundary conditions at the top and side boundaries.

%

\acknowledgments

This work was supported in part by NSF grants  AGS-1250374, AGS-1262772,  and AST-1312802,
and NASA grants NNX14AK66G, and NNX14AC87G to New Jersey
Institute of Technology and RFBR  grants 15-02-01077, 15-02-01089, 15-02-03717, 15-02-03835,  15-02-08028, 16-02-00254, 16-02-00749, and 16-32-00315.
This study was supported
by the Program of basic research of the RAS Presidium No. 9. Authors
acknowledge the Marie Curie PIRSES-GA-2011-295272 RadioSun project.

\bibliographystyle{apj}
\bibliography{NLFFF}

\end{document}